\documentclass[%
aip,
reprint,%
]{revtex4-1}
\pdfoutput=1
\usepackage{graphicx}     
\usepackage{amsmath}
\usepackage{esint}
\usepackage{tocbasic}
\usepackage[T1]{fontenc}
\usepackage{mathptmx}
\usepackage{amssymb, amsfonts}
\usepackage{array}
\usepackage{braket}
\usepackage{cases}
\usepackage[utf8]{inputenc}
\usepackage[overload]{empheq}
\usepackage{subfig} 
\usepackage{float}
\usepackage{floatflt}     
\usepackage{relsize}  
\usepackage{multirow}
\usepackage{dcolumn}  

\newcolumntype{P}[1]{>{\hspace{0pt}}p{#1}}
\newcolumntype{d}[1]{D{.}{.}{#1}}
\newcolumntype{K}[1]{>{\centering\arraybackslash}p{#1}}
\newcommand{\DS}{\displaystyle}
\usepackage{floatflt}     
\usepackage{wrapfig}
\usepackage{color}
\usepackage{pstricks}
\usepackage{hyperref}
\hypersetup{
	colorlinks   = true, 
	urlcolor     = blue, 
	linkcolor    = blue, 
	citecolor    = blue 
}

\makeatletter
\newcommand*{\rom}[1]{\expandafter\@slowromancap\romannumeral #1@}
\usepackage{color}
\definecolor{DarkRed}{rgb}{0.35,0.01,0.01}
\definecolor{Linen}{rgb}{0.98,0.98,0.94}
\definecolor{Blue}{rgb}{0.,0.,1.0}
\definecolor{DarkBlue}{rgb}{0.099,0.099,0.44}
\definecolor{DarkGreen}{rgb}{0.0,0.4,0.0}
\definecolor{Turquoise}{rgb}{0.0,0.9,0.7}
\begin{document}

\title{Electromagnetic calculations for multiscale and multiphysics simulations: a new perspective}

\author{Dung N. Pham}
\email{dnpham@wpi.edu}
\affiliation{Department  of   Physics,  Worcester
	Polytechnic Institute, Worcester, MA 01609, USA.}
\affiliation{Center for Computational NanoScience,  Worcester
	Polytechnic Institute, Worcester, MA 01609, USA.}
\author{Sathwik Bharadwaj}
\email{sathwik@wpi.edu}
\affiliation{Department  of   Physics,  Worcester %
  Polytechnic Institute, Worcester, MA 01609, USA.}
\affiliation{Center for Computational NanoScience,  Worcester
	Polytechnic Institute, Worcester, MA 01609, USA.}
\author{L. R. Ram-Mohan}
\email{lrram@wpi.edu}
\affiliation{Department  of   Physics,  Worcester
	Polytechnic Institute, Worcester, MA 01609, USA.}
\affiliation{Center for Computational NanoScience,  Worcester
	Polytechnic Institute, Worcester, MA 01609, USA.}
\affiliation{Department of Electrical and Computer
	Engineering, Worcester
	Polytechnic Institute, Worcester, MA 01609, USA.}
\affiliation{Department of Mechanical Engineering, Worcester
	Polytechnic Institute, Worcester, MA 01609, USA.}

\pacs{
	84.40.Az, 
	03.50.De, 41.20.-q,  
	43.20.Mv, 
	42.82.Et, 
	41.20.Jb, 84.40.-x, 
	02.70.Dh, 
	42.55.Tv, 
	78.67.Pt
}

\keywords{cavity resonators, photonic crystals, waveguides, multiscale modeling,
	Hermite interpolation polynomials, finite element method, 
	accidental degeneracy
}

\begin{abstract}
  Present   day   electromagnetic   field  calculations   have   basic
  limitations  since they employ  techniques based on  edge-based
  discretization methods.   While these vector finite  element methods
  (VFEM) solve the issues of tangential continuity of fields and the
  removal of  spurious solutions, the  resulting fields do not  have a
  unique directionality at nodes in the discretization mesh.
  This review  presents three calculations of  electromagnetic fields:
  ($i$)  waveguides,  ($ii$)  cavity   fields,  and  ($iii$)  photonic
  crystals.  We  have developed Hermite interpolation  polynomials and
  node-based finite  element methods  in the framework  of variational
  principles.  We  show that the Hermite-finite  element method (HFEM)
  better accuracy  with lower  computational cost, and  provides field
  directional continuity across the discretized space. It also permits
  multiscale  calculations  with  mixed physics.  For example,  the
  nanoscale modeling  of quantum  well semiconductor  laser structures
  has to  be done together  with cavity electrodynamics at  the micron
  scale  in   vertical-cavity  surface   emitting  lasers   and  other
  optoelectronic  systems.  These  can be  treated with  high accuracy
  with the new HFEM, which is also applicable to quantum mechanical
  simulations in meso- and nano-scale systems.
  We use group representation theory  to derive the
  HFEM polynomial  basis set in  two dimensions. In three dimensions
  we derive these polynomials using usual methods.
  We  show that  degeneracies in  the  frequency spectrum  in a  cubic
  cavity  can be  denumerably large  even though  the symmetry  of the
  cube,  $O_h$, supports  only  singlets, doublets,  or triplets.  The
  additional operators  available for the problem  explains the origin
  of  this  ``accidental  degeneracy.''  We  discuss  this  remarkable
  degeneracy and its reduction in detail.
  We  consider photonic  crystals corresponding  to a  2D checkerboard
  superlattice structure,  and the Escher drawing  of ``The Horsemen''
  which satisfies the  nonsymmorphic group $pg$. We show  that HFEM is
  able to  deliver high  accuracy in  such spatially  complex examples
  with far less computational effort than Fourier expansion methods.
  Finite element  analysis employs geometric discretization  and hence
  transcends geometrical limitations. Techniques explained here can be
  immediately   extended  to   realistic  and   geometrically  complex
  structures. The  new algorithms developed  here hold the  promise of
  successful modeling  of multi-physics systems.  This  general method
  is  applicable  to a  broad  class  of  physical systems,  e.g.,  to
  semiconducting  lasers   which  require  simultaneous   modeling  of
  transitions  in  quantum  wells  or dots  together  with  EM  cavity
  calculations, to modeling plasmonic structures in the presence of EM
  field  emissions,  and  to  on-chip  propagation  within  monolithic
  integrated circuits.
\end{abstract}

\maketitle
\begin{center}
	\today
\end{center}
{\protect\NoHyper\tableofcontents
\makeatletter
\def\@pnumwidth{3em}
\def\@tocrmarg {3.5em} 
\let\toc@pre\relax
\let\toc@post\relax
\makeatother}
\section{Introduction}\label{sec:Intro}
\renewcommand{\thefootnote}{\fnsymbol{footnote}}
\setcounter{footnote}{1}

  The idea  of
guiding  electromagnetic  waves  along  conducting rods  has  been  of
interest  for  more  than  a  century.\cite{avs:sommerfeld1899}
During the 1940s, the need to  design radars for detection of aircraft
and  warships provided  a new impetus  for the  analysis of  waveguides.
Schwinger\cite{avs:schwinger1968,avs:schwinger2006}  was 
responsible  for the  theoretical framework  for designing  waveguides
with complex shapes  and embedded dielectrics.   With the
advent of digital computation in the past 60 years, it became
possible   to  consider   problems  with   waveguide  geometries   and
characteristics  that had  no  closed-form  analytical solutions.   In
particular,   the   numerical   simulation    of   the   behavior   of
electromagnetic  waves at  the  microwave  frequencies attracted  much
interest from their use in  transmitters and receivers of radio waves.
High-accuracy calculations of  electromagnetic
fields through geometry discretization  has been the focus of intense
investigation  over this period.  The approach  offers detailed  real-space
information appropriate  for the analysis  of fields not only  in open
domains but  in closed regions  of complicated structure, such  as the
dielectric, magnetic  and semiconducting  layers encountered  in radio
frequency  integrated circuits.  The impact  of predictive  capability
with this approach in electromagnetic simulation has been profound for
problems ranging  from antenna  design to on-chip  signal propagation.
Typically,  the  on-chip  electromagnetic   solvers  are  invoked  for
modeling  ports  and  guiding   structures,  the  device  solvers  are
parameterized models, and the passives are lumped elements. While this
design philosophy  has been very  successful at lower  frequencies and
for large devices, the  eventual addition of full-wave electromagnetic
solvers to  fully integrated  chip layout  and device  design modeling
tools is a foregone conclusion.

\begin{figure}[b!] 
	\begin{center}
		\includegraphics[width=2.5in]{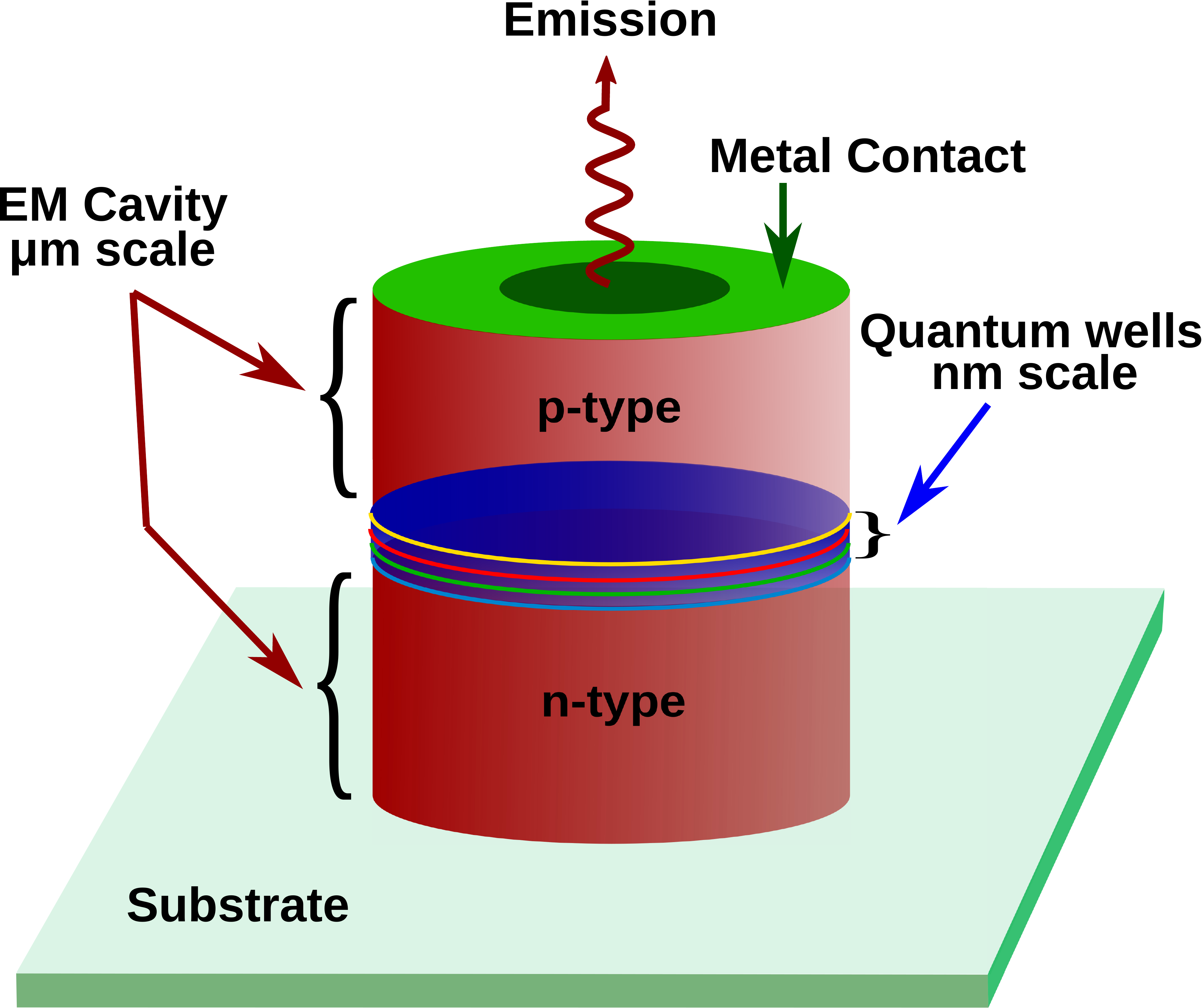}\\ \vspace{0.2in}
	\end{center}
	\vspace{-0.2in}
	\caption{A schematic of a vertical cavity surface emitting
          laser (VCSEL) is shown. The quantum well is sandwiched two semiconducting
          cylinders. The quantum well region has the geometry at the
          nanometer scale, while the photonic cavity is at the micron scale.}
	\label{avsfig:multiscale_physics}
	\vspace{-0.2in}
\end{figure}

This leads to the critical issue of obtaining field calculations that
provide the spatial resolution adequate for multi-scale problems, for
example, field propagation into and out of active electronic devices
that are much smaller than the wavelengths of the propagating signals
they manipulate. For RF circuits operating at sub-millimeter wave
frequencies, the designs of transitions and interactions of signals at
the transistor level are intensive engineering exercises to obtain the
critical matching conditions that make useful circuits possible at
these demanding frequencies.\cite{avs:LeongDeal} Given the geometric
complexities and the material loss components that are relevant with
increasing frequency above 300\,GHz, most design cycles require
extensive back-fitting to measurement and redesign. This iterative
process is very expensive in time and fabrication costs compared with
design methods at lower frequencies where the full-wave modeling can
be restricted to port transitions and guiding structures of minimal
loss. Therefore, a solver methodology that can obtain high spatial
resolution without poor computational scaling and discretization error
behaviors is a key element of integrating full-wave electromagnetics
when small (10's of nanometers) active devices interact with metals and
dielectrics that are far from perfect conductors or lossless. 
Figure~\ref{avsfig:multiscale_physics} shows a schematic of a vertical
cavity surface emitting laser (VCSEL) as  an example of a
multiscale application in which the active quantum well region is of
nanometer scale while the photonic cavity is at the micron scale.

In this review, we also consider cavity electrodynamics, level
degeneracies in a cubic cavity, and their identification. We show that
the electrodynamic cubic cavity exhibits ``accidental degeneracy.'' We
explore the dynamic interaction between symmetry, level degeneracy,
and its removal.

Finally, we demonstrate the efficacy of HFEM in the modeling of
photonic crystals (PC). We consider a checkerboard superlattice wires and
a beautiful application to Escher's drawing, ``The Horsemen,'' treated
as a photonic crystal having a nonsymmorphic symmetry group $pg$.

\section{Earlier approaches}
\noindent{\bf The Vector finite element method}

The most favorable trade-off from a
numerical standpoint has been to sacrifice polynomial degrees of
freedom for the representation of either the normal or tangential
fields in exchange for compliance with Nedelec conditions. The
mainstay of this approach is the use of Nedelec-compliant vector
finite elements (VFEM).
\cite{avs:Nedelec1980,avs:SunCendes1995,avs:paulsen1991B,avs:JFLee1991b,avs:Peterson1994,avs:webb1995}
The most widely implemented element basis functions share a common
attribute that tangential field boundary conditions are explicit at
triangle boundaries and that normal field representations are one
polynomial order less than that of the tangential fields. These
conditions and the mixed-order field representations are well-documented 
and effort has been made to suppress potentially unphysical solutions
through the VFEM basis choices. \cite{avs:Peterson1994}

The disadvantage of mixed-order  polynomial approaches is the inherent
imbalance in discretization error  with increasing mesh density. While
every  discretization method  introduces  errors  with arbitrary  mesh
scaling, $h$-convergence  is achieved in  a well behaved  finite element
calculation, and  it would  be expected  that increasing  mesh density
where solutions change  rapidly should provide much  better real space
functions until extremely dense  conditions prevail. However, in mixed
order elements,  mesh refinement  toward a  dense grid  of equilateral
triangles  (in 2D)  can  decrease  the overall  quality  of the  field
representation  in polynomials  because  increasing  portions of  real
space are described by lower  order polynomials since field components
with projections normal to the triangle boundaries occur ubiquitously.

\begin{figure}[t!] 
 \begin{center}
 \includegraphics[width=2.5in]{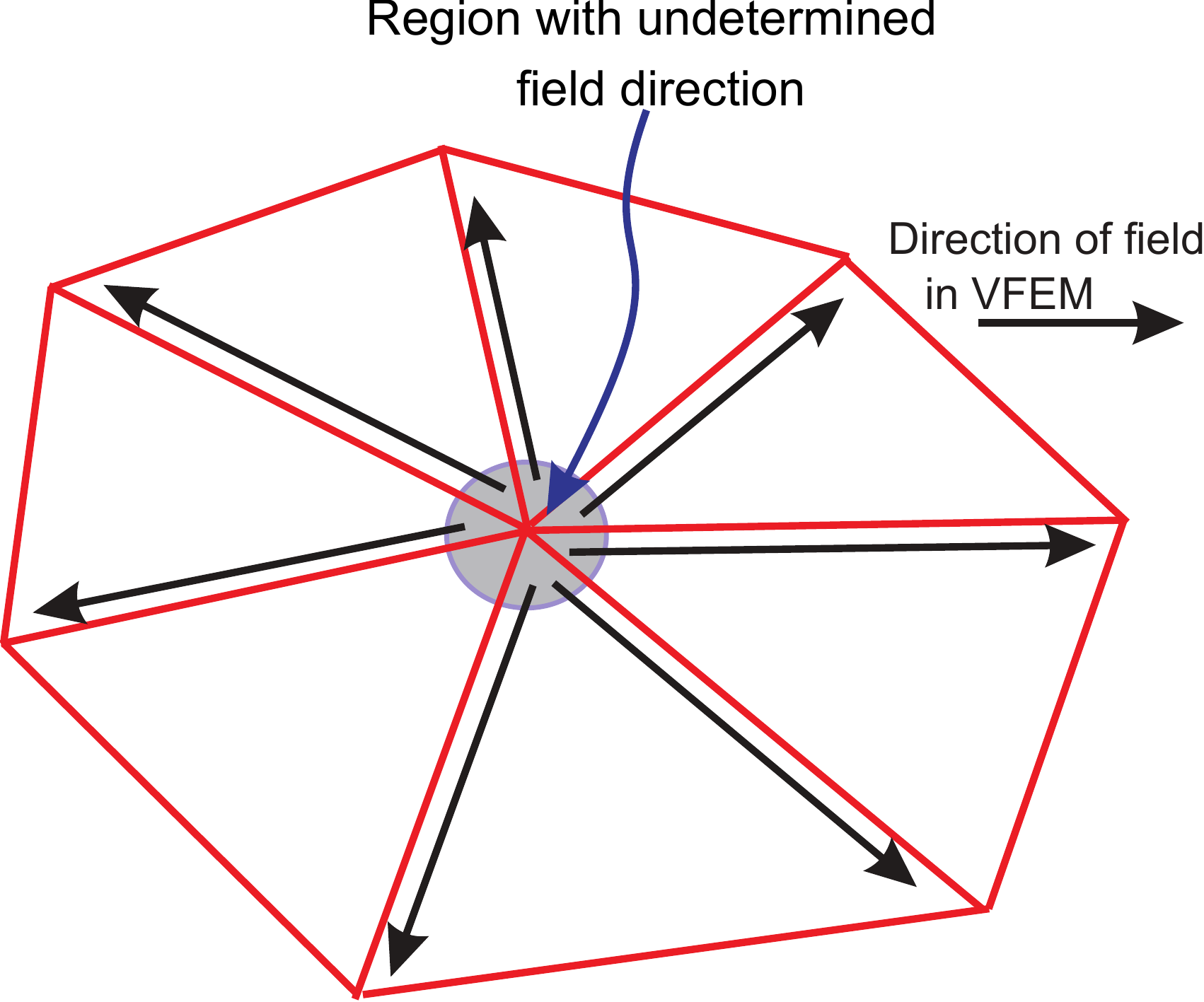}\\ \vspace{0.2in}
\end{center}
   \vspace{-0.2in}
\caption{We show the vector fields at vertices in the edge-elements
  in VFEM. These are
  directed along the edges, leading to a region of ill-defined
  direction for the field at each vertex. Thus the more the number of
  vertices the larger is the region of poorly defined fields.
   }
   \label{avsfig:VFEMvertices1}
\end{figure}

In the  limit of  a very  dense mesh,  the entire  solution can  be no
better  than the  lowest order  description because  the inter-element
boundary regions dominate over  the vanishing triangle interior.  What
is  worse is  that the  as-written basis  functions produce  ambiguous
vector fields  at triangle  vertices in the  sense that  approaching a
point  in space  that is  shared as  a common  corner node  of several
triangles produces fields  that are unique to  each triangle, creating
an  overdetermined basis  representation for  the fields  at vertices.
(See  Fig.~\ref{avsfig:VFEMvertices1}.)  This  occurs  because the  basis
decomposes the field there as projections orthogonal to adjacent edges
that  are  not  shared  among  all the  triangles  sharing  the  node.
Therefore, an extremely dense mesh results in at best a constant value
description  if additional  numerical techniques  are not  deployed to
mitigate the  problem of multiple  definitions of the vector  field at
shared vertices. In  addition, when such fields  calculated using VFEM
are  employed in  further applications,  such as  determining electron
trajectories  in accelerators  or in  high power  vacuum tube  design,
these  regions around  vertices  inject uncertainties  in the  charged
particle trajectories.

Given   the   inherent   side   effects  of   vector   basis   element
discretization, it  is not surprising  that most  of the work  on VFEM
solution enhancement  and refinement  has focused on  the construction of
higher order polynomial basis functions to increase spatial resolution
rather  than  dense  meshing ($p$-convergence).   Hierarchical approaches  are  used  in
practice      and      have      been     published      in      great
detail.\cite{avs:AndersonVolakis,avs:Webb1999}  In
practice,  the  difficulty 
with this route  to better spatial resolution is  that mesh refinement
or  the discretization  of  disparately sized  physical regions  often
requires the mating of finite elements with different basis orders.

The enormous advantages to finding a nodal basis that scales well with
meshing refinement  make this  an attractive  topic to  re-engage. The
spectral pollution  that arises from  nodal based basis  functions has
been  analyzed  elegantly  from  a  mathematical  point  of  view  for
traditional choices of polynomials.\cite{avs:Boffi}

The present article analyzes, for comparable degrees of freedom (DOF),
the impact of choosing an  alternative nodal basis formed from Hermite
interpolation  polynomials {\it  versus}   the use  of vector  finite
elements. The advantages  of the Hermite finite  element method (HFEM)
are shown for  canonical waveguide problems and  compared to published
treatments.   The nodal  representation with  function and  derivative
continuity (the core  of the Hermite approach)  results in unambiguous
field  descriptions at  shared nodes  among triangles  and treats  the
field  components with  uniform polynomial  degree (not  mixed-order.)
The built-in access to derivative quantities should offer ready access
to     quantities     that      are     useful     for     sensitivity
analysis.\cite{avs:Webb2001} This  treatment results  in global
functions  without  severe  coarsening  which  will  be  suitable  for
analyzing interactions  with small  features, such  as those  found in
high  frequency  transistor  circuits.   In  an  on-chip  calculation,
including transitions to coplanar  waveguide and micro-strip waveguide
structures,  some thin  film  layers  will set  a  mesh  size that  is
incompatible with  the overall  structure size and  propagating signal
wavelength. At  present, these  treatments in circuit  design software
are       dominated       by       finite-difference       time-domain
methods.\cite{avs:Zheng2010}

\noindent{\bf The boundary element method}

For external  regions,  boundary integral
techniques have  been combined with finite element  methods to capture
the  asymptotic  behavior   in  open  systems.\cite{avs:Tuncer2010}  While
offering  surface  vs.   volume  discretization  advantages,  boundary
integral formulations are self-consistency  relations and do not enjoy
the advantages of variational quadratic convergence. The use of dyadic
Green's  functions  in  EM  involves  derivatives  of  these  singular
functions  at the boundaries  making the  evaluation of  the integrals
more complex.  Dimensional continuation techniques applied
for the evaluation of  such hypersingular integrals will substantially
enhance  the computational  accuracy.\cite{avs:LRR_Dimcontinuation} However, the  matrices generated in this
method are dense, and are typically nonlinear in the eigenfrequencies.

\noindent{\bf Other methods}

Finally,   meshless  methods\cite{avs:Shankar2004,avs:SladekPan2013,avs:LiuGuBook}
have been  used to treat a  number of physics  problems, including EM,
and are at the opposite  extreme in terms of numerical representations
to the method  given in this paper where we develop  a higher order of
derivative  continuity  in spatial  basis.   Meshless  schemes can  be
interpreted in many  ways, but often  involve the use of  basis functions
that  are not  restricted to  a finite  volume.  The  resulting matrix
formulations  have the  advantage  of being  fully block-diagonal  but
require a separate treatment of continuity boundary conditions. In these
situations the boundary conditions  are enforced as a constraint which
can  be a penalty  method or  numerical flux  minimization. Derivative
information  is rarely  treated when  forming eigenfunctions  in these
methods.

\section{The source of spurious solutions}
EM fields in  a  physical system should satisfy the wave equation
along with a  divergence-free condition. For example,  in the electric
field  formulation  (E-field),  we   solve  Maxwell's equations
\begin{align}
\nabla \times \nabla \times \mathbf{E} &= \epsilon \, \mu \, \omega^2
\, \mathbf{E}; \label{eq:wave1}  \\ 
\label{eq:div}
\nabla \cdot \mathbf{\epsilon E} &= 0,
\end{align}
where $\omega$ is the eigenfrequency  and $\epsilon$ and $\mu$ are the
permittivity and permeability of  the material.  Computationally if we
simply attempt  to solve the  wave equation Eq.   (\ref{eq:wave1}), we
are   not  guaranteed   that  the   obtained  solutions   satisfy  the
divergence-free  condition  of   Eq.~(\ref{eq:div}).   Solutions  with
either  zero frequency  ($\omega  = 0$)  or  with non-zero  divergence
($\nabla \cdot  \epsilon \mathbf{E} \neq  0$) that are  obtained while
solving Eq.~(\ref{eq:wave1})  are known  as the  spurious (unphysical)
solutions.   Such  spurious   solutions  also   corrupt  the   desired
eigenspectrum.

The   vector  basis   functions  can   be  constructed   in  VFEM   to
systematically eliminate  spurious solutions by casting  them into the
null             space             of             the             curl
operator.\cite{avs:SunCendes1995,avs:Peterson1994,avs:Webb1988}   These  solutions
correspond  to  the  zero-frequency   (static)  solutions  in  the  EM
problems.  For matrix dimensions of $10^3$ in typical EM calculations,
nearly 20-30\%  of the solutions belong  to this class and  are thrown
away,  being  unusable  solutions.\cite{avs:Peterson1994}
Carrying  this 
overhead in the calculation  is computationally expensive when scaling
to  sophisticated structures.

In large scale  computations where the method  of domain decomposition
is often employed, all solutions for each sub-domain are calculated so
that the solutions for the entire domain can be constructed from those
of the sub-domains.  The zero-frequency modes of a sub-domain are also
needed  as  Fourier  components  in  order  to  construct  the  global
solutions.   However,  it  is   not  simple  to  discriminate  between
acceptable zero frequency solutions and the pollution of the nullspace
with spurious solutions.

\section{The proposed Hermite finite element method}
We propose the use of the  Hermite basis functions because they employ
spatial derivative degrees of freedom  that directly coincide with the
operators  in  the  curl,  and   are  completely  consistent  with  EM
theory. We show that the approach  yields better accuracy, with a more
physical (smoother)  representation of  fields, than from  VFEM.  In
2D calculations, this
method does  not generate  the spurious  solutions that  plagued nodal
based Lagrange FEM encountered earlier  in the 1970's, even though the
${C}_1$-continuous  Hermite  polynomials   are  also  scalar  in
nature. Our alternative set of  polynomial basis functions for 2D, the scalar
fifth-order  Hermite  interpolation   polynomials  for  the  numerical
calculations of EM  fields removes all the  above difficulties.  These
polynomials are associated  with degrees of freedom  that include both
function value and spatial derivatives up  to second order.
 Recently,    Kassebaum,   Boucher   and   Ram-Mohan
(KBR)\cite{avs:pkass}   have  shown   how  to   derive  these
polynomials  using group  representation theory,  giving a  comparison
with    the    earlier    basis    functions    occurring    in    the
literature,\cite{avs:bell1969,avs:HolandBell,avs:argyris,avs:Dhatt1984,avs:LRR_book}
which use the same derivative degrees of freedom but are distinct from
existing sets of polynomials.

In our approach,  each in-plane component of the  field is represented
by scalar  Hermite shape functions. The method is equally effective
with E-fields and H-fields, as we will demonstrate. These  functions ensure tangential
continuity  along  shared  sides  of  triangles,  thereby  eliminating
spurious solutions.  The ${C}_1$-continuity perpendicular  to the
sides of the  triangular elements leads to  smoother reconstruction of
solutions.  This representation  guarantees consistency  in the  field
direction at  the vertices  of triangles.  These properties  allow for
more  accurate  solutions  of  electrodynamics  problems  with  faster
h-convergence because of the higher  derivative shape functions. We show
here that the  scalar HFEM yields four orders of magnitude higher accuracy with
fewer elements  than those needed  in the presently  prevalent methods
for waveguides.\cite{avs:JFLeeHierarchical2003}

In  two dimensional  (2D)  waveguides, we circumvent  the issue of spurious solutions by
solving for  only one of  the field  components ($H_z$ or  $E_z$), and
obtain the  other two components  using boundary conditions  (BCs) and
the divergence-free condition.\cite{avs:Hayata,avs:Cendes1971,avs:Ahmed} However, such freedom does
not exist  in three dimensions  (3D). In a numerical  calculation, the
anticipated zero  frequency solutions  can have  non-zero frequencies
due  to  discretization,  and  will  occur  inter-mixed  with the  physical
spectrum.\cite{avs:jin2002, avs:Peterson1994}

Nodal  representation of  field components  with scalar  functions and
their  derivatives  results in  an  unambiguous  field description  at
shared nodes among  adjacent elements and treats  the field components
on  a uniform  footing.\cite{avs:HFEM_JAP2016} The  built-in derivative  degrees of
freedom (DOFs) allow us to  readily calculate the additional quantities
such  as surface  currents, while  providing smoother  solutions. This
treatment results in global  functions without severe coarsening which
will be suitable for analyzing  interactions with small features, such
as those found in high frequency transistor circuits and VCSELs. Within this
scheme, dielectric discontinuities along interfaces can be handled
cleanly using Fermi smoothing functions.\cite{avs:RemAccDegen17} 

For the brick element in 3D,  the polynomials are constructed from the
outer product of the  1D Hermite polynomials.\cite{avs:LRR_book} For a
tetrahedral element,  the Hermite  polynomials are  generated directly
based on  the geometry of  the tetrahedron and conditions  on function
values and derivatives at the vertex nodes and mid-face nodes.  {\it We then have for both types
  of element  polynomials with  both tangential and  normal derivative
  continuity across the  element boundary.}  We note  that the HFEM
approach  yields better  accuracy, with  a smoother  representation of
fields than those obtained using VFEM.  The HFEM scheme yields several
orders of higher accuracy with fewer elements than those needed in the
presently        prevalent         3D        implementations        of
VFEM.   \cite{avs:comsol,avs:hfss,avs:mfem}    The   node-based   HFEM
representation guarantees  consistency in  the direction of  fields at
the vertices of elements.\cite{avs:LRRCavityEM2018}
We impose  the divergence-free  condition through a  constant Lagrange
multiplier  term introduced  into  the action  integral in 3D,  and also  by
explicitly requiring a zero-divergence  condition at each node through
the derivative  DOFs available  at each node  in our  formulation. Any
surviving  $\omega\neq 0$  spurious solutions  are then  eliminated by
identifying          them          using          their          large
$|\nabla\cdot  {\bf E}|/|\nabla\times{\bf  E}|$ ratio.  We demonstrate
that this  procedure does not alter  or influence the accuracy  of the
physical solutions.


In VFEM  implementations, the  eigenfrequencies of  spurious solutions
are pushed to  zero, either through the Nedelec  conditions or through
their  removal at  each iteration,  as mentioned  earlier.  In  either
case, this  is an expensive  numerical procedure.  In  the literature,
the  first approach  for  eliminating spurious  solutions with  scalar
polynomials was  the penalty factor  method ({\it i.  e.},  a Lagrange
multiplier scheme).\cite{avs:RahmanDavies} However, a  fixed choice of the
penalty  factor   fails  to   impose  the   zero-divergence  condition
adequately  for   all  frequencies.\cite{avs:Webb1988}   Furthermore,  the
penalty  term   itself  can introduce  an  additional   set  of  spurious
solutions.\cite{avs:Webb1985}

Within the HFEM framework for 3D  calculations, we now have the luxury
of explicitly imposing a zero-divergence  condition at each node while
using  Hermite  interpolation  polynomials since  we  have  derivative
degrees                           of                           freedom
there.\cite{avs:Konrad1989,avs:Pinciuc2014,avs:PinciucPhD}  While this
does not ensure the complete removal of the divergence in the interior
of   the  finite   element  through   interpolation,  it   reduces  it
substantially, especially as  the size of the element  is reduced.  In
this  article,   we  use  a   constant  penalty  factor,   and  impose
zero-divergence at  all nodes to  identify the spurious  solutions for
elimination. Either brick elements (216 DOFs) or  tetrahedral elements
(56 DOFs) can be used in the calculations.

The  full  spectrum  of  propagating   modes  in  a  partially  filled
rectangular waveguide is obtained with HFEM and compared with analytic
solutions.  We plot the eigenfrequencies  of the different modes which
shows the  existence of cut-offs  in the propagating  mode frequencies
which agree perfectly with the analytic solutions. The magnetic fields
for various modes are plotted to  show the continuous character of the
solutions  in  HFEM.   The  capturing of  eigenstates  in  the  higher
dielectric region as dielectric contrast increases is analogous to the
evolution of the electronic states  in an asymmetric quantum well from
above-barrier  states to  states  bound  in the  quantum  well as  the
barrier energy in the well region is lowered.

In the following sections we consider cavity electrodynamics within a
finite element framework using Hermite brick elements and also employ
tetrahedral elements for comparison.

Finally, we show the advantages of HFEM to determine dispersion
relation in photonic crystals. The lattice of dielectric cylinders is
first treated in order to benchmark the method we are proposing. The
application to a checkerboard superlattice is to show how dielectric corners are
treated. The modeling of the Escher drawing as a final example
illustrates the advantage of HFEM when the spatial complexity of the
distribution of dielectric is so high that the plane wave methods are
not as appropriate.

\section{HFEM formulation of electromagnetic fields} 
\label{section:homogwaveguides}
 \begin{figure}[t!] 
 \begin{center}
 \includegraphics[width=2.2in]{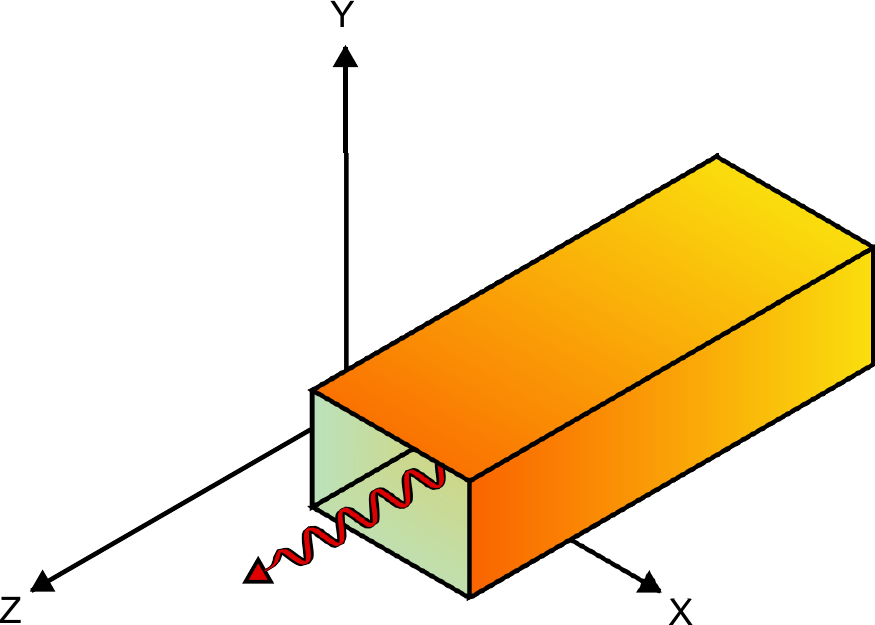}\\ \vspace{-0.2in}
\end{center}
\caption{A homogeneous waveguide with a rectangular cross-section
  $a\times b$ is shown. The propagating wave moves along the $z$-axis.
}
\vspace{-0.2in}
\label{avsfig:Emptywaveguide1}
\end{figure}

We begin with Maxwell's equations expressed in MKS units,

\begin{align}
\nabla \cdot {\bf{D}} &= \rho,
\label{eq:max1}\\
\nabla \times {\bf{H}} - \dfrac{\partial {\bf{D}}}{\partial t} &=
{\bf{J}}, 
\label{eq:max2}\\ 
\nabla \times {\bf{E}} + \dfrac{\partial {\bf{B}}}{\partial t} &= 0,
\label{eq:max3}\\ 
\nabla \cdot {\bf{B}} &= 0.
\label{eq:max4}
\end{align}%

\noindent In  the above,  the displacement  vector ${\bf{D}}$  and the
magnetic  flux  density  ${\bf{B}}$  are expressed  in  terms  of  the
electric and magnetic fields ${\bf{E}}$ and ${\bf{H}}$,
\begin{equation}
{\bf{D}} = \epsilon{\bf{E}}; \mbox{\hspace{0.3in}} {\bf{B}} 
= \mu{\bf{H}}.
\label{eq:epsmu}
\end{equation}
If  the   medium  is  isotropic,  $\epsilon$  and   $\mu$  are  scalar
quantities,  rather  than  second-rank  tensors.  Let  us  define  the
dimensionless quantities $\epsilon_r$ and $\mu_r$ so that
\begin{equation}
\epsilon = \epsilon_r \epsilon_0, \mbox{\hspace{0.3in}} 
\mu = \mu_r \mu_0,
\label{eq:epsmu2}
\end{equation}
with $\epsilon_0$ and $\mu_0$ being  the permittivity and permeability
of free space, respectively.

We assume that the dielectric regions of the waveguide are charge-free
and current-free.

The Maxwell's equations are combined to form  the
wave equations for the $E$-field and $H$-field,
\begin{align}
\nabla \times \left(\dfrac{1}{\epsilon_r} \nabla \times {\bf H}\right)
 - k_0^2 \mu_r {\bf H} &= 0, \label{eq:lag1}\\
 \nabla \times \left(\dfrac{1}{\mu_r} \nabla \times {\bf{E}} \right)
 - k_0^2 \epsilon_r {\bf{E}} &= 0,
 \label{eq:subse4}
\end{align}
where $k_0=\omega/c$.
We observe that either equation may be used to set up the action
integral. In order to define the action, we  begin by multiplying the
differential equation, Eq.~(\ref{eq:lag1}), by $\delta {\bf H}^*$
and integrating over the 
physical domain. 
We use the vector identity
\begin{equation}
\begin{split}
\nabla \cdot \left( {\bf{P}} \times {\bf{R}} \right) 
&= \bigg[\epsilon_{ijk} \left(\partial_i P_j \right) R_k - P_j \epsilon_{jik} \partial_i R_k\bigg]\\
&= \left(\nabla \times {\bf{P}} \right) \cdot {\bf{R}} - {\bf{P}} \cdot \left( \nabla
  \times {\bf{R}} \right). 
\end{split}
\label{eq:lagdev1}
\end{equation}
Now let ${\bf{R}} = \alpha \nabla \times {\bf{Q}}$. Then from
Eq.~(\ref{eq:lagdev1}), 
\begin{align}
\nabla \cdot \left( {\bf{P}} \times \left(\alpha \nabla \times {\bf{Q}} \right)
\right) 
 =& \left(\nabla \times {\bf{P}} \right) \cdot \left(\alpha \nabla \times
   {\bf{Q}} \right) \nonumber \\
 &- {\bf{P}} \cdot \left( \nabla \times \left( \alpha \nabla \times{\bf{Q}} 
   \right) \right). 
\label{eq:lagdev2}
\end{align}
This relation, along with Gauss's theorem and the substitutions ${\bf{Q}}  =  {\bf{H}}$,  
${\bf{P}}  = \delta {\bf{H}}^{*}$,  and  $\alpha  =
\epsilon_r^{-1}$, leads to the integrals
\begin{eqnarray}\label{eq:lagdev4}
\int_V\!\! d^3 r\, \delta{\bf{H}}^{*}\!\cdot\! [\nabla \!\times\! \dfrac{1}
    {\epsilon_r} ( \nabla \!\times\! {\bf{H}} ) ] 
 \!&=&\!\!\int_V\!\!d^3 r \left(\nabla \!\times\! \delta{\bf{H}}^{*} \right) \!\cdot\!
 \dfrac{1} 
   {\epsilon_r} \left( \nabla \times {\bf{H}} \right)\nonumber\\ 
& &\hspace{-0.5in}- \oint_S ds\, {\bf{\hat{n}}} \cdot \left[\delta{\bf{H}}^{*} \times
   \dfrac{1}{\epsilon_r} 
   \left( \nabla \times {\bf{H}} \right) \right]. 
\end{eqnarray}
The integrand of the  surface term may be rewritten as
\begin{equation}
{\bf{\hat{n}}} \cdot \left[ \delta{\bf{H}}^{*} \times \dfrac{1}{\epsilon_r} \left(\nabla
    \times {\bf{H}} \right) \right] 
= -\delta{\bf{H}}^{*} \cdot \left[ {\bf{\hat{n}}} \times \dfrac{1}{\epsilon_r}\left(\nabla
    \times {\bf{H}} \right) \right]. 
\label{eq:lagdev6}
\end{equation}

We are interested in solving for time-harmonic fields, so
that
${\bf H}\!\left({\bf  r},t\right)= {\bf H}\!\left({\bf  r}\right) \exp
\!\left(-i  \omega t\right)$  and similarly  for $\bf  E$.
For time-harmonic  fields, the relations between ${\bf{E}}$ and
${\bf{H}}$ is given by
\begin{equation}
{\bf H} =  -\dfrac{i}{\mu \omega} \nabla \times {\bf E},\hspace{0.3in}
{\bf E} = \dfrac{i}{\epsilon \,\omega} \nabla \times {\bf H}.
\label{eq:cross4}
\end{equation}
Therefore, ${\bf{E}}$ is  proportional to $\epsilon_r^{-1}\left(\nabla \times{\bf{H}} 
\right)$.   Assuming the waveguide  to be  enclosed by  a perfectly
conducting material, one of the boundary conditions is that ${\hat{\bf
    n}} \times {\bf{E}} =  0$. Thus, the surface integral in
Eq.~(\ref{eq:lagdev4}) is exactly zero.

Note  that it  is possible  to work  instead with  the  electric field
formulation of  Eq.~(\ref{eq:subse4}). In this case,  the surface term
takes the form
\begin{equation}
\oint_S ds\, {\bf{\hat{n}}} \cdot \left[\delta{\bf{E}}^{*} \times
  \dfrac{1}{\mu_r} \left( 
    \nabla \times {\bf{E}} \right) \right]. 
\label{eq:lagdev8}
\end{equation}
We see that for the perfectly conducting boundary, the surface term
arising from an integration by parts vanishes.

We thus see that the integrated form of  Eq.~(\ref{eq:lag1}) is
expressible as  
\begin{equation}
  \label{eq:getL38}
\delta\int_V d^3 r \,\dfrac{1}{2}\left[ \left(\nabla \times
    {\bf{H}}^{*} \right) 
\cdot \dfrac{1}{\epsilon_r} \left(\nabla \times {\bf{H}} \right)
 - k_0^2 \mu_r {\bf{H}}^{*} \cdot {\bf{H}} \right] = 0. 
\end{equation}

Equation~(\ref{eq:getL38}) is now interpreted as the functional
variation of the action integral. (Usually the action is the
  time integral of the Lagrangian. Here the Lagrangian is independent
  of time since a harmonic solution in time is assumed. Hence the ${\cal
    A}/T$ is appropriate). We may write it as
\begin{equation}
  \label{eq:getL39}
  \delta {\cal A}/T = {\bf L} =0.
\end{equation}
The principle of stationary action then  identifies the action to be


\begin{equation}
{\bf L} = \dfrac{1}{2}\int_V d^3 r \left[ \left(\nabla \times
    {\bf{H}}^{*} \right) 
\cdot \dfrac{1}{\epsilon_r} \left(\nabla \times {\bf{H}} \right)
 - k_0^2 \mu_r {\bf{H}}^{*} \cdot {\bf{H}} \right]. 
\label{eq:getL4}
\end{equation}

\noindent Similarly, using the electric field formulation of Maxwell's
equations yields


\begin{equation}
{\bf L} = \dfrac{1}{2}\int_V d^3 r \left[ \left(\nabla \times
    {\bf{E}}^{*} \right) 
\cdot \dfrac{1}{\mu_r} \left(\nabla \times {\bf{E}} \right)
 - k_0^2 \epsilon_r {\bf{E}}^{*} \cdot {\bf{E}} \right]. 
\label{eq:getL5}
\end{equation}

\begin{figure}[t!] 
 \begin{center}
 \includegraphics[width=2.5in,height=4.in]{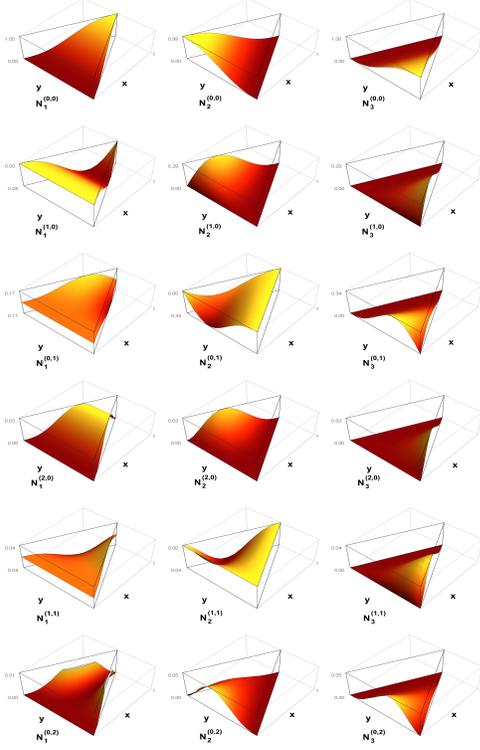}\\ \vspace{-0.2in}
\end{center}
\caption{The $C_{1}$-continuous 18 DOF quintic Hermite interpolation polynomials that have
tangential and normal derivative continuity across the element are
plotted on an equilateral triangle. These polynomials were first
derived using group 
representation theory by KBR.\cite{avs:pkass} The superscripts on the
shape function $N_i^{(m,n)}$ denote the order of the $x$ or $y$
derivative value set to unity at its associated node $i$. 
}

\vspace{-0.2in}
\label{avsfig:KBRpolyfigs}
\end{figure}

In  the  FEM  framework,  the  action  integral  is  discretized  into
triangular  elements  in  2D,  and either  hexahedral  or  tetrahedral
elements in 3D.  For tetrahedral elements in 2D,  the Hermite elements
exhibit $C_1$  continuity throughout  the finite  element mesh.  For a
given Cartesian component of  $\bf{H}$, say $f(x,y)$, the interpolated
value in terms of the 2D Hermite basis functions $\phi_i$ is given by
\begin{equation}
f(x,y)=\sum\limits_{i=1}^{18} f_i \phi_i(x,y),
\end{equation}
where the $f_i$ are degrees of freedom assigned to the function value
and its derivatives at the vertices of the triangle. This allows the
enforcement of either derivative continuity or EM boundary conditions
depending upon the material composition of adjacent triangles. We have
provided the basis for an equilateral triangle because that is the
optimization goal of most mesh refinement and it is also the rationale
for the KBR group theoretical development of the basis. The basis functions $\phi_i$ for
a reference triangle and the numbering sequence for the assignment of
function and derivative values at the nodes have been published 
elsewhere.\cite{avs:pkass,avs:HFEM_JAP2016} In
Fig.~\ref{avsfig:KBRpolyfigs},
we plot the KBR basis polynomials, and we observe that they satisfy
triangular $C_{3V}$ symmetry.

For a hexahedral element in  3D calculations, the HFEM polynomials are
constructed through the  products of 1D Hermite  polynomials in $x,y,$
and $z$  directions. This  guarantees continuity  of the  field values
$f$,    first    derivatives    $\partial_xf,\partial_yf,\partial_zf$,
cross-term                      second                     derivatives
$\partial^2_{xy}f,\partial^2_{yz}f,\partial^2_{xz}f$,              and
$\partial^3_{xyz}f$.   One  can  also employ  HFEM  using  tetrahedral
elements  instead,   as  in   many  cases  tetrahedra   offer  more
flexibility in discretizing the geometry  of the problem.  The quintic
Hermite  interpolation polynomials  for  tetrahedral elements  assures
continuity  for   the  function  value  $f$,   the  first  derivatives
$\partial_xf,\partial_yf,\partial_zf$, and all  the second derivatives
$\partial^2_{xx}f,\partial^2_{yy}f,\partial^2_{zz}f,
\partial^2_{xy}f,\partial^2_{yz}f,\partial^2_{xz}f,$ at each vertex of
the  tetrahedral element.  At  the face  centers  of the  tetrahedron,
continuity  in  the  function  value and  the  first  derivatives  are
guaranteed.

\section{Waveguides}
\subsection{Boundary conditions}

   In this section, we consider a rectangular conducting
waveguide with  sides parallel  to the $x$-  and $y$-axes (see
Fig.~\ref{avsfig:Emptywaveguide1}).
The boundary
conditions are given by 
\begin{align}
{\hat{\bf{n}}} \cdot {\bf{H}} &= 0,
\label{eq:BC1} \\
{\hat{\bf{n}}} \times {\bf{E}} &= 0.
\label{eq:BC2}
\end{align}
Expanding Eq.~(\ref{eq:BC1}) we have
\begin{equation}
n_x H_x + n_y H_y = 0.
\label{eq:BC3}
\end{equation}
For the left and right boundaries, which are parallel to the $y$-axis,
Eq.~(\ref{eq:BC3}) simplifies to
\begin{equation}
H_x = 0 {\mbox{ for }} x \in \{0,d\}.
\label{eq:BC4}
\end{equation}
For the top and bottom boundaries, we get
\begin{equation}
H_y = 0 {\mbox{ for }} y \in \{0,h\}.
\label{eq:BC5}
\end{equation}
In  addition,  Eq.~(\ref{eq:BC2})  may  be used  to  generate  derivative
boundary  conditions  at  the  edges.  Recall that the differential form
of the Maxwell-Amp{\` e}re equation is given by
\begin{equation}
\nabla \times {\bf{H}} - \dfrac{\partial {\bf{D}}}{\partial t} = {\bf{J}}.
\tag{\ref{eq:max2}}
\end{equation}
Assuming that dielectric properties do not vary over time and there is
no current density at the boundary, we have
\begin{equation}
\nabla \times {\bf{H}} = \epsilon \dfrac{\partial {\bf{E}}}{\partial t} 
= -i \omega \epsilon {\bf{E}}
\label{eq:BC7}
\end{equation}
for a time-harmonic field.
Substituting    Eq.~(\ref{eq:BC7})  into
Eq.~(\ref{eq:BC2}) yields
\begin{equation}
{\hat{n}} \times \left(-\dfrac{1}{i \omega \epsilon} \nabla \times
  {\bf{H}} \right) = 0. 
\label{eq:BC9}
\end{equation}
Expand the cross product to obtain

\begin{equation}
\hspace{-0.1in}\begin{pmatrix}
n_y \left(\partial_x H_y - \partial_y H_x\right) \\
-n_x \left(\partial_x H_y - \partial_y H_x\right) \\
n_x \left(\partial_z H_x - \partial_x H_z \right) -
n_y \left(\partial_y H_z - \partial_z H_y \right)
\end{pmatrix} =
\begin{pmatrix}
0\\0\\0
\end{pmatrix}.
\label{eq:BC10}
\end{equation}

\noindent  The  first and  second  vector components  of the  cross
product each yield
\begin{equation}
\partial_x H_y = \partial_y H_x.
\label{eq:BC11}
\end{equation}
On the  left and right boundaries,  we have already shown  that $H_x =
0$.  Since  these edges  are parallel  to the $y$-axis  and $H_x  = 0$
along  these  edges,  it  follows  that $\partial_y  H_x  =  0$.  From
Eq.~(\ref{eq:BC11}) we then obtain  derivative conditions on $H_y$ and $H_x$:
\begin{align}
\partial_x H_y &= 0, {\mbox{ for }} x \in \{0,d\}.\label{eq:BC12} \\
\partial_y H_x &= 0, {\mbox{ for }} y \in\{0,h\}.
\label{eq:BC13}
\end{align}
\noindent                Together,                Eqs.~(\ref{eq:BC4}),
(\ref{eq:BC5}),(\ref{eq:BC12}) and (\ref{eq:BC13})  constitute all the
boundary conditions for a conducting waveguide of width $d$ and height
$h$.

It is straightforward to show that in 2D the divergence condition
$\nabla\cdot{\bf E}=0$
is automatically satisfied in calculations for the fields
in the cross-section of the  waveguide. We refer the reader to
Nayfeh's treatment.\cite{avs:nayfeh_book}

\subsection{Results for a homogeneous waveguide}
Our first example is a homogeneous
rectangular ($d \times  h$) waveguide with  $\epsilon_r = 1,
\mu_r =  1$, and $k_z  = 1$.  The dimensions  of the cross-section were
chosen to be $d=20$ and
$h=10$.  
There is no inherent separation of
TE and TM modes in the HFEM formulation and therefore the eigenproblem
returns all  physical solutions. The resulting  $H_z$ field components
were calculated  in post-processing where those with  finite magnitudes
are  TE modes  and those  with $H_z$  approaching the  numerical noise
floor  are  TM modes.  The eigenfunctions for  the  degenerate TE  and  TM  
modes of  the homogeneous waveguide were
separated in  post-processing since  arbitrary linear  combinations of
degenerate modes result from a typical numerical diagonalization.

To generate pure TE modes  during the postprocessing stage, we enforce
the condition that
\begin{equation}
\frac{\partial H_y}{\partial x} -
\frac{\partial H_x}{\partial y} = 0.
\label{eq:TEcond}
\end{equation}
Since  the electric field  is derived  from the  curl of  the magnetic
field, this expression forces the $z$-component of the electric field
to equal zero, creating a pure TE mode.

To create  a pure TM mode, we  recall that $H_z$ is  obtained from the
in-plane  components using  the  divergence condition.   We can  force
$H_z$ to equal zero by forcing
\begin{equation}
\frac{\partial H_x}{\partial x} +
\frac{\partial H_y}{\partial y} = 0.
\label{eq:TMcond}
\end{equation}
Either Eq.~(\ref{eq:TEcond}) or  Eq.~(\ref{eq:TMcond}) may be enforced
after  calculating  the  eigenfunctions  by  multiplying  one  of  the
in-plane components,  either $H_x$ or  $H_y$, by a scale  factor until
one of the equations is satisfied. 

\begin{figure}[t] 
  \begin{center}
\includegraphics[width=3.0in]{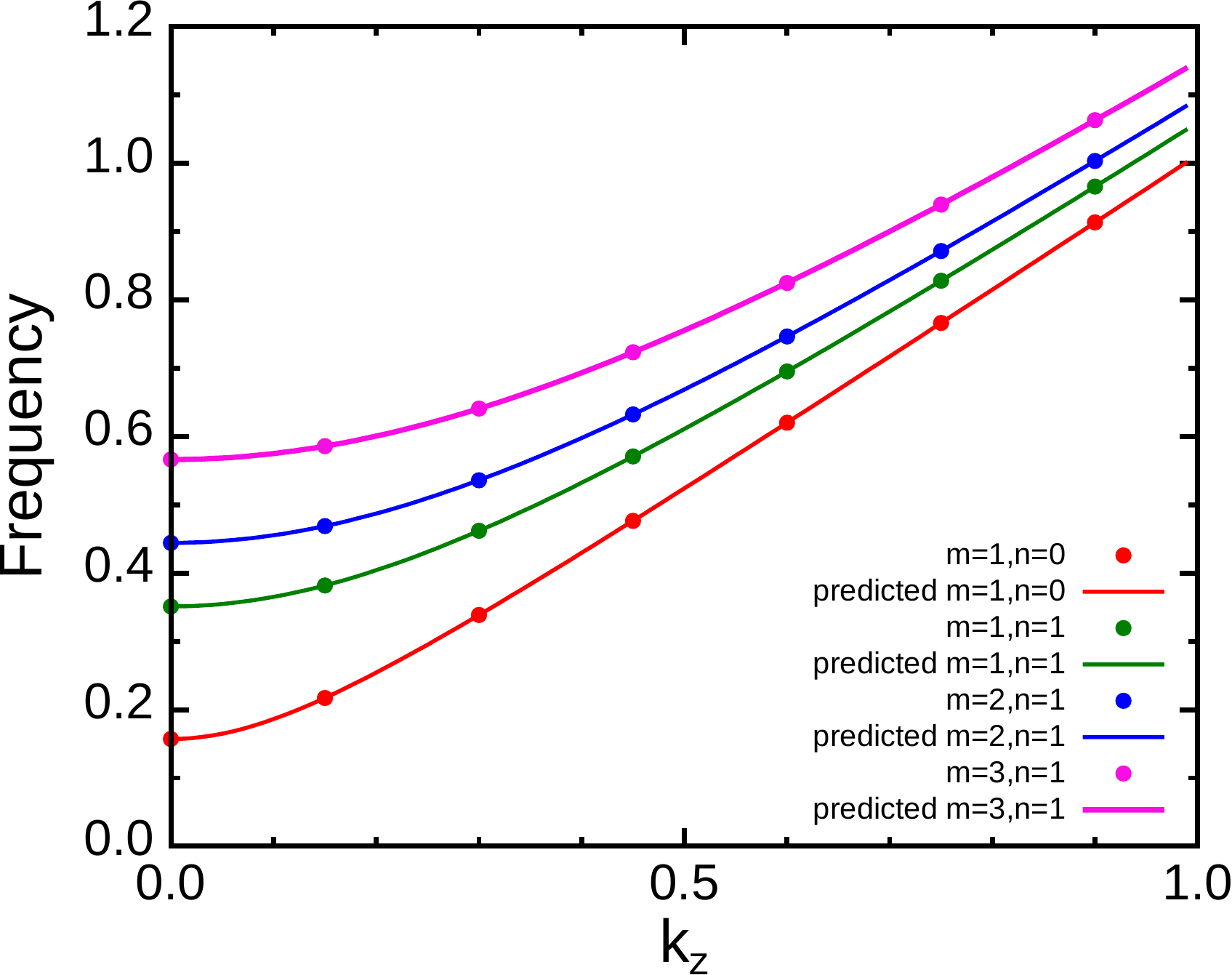}
\end{center}
\vspace{-0.2in}
\caption{Frequency eigenvalues $(\omega/c)$ of several
  $\text{TE}_{mn}$ modes of a hollow $(\epsilon_r=1)$ rectangular
  waveguide with 2:1 aspect ratio are shown as functions of $k_z$. The
  analytical   values (curves) are compared with those calculated using HFEM
 (dots), and the two are essentially identical (see
 Table~\ref{table:evaltab}). (From  Boucher {\it et al.}, Ref. \onlinecite{avs:HFEM_JAP2016}.)}
  \vspace{-0.1in}
 \label{avsfig:eigplot1}
\end{figure}
The eigenvalues of a few propagating  modes are plotted
as a function of $k_z$  in  Fig.~\ref{avsfig:eigplot1}.   Relative
errors in  the eigenvalue  calculations for the  homogeneous waveguide
varied from $10^{-14}$ for the lowest state to $10^{-10}$  with approximately $2700$ DOF in
the global matrix, as shown  in Table~\ref{table:evaltab}. 

\begin{table*}[t!]
\begin{center}
  \caption{\label{table:evaltab} Numerically calculated eigenvalues versus
 their predicted values for the ten lowest-energy modes of the
 homogeneous rectangular waveguide. The number of Hermite finite elements used was
406 with a global matrix size of 2784. (Adapted from Boucher {\it et
  al.}, Ref. \onlinecite{avs:HFEM_JAP2016}.)}
\vspace{-0.1in}
\begin{tabular}{c|c|c|c} \hline \hline
\hspace{0.2in} Mode \hspace{0.2in} & \hspace{0.3in} Theoretical
\hspace{0.3in} & \hspace{0.2in} HFEM \hspace{0.2in} &
\hspace{0.2in} Error\hspace{0.2in} \\
\hline
$TE_{10}$ & 1.024\,674\,011\,002\,72 & 1.024\,674\,011\,002\,69 & $3.0 \times 10^{-14}$\\
$TE_{20}$ & 1.098\,696\,044\,010\,89 & 1.098\,696\,043\,949\,03 & $6.2 \times 10^{-11}$\\
$TE_{01}$ & 1.098\,696\,044\,010\,89 & 1.098\,696\,044\,076\,11 & $6.5 \times 10^{-11}$\\
$TE_{11} + TM_{11}$ & 1.123\,370\,055\,013\,61 & 1.123\,370\,054\,915\,78 & $9.8 \times 10^{-11}$\\
$TE_{11} + TM_{11}$ & 1.123\,370\,055\,013\,61 & 1.123\,370\,055\,117\,61 & $1.0 \times 10^{-10}$\\
$TE_{21} + TM_{21}$ & 1.197\,392\,088\,021\,78 & 1.197\,392\,087\,905\,68 & $1.1 \times 10^{-10}$\\
$TE_{21} + TM_{21}$ & 1.197\,392\,088\,021\,78 & 1.197\,392\,088\,225\,77 & $2.0 \times 10^{-10}$\\
$TE_{30}$ & 1.222\,066\,099\,024\,51 & 1.222\,066\,099\,118\,76 & $9.4 \times 10^{-11}$\\
$TE_{31} + TM_{31}$ & 1.320\,762\,143\,035\,40 & 1.320\,762\,143\,075\,83 & $4.0 \times 10^{-11}$\\
$TE_{31} + TM_{31}$ & 1.320\,762\,143\,035\,40 & 1.320\,762\,143\,933\,56 & $9.0 \times 10^{-10}$\\
\hline \hline
\end{tabular}
\end{center}
\end{table*}
\begin{figure}[b!] 
\begin{center}
  \includegraphics[width=2.4in]{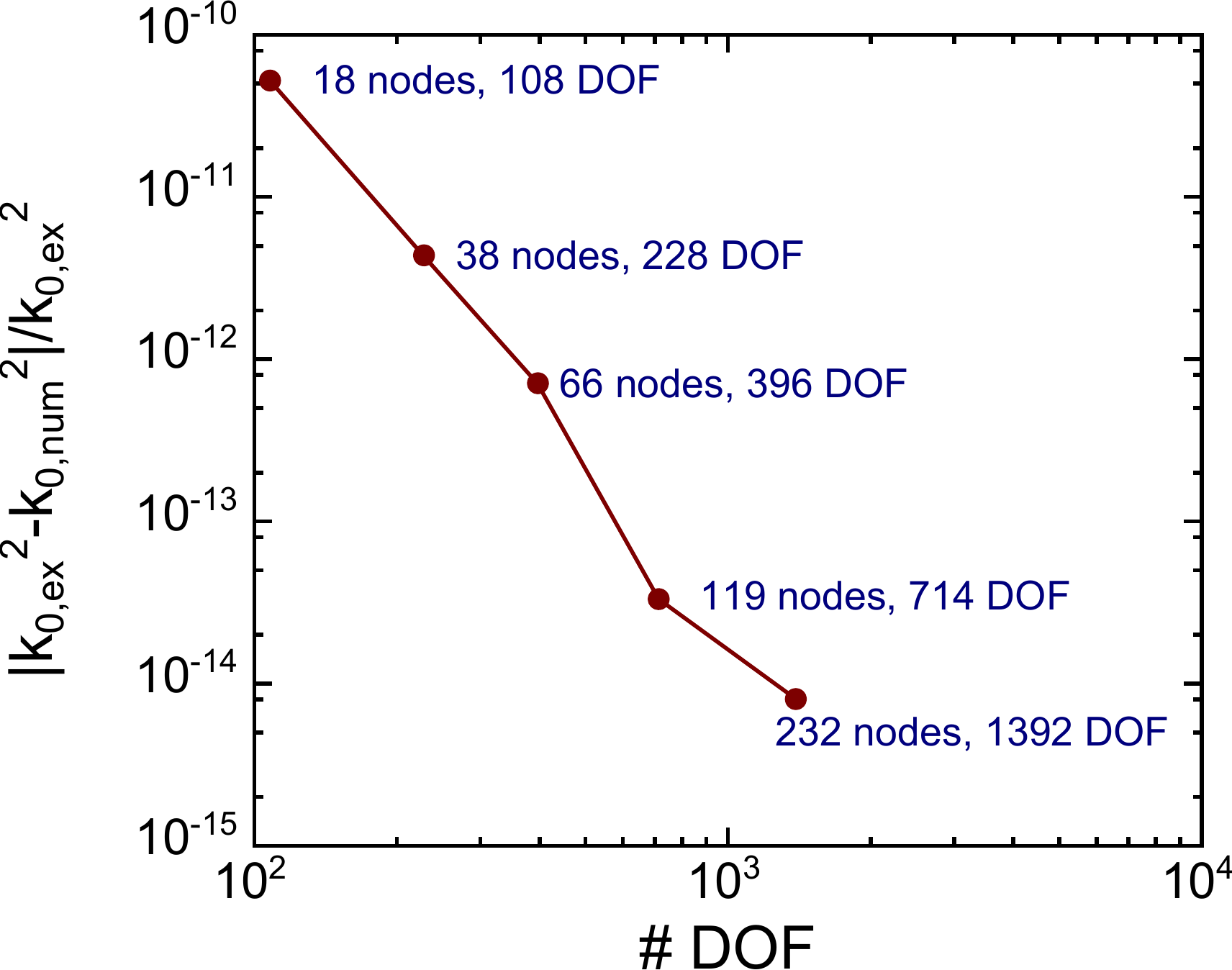}\\\vspace{0.1in}
  \includegraphics[width=2.4in]{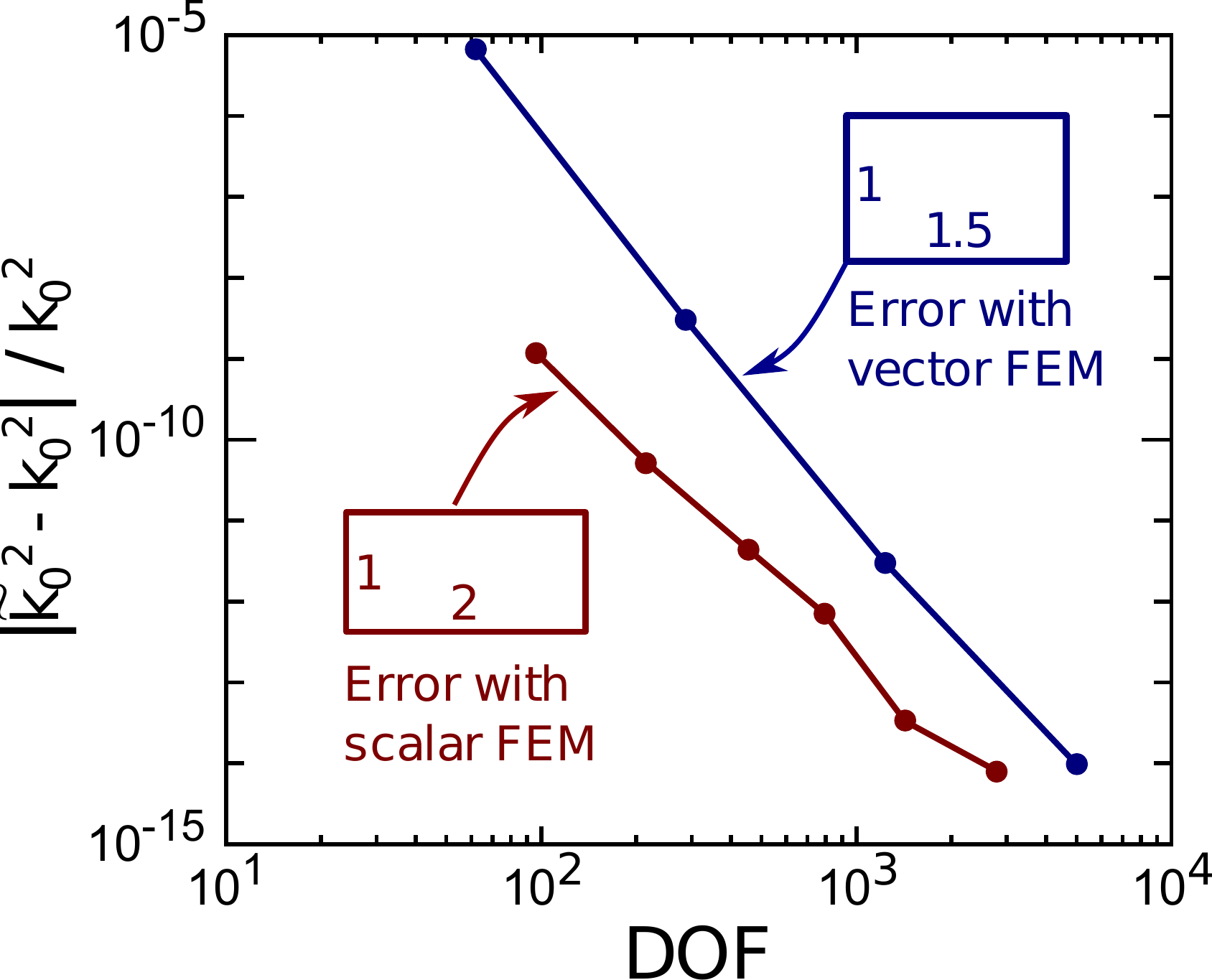}
\caption{The  convergence  of   the  eigenvalue  for  the  lowest
  frequency,  $TE_{10}$-mode  is displayed  as  the  global number  of
  degrees of freedom (DOF) is  increased through mesh refinement for a
  homogeneous waveguide.  The solid  curve (red) is obtained with
  HFEM     and    compared     with     published    results     using
  VFEM.\cite{avs:JFLeeHierarchical2003}\label{avsfig:ErrorComparisonJFLee}
  (From Boucher {\it et al.}, Ref. \onlinecite{avs:HFEM_JAP2016}.)}
\end{center}
\end{figure}

A few comments are in order:
\begin{enumerate}
\item 
From Table~\ref{table:evaltab},  it is  clear that the  eigenvalues of
the  homogeneous waveguide, particularly  the lowest  eigenvalue, show
close  agreement  with  analytical   values.  We  note  that  Lee
{\it  et al}.\cite{avs:JFLeeHierarchical2003}    have  made   an  analysis   of  the
convergence of eigenvalues for hierarchical vector finite elements for
waveguides. The  lowest eigenvalue  reported by them  has an  error of
$10^{-12}$ for approximately the same  number of degrees of freedom as
compared  with  our   lowest  eigenvalue  which  has  double-precision
accuracy as seen in Table~\ref{table:evaltab}.

The lowest frequency propagating mode is the $TE_{10}$ mode. 
The $TE_{mn}$ and $TM_{mn}$ modes with $m,n \neq 0$ are  degenerate.  

\item The diagonalizer delivers a linear combination of degenerate TM-
and TE-modes.   In order to resolve the eigenfunctions  into  distinct
TM- and TE-modes, it  is necessary to
rescale one of the in-plane magnetic field components before using the
in-plane  components  to  calculate   $H_z$  and  the  electric  field
components.   The  need  to  rescale  one of  these  components  is  a
consequence of the fact that the in-plane field components are used to
construct the global matrix, without explicitly setting $H_z$ or
$E_z$ to  zero for  TM and TE  modes, respectively.  While  for
every TM  mode, there exists  a TE mode  with the same  frequency,  the
resulting degeneracy  cannot be resolved  by a simple  perturbation of
the waveguide cross-sectional dimensions or global matrix elements.

\item Accidental degeneracies, which involve different modes having the same
frequency,  may occur, especially  if one  dimension of  the waveguide
cross-section  is  commensurate  with  the  other.   These  accidental
degeneracies may be removed by introducing a small perturbation in the
global matrix, and do not require any additional postprocessing.

\item We note that the transverse nature of the modes is
  demanded in vector finite element analysis for every element. Here we impose
  the transversality condition at the end of the calculation for the global
  eigenstates. 

\end{enumerate}
\vspace{0.1in}

We now determine the convergence  properties of the HFEM solutions. In
Fig.~\ref{avsfig:ErrorComparisonJFLee}(a), we show the convergence of the
eigenfrequency to  its analytically determined value  in a homogeneous
waveguide.  For the  lowest  frequency $TE_{10}$-mode,  as the  global
number  of DOF  is  increased through  mesh  refinement, the  accuracy
improves steadily until  2400 DOF when the curve in  red (lower curve)
obtained with  HFEM reaches down  to $10^{-14}$. The HFEM  delivers an
accuracy of  $10^{-9}$ with just 96  DOFs (8 nodes). These  results are
extraordinary in terms of how quickly the frequency of the lowest mode
is determined  accurately. In  the same figure,  we have  overlaid the
VFEM  data\cite{avs:JFLeeHierarchical2003}  in  the  blue  curve  (upper
curve). At  the low  end of  mesh refinement,  with $\sim$100  DOF, the
hierarchical VFEM employing a comparable order quintic polynomial
basis  has an  error  of  $\simeq 10^{-5}$  for  a hollow  rectangular
waveguide. With further mesh refinement  leading to $\sim$5000 DOF the
VFEM has an  error comparable to our HFEM  formulation with $\sim$2400
DOF.

We  can also  consider  the error  in the  eigenvalues  of the  higher
states.  The eigenvalues  of solutions  above the  ground state  have
errors approximately  three to  four orders  of magnitude  higher, but
converge at  the same rate as  the ground state.  This  indicates that
the derivative continuity of the fields is allowing for a high-quality
variational solution  even at  modest discretization levels,  which is
the essence of FEM.

\subsection{HFEM results for the inhomogeneous waveguide}
\label{sec:waveguide_results2}
\begin{figure}[t] 
\begin{center}
\includegraphics[width=2.20in]{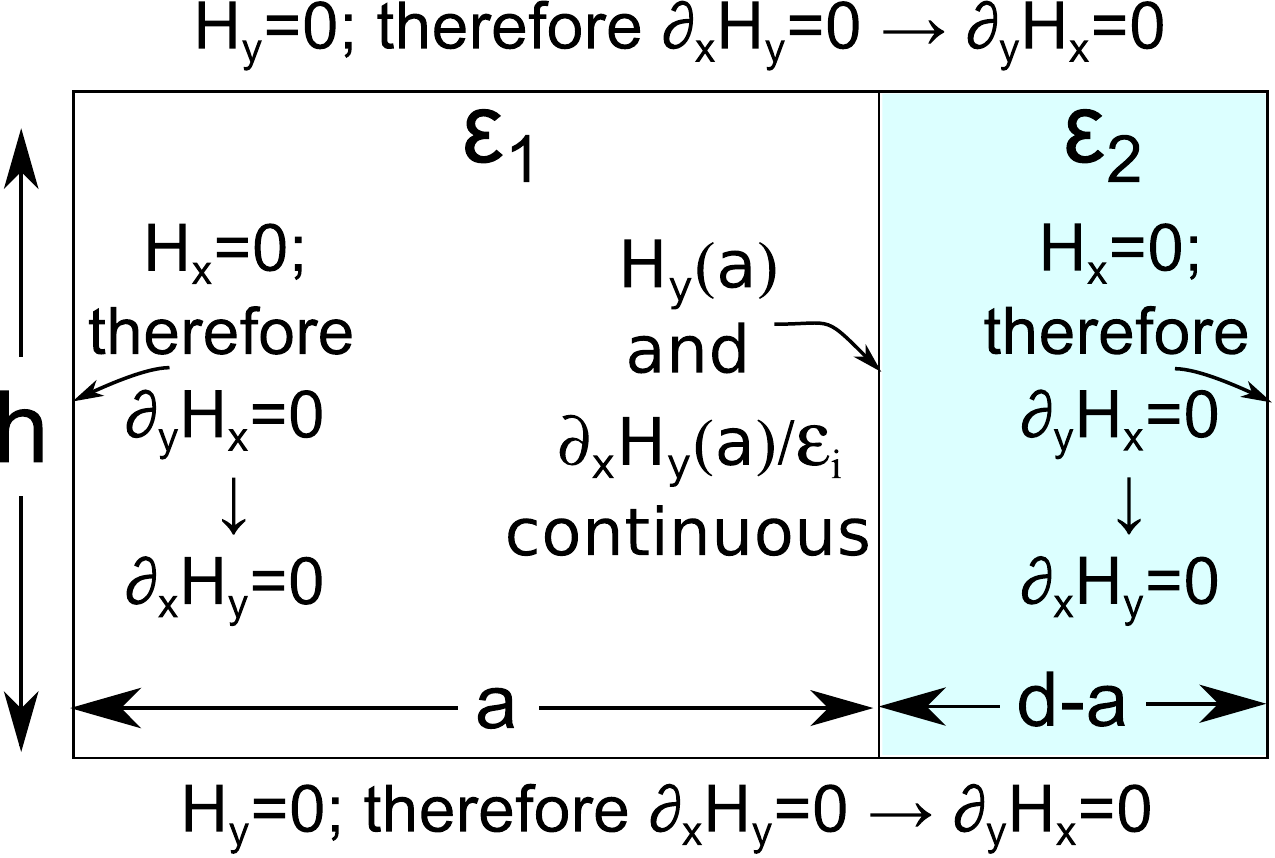}
\end{center}
\vspace{-0.2in}
\caption{The boundary conditions and dimensions for a partially filled
 perfectly electrically conducting (PEC) waveguide are shown.  These boundary conditions were implemented
  in the finite element simulations.}
\label{avsfig:inhbcs}
\vspace{-0.2in}
\end{figure}

We  now  consider  a  partially filled  waveguide  with  geometry  and
boundary  conditions as  shown  in Fig.~\ref{avsfig:inhbcs}.  
This is  an
especially  attractive  test case  for  HFEM  because it  addresses  a
canonical problem for which VFEM  was created, namely, the presence of
spurious solutions in other scalar FEM formulations. In addition, this
particular waveguide configuration shows the efficacy of using HFEM to
resolve spatially  varying fields  of both sinusoidal  and exponential
(sinh  and  cosh) dependences.  Fields  of  both types  have  analytic
solutions as shown  below, and therefore we  can compare with
analytic solutions. These field concepts are technologically important
to capture with  the HFEM technique because they appear  in the design
of slow-wave structures, on-chip waveguides, and dielectrically-loaded
leaky-wave antennas.
\begin{figure}[b!] 
\begin{center}
  \includegraphics[width=2.3in]{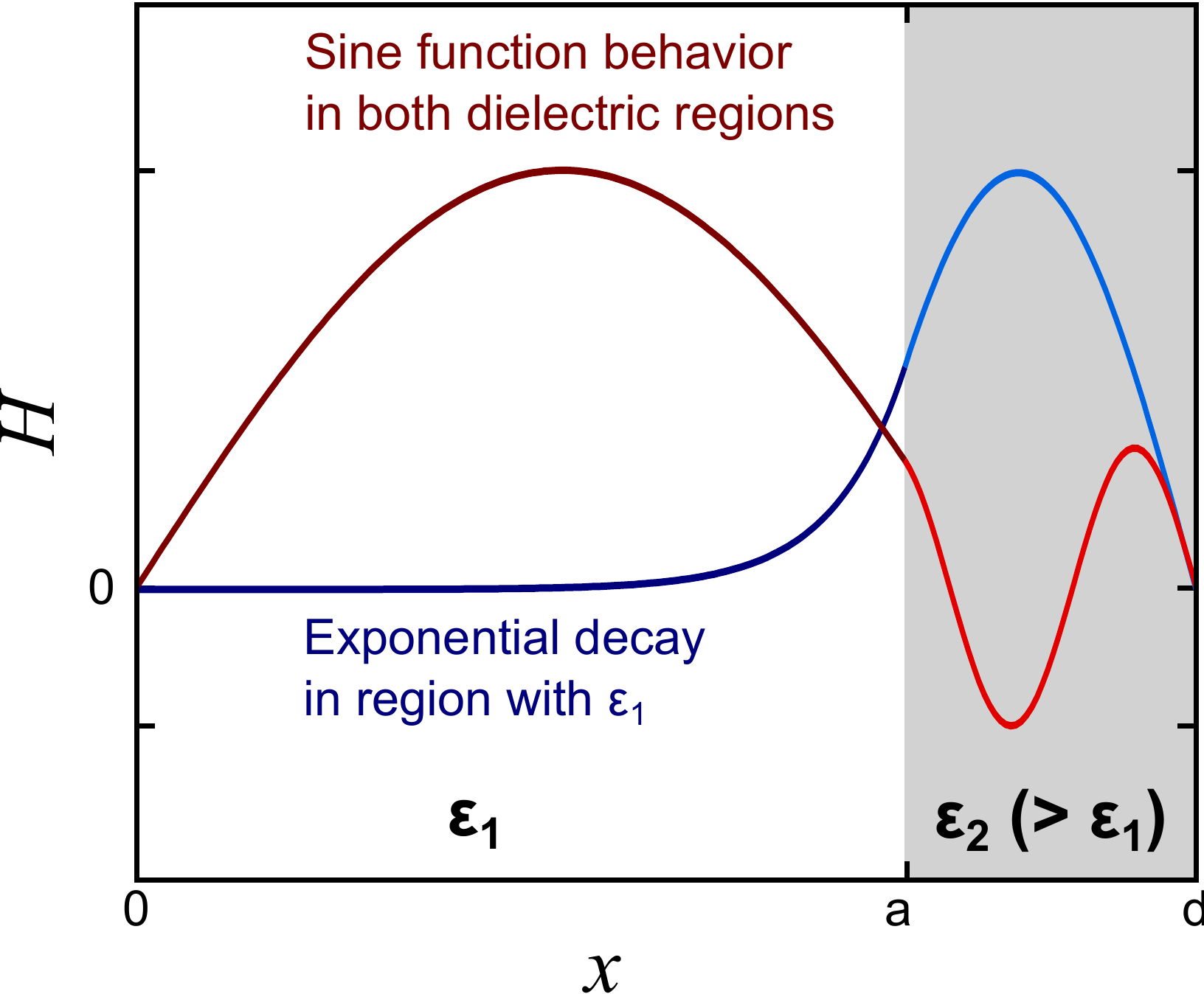}\\ \vspace{0.1in}
    \includegraphics[width=2.30in]{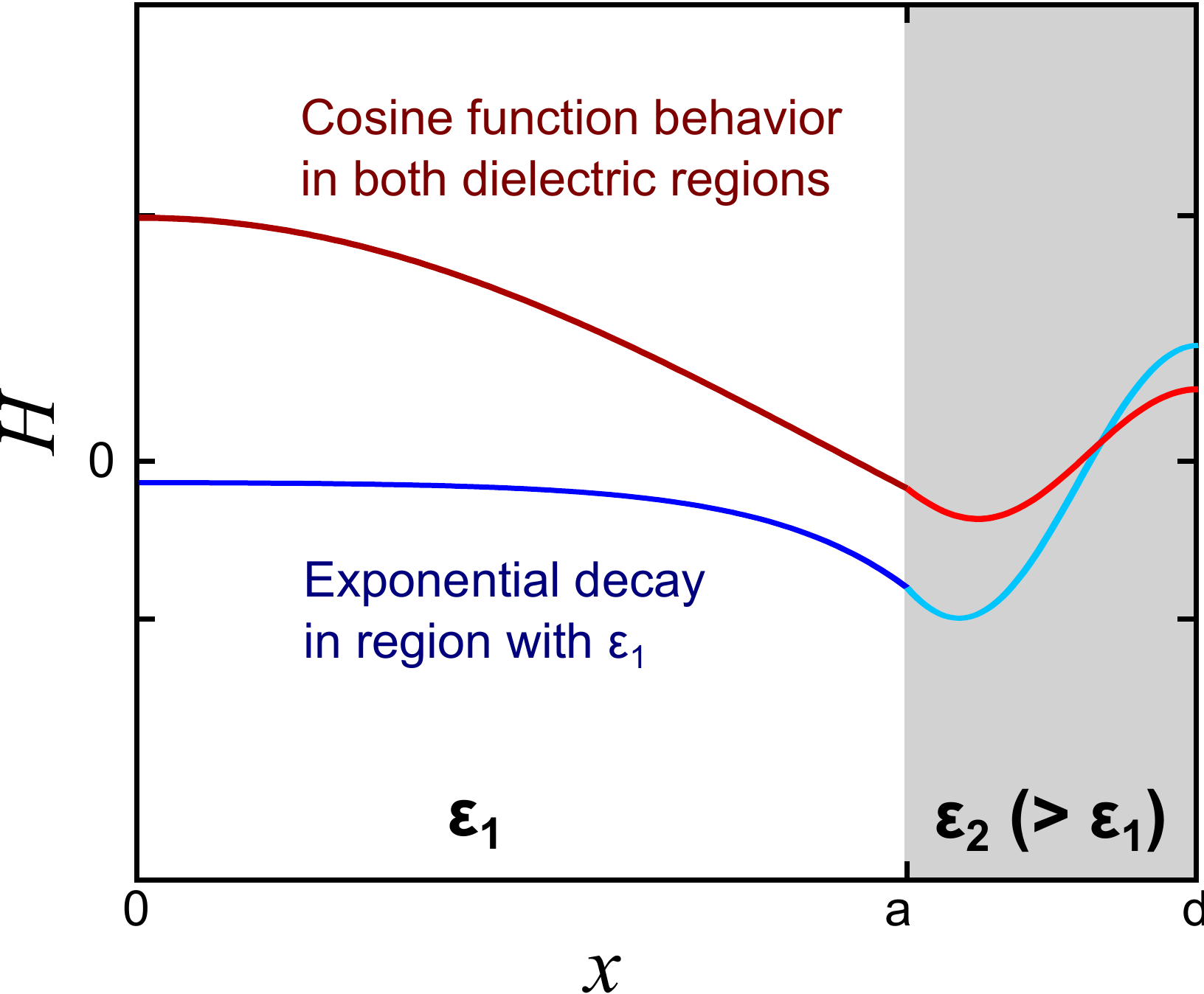}
\caption{Variation of the mode frequency as the dielectric ratio is
 increased is shown in an inhomogeneous dielectrically loaded
 waveguide. (a) The waveguide sustains sine/sine and sine/sinh type
 solutions for $H_x$.
(b) The waveguide
sustains cosine/cosine and cosine/cosh type solutions for $H_y$. (Adapted from Boucher {\it et al.}, Ref.~\onlinecite{avs:HFEM_JAP2016}.) 
 \label{avsfig:hxysolns2}}
\end{center}
\end{figure}
It  was our  expectation that  HFEM  would perform  extremely well  in
resolving both  types of eigenmodes  because these issues  are readily
encountered  when  using  scalar  HFEM  in  the  solution  of  quantum
mechanical wavefunctions. The  combination of a sine  and a hyperbolic
sine field solution  is analogous to the quantum calculation for an asymmetric  quantum well. When
$(k_z^2+(n\pi/h)^2)/\epsilon_1  < k_0^2$  this  is  equivalent to  the
potential energy of  an electron in such a well  being higher than the
electron energy in the barrier region  leading to an  exponentially falling solution  in the
``barrier  region''   which  is  analogous  to   the  dielectric  with
permittivity         $\epsilon_1$.        When         $k_o^2        >
(k_z^2+(n\pi/h)^2)/\epsilon_2$  sinusoidal   solutions  are  obtained,
corresponding   to  the   confined  state   solutions  in   a  quantum
well. Similar analogies have  been drawn previously. Not surprisingly,
the symmetric potential well problem  in 1D quantum mechanics has been
compared  with the  electromagnetic confinement  in a  dielectric slab
waveguide surrounded by  air\cite{avs:Portis1978} as this  amounts to the
same problem.

\begin{figure}[t!] 
\begin{center}
\includegraphics[width=3.05in]{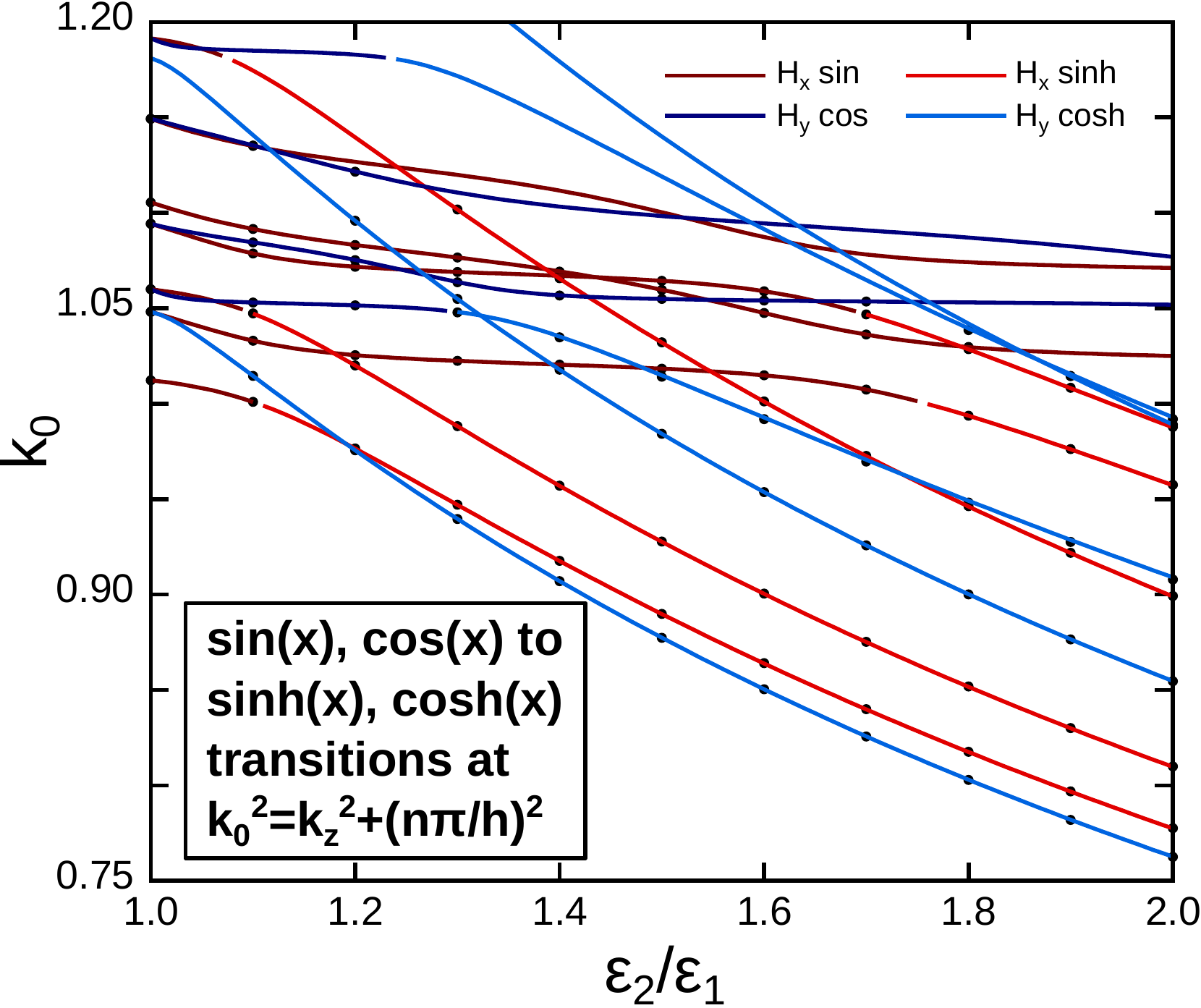}
\caption{Eigenvalues of several propagating modes obtained from
  analytic dispersion relations, are plotted along with their HFEM
  computed values (points) as functions of the dielectric ratio
 $\epsilon_2 / \epsilon_1$, which is here limited from 1.0 to
 2.0. the sine (cosine) -like solutions evolve into their hyperbolic
 form in the lower dielectric region as the dielectric ratio
 increases. \label{avsfig:inh_eigplot} (From Boucher {\it et al.}, Ref.~\onlinecite{avs:HFEM_JAP2016}.)}
\end{center}
\vspace{-0.3in}
\end{figure}

We plot fields of both  sinusoidal and exponential spatial variation in
Fig.~\ref{avsfig:hxysolns2}(a). The predicted longitudinal section electric 
(LSE, or  TE to $\hat{x}$) eigenvalues are shown as
functions           of            $\epsilon_2/\epsilon_1$           in
Fig.~\ref{avsfig:hxysolns2}(b). The quantum well  analogy suggests that as
$\epsilon_2$  is  increased more  modes  are  captured by  the  higher
dielectric region  leading to  the sinusoidal  behavior in  the larger
dielectric region and  an exponential decay into  the lower dielectric
region, as shown schematically in Fig.~\ref{avsfig:hxysolns2}(a).

\begin{figure*}[t] 
\begin{center}
\includegraphics[width=4.8in]{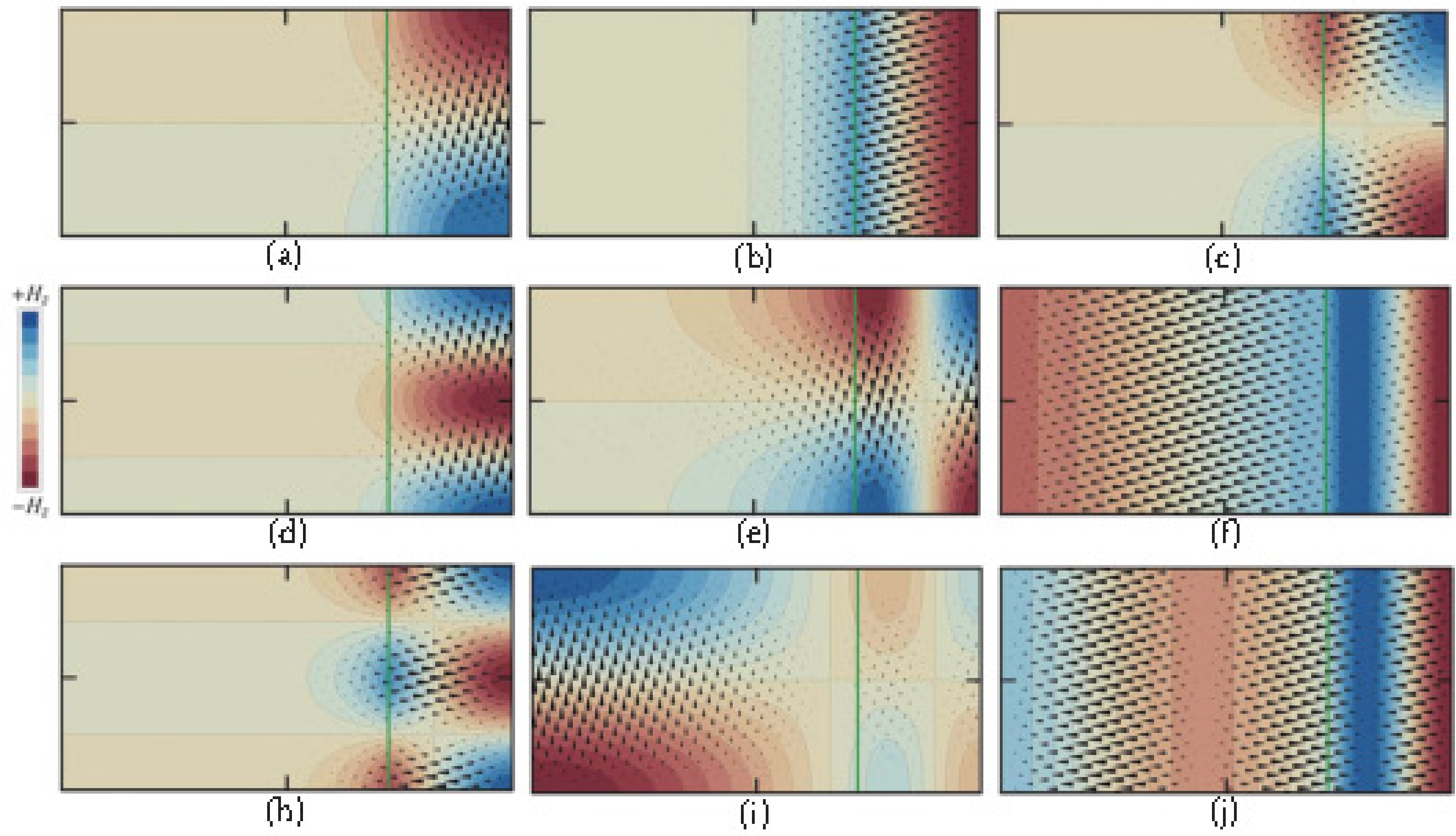}
\caption{\label{avsfig:e2-1p5-plot1}Field  patterns for  the nine  lowest
  frequency LSE modes an inhomogeneous  2:1 aspect ratio  
 PEC waveguide
  are plotted in (a)-(j) in increasing order. The dielectric ratio was
  $\epsilon_2/\epsilon_1  = 1.5$  (region 1  is left  of the  vertical
  line, region 2  to the right) and $k_z =  1.0$. The arrows represent
  the  in-plane  $H$-fields,  while  the   shading  shows  the
  $H_z$-field pattern. (From Boucher {\it et al.}, Ref. \onlinecite{avs:HFEM_JAP2016}.)}
\end{center}
\vspace{-0.2in}
\end{figure*}
The predicted eigenvalues which correspond to longitudinal section magnetic 
(LSM, or TM to $\hat{x}$) modes are also shown
as  functions of  $\epsilon_2/\epsilon_1$.  Note  that every  solution
undergoes a transition  from the cosine-like regime  to the hyperbolic
regime, as shown  by the color change in the  plot of each eigenvalue,
at  a certain  threshold value  of the  permittivity. These  threshold
values  depend  on  the  value  of $k_z$  and  the  frequency  of  the
eigenfunctions   in    the   $y$-direction.    This   is    shown   in
Fig.~\ref{avsfig:hxysolns2}.

The analytically  determined eigenvalues were compared  to the results
found  using  the  HFEM.   Eigenvalues of  several  propagating  modes
obtained from  analytic dispersion  relations, are plotted  along with
their computed  values (points) as  functions of the  dielectric ratio
$\epsilon_2 / \epsilon_1$ in Fig.~\ref{avsfig:inh_eigplot}. The agreement
between the theoretically determined  eigenvalues and those calculated
using  HFEM  is  excellent.   The  first   10  eigenvectors   in  the
inhomogeneous waveguide, with  a dielectric constant $\epsilon_2=1.5$,
are  shown  in  Figs.~\ref{avsfig:inh_eigplot}.  The  eigenmodes  for  a
dielectric constant $\epsilon_2=2.0$  lead to more of  the modes being
confined in  the higher  dielectric region as  compared with  the case
where $\epsilon_2=1.5$.

The  HFEM correctly  solves  for the  modes  supported  by  partially
dielectric-loaded waveguides  (both slow- and fast-wave  regimes; both
LSE  and  LSM  modes).  Of  particular  importance  is  that  the
formulation  can  solve  for  these  various  dielectric-loaded  modes
without spurious solutions.

 \section{Cavity electrodynamics and accidental degeneracies}

 To demonstrate the  capabilities of HFEM in 3D  calculations, in this
 section we solve Maxwell's equations in a cubic cavity. Since HFEM is
 equally applicable to  {\bf E}-field calculations as it is to  {\bf H}-field, here
 we choose to work with {\bf E}-field. Recall that the functional
 integral to be optimized  is then of the form
\begin{align}
\label{eq:actionint0}
{\bf L}\! = \!\!\int_{V} \!dV\
\Big[\nabla \times \mathbf{E^*} \cdot \mu_r^{-1} \nabla \times \mathbf{E}
- k_0^2 \, \mathbf{E^*} \cdot \epsilon_r \, \mathbf{E} \,\Big].
\end{align}
We  note  that  in 3D  we  do  not  have  the freedom  to  impose  the
divergence-free condition on the fields since we need to solve for all
three field components as opposed to  only one of the components as in
waveguides.  So  to minimize the divergence,  we introduce a
penalty term
$\lambda  \, \big|\nabla  \cdot \epsilon_r  \mathbf{E}\big|^2$ in  the
functional ${\bf L}$, given by
\begin{eqnarray}
\label{eq:actionint}
{\bf L}\! &=& \!\!\int_{V} \!dV \!
\Big[\nabla \! \times \! \mathbf{E^*} \! \cdot \mu_r^{-1} \nabla \! \times \! \mathbf{E}
 \!- \! k_0^2 \, \mathbf{E^*} \! \cdot \epsilon_r \, \mathbf{E} \nonumber \\
&&\hspace{0.3in}+ \lambda \, \big|\nabla \cdot \epsilon_r \mathbf{E}\big|^2 \, \Big],
\end{eqnarray}
where $\lambda$ is the Lagrange multiplier. The fields are represented
by Hermite interpolation polynomials on hexahedral elements multiplied
by the  values of fields  and their  derivatives at  the vertices
(nodes)  of each  element. The  integral can  be discretized  over the
elements  to  obtain   a  matrix  equation  in  terms   of  the  nodal
parameters. We invoke the principle  of stationary action, and set the
variation of ${\bf L}$ with respect to $\mathbf{E}^*$  equal to zero. This yields a
generalized  eigenvalue   problem  which  is  solved   to  obtain  the
frequencies  and field  distributions  of the  resonating modes.   The
magnetic fields  are readily obtained  from the electric  fields using
Maxwell's equations.

We consider  a cubic cavity with perfectly conducting
metallic  boundaries. We  assume that  the dielectric  regions of  the
cavity are charge-free  and current-free. At the surface  of a perfect
electrical conductor, the electric and  magnetic fields satisfy
the boundary conditions (BCs)\cite{avs:jackson_book,avs:griffiths_book,avs:Harrington}
\begin{align}
\label{eq:BCinmetal}
{\hat{\bf n}}\times {\bf E} = 0 \hspace{0.25in} \mbox{and}
\hspace{0.25in} {\hat{\bf n}}\cdot {\bf H}=0.
\end{align}
These relations  give the  BCs on  the periphery  of the  cavity. When
working with electric  fields, the tangential components  of the field
are set to zero at the boundary, while the normal components are
determined variationally. 
\subsection{Origin and nature of spurious solutions}
\begin{figure}[!tbhp]   
\includegraphics[width=1.60in]{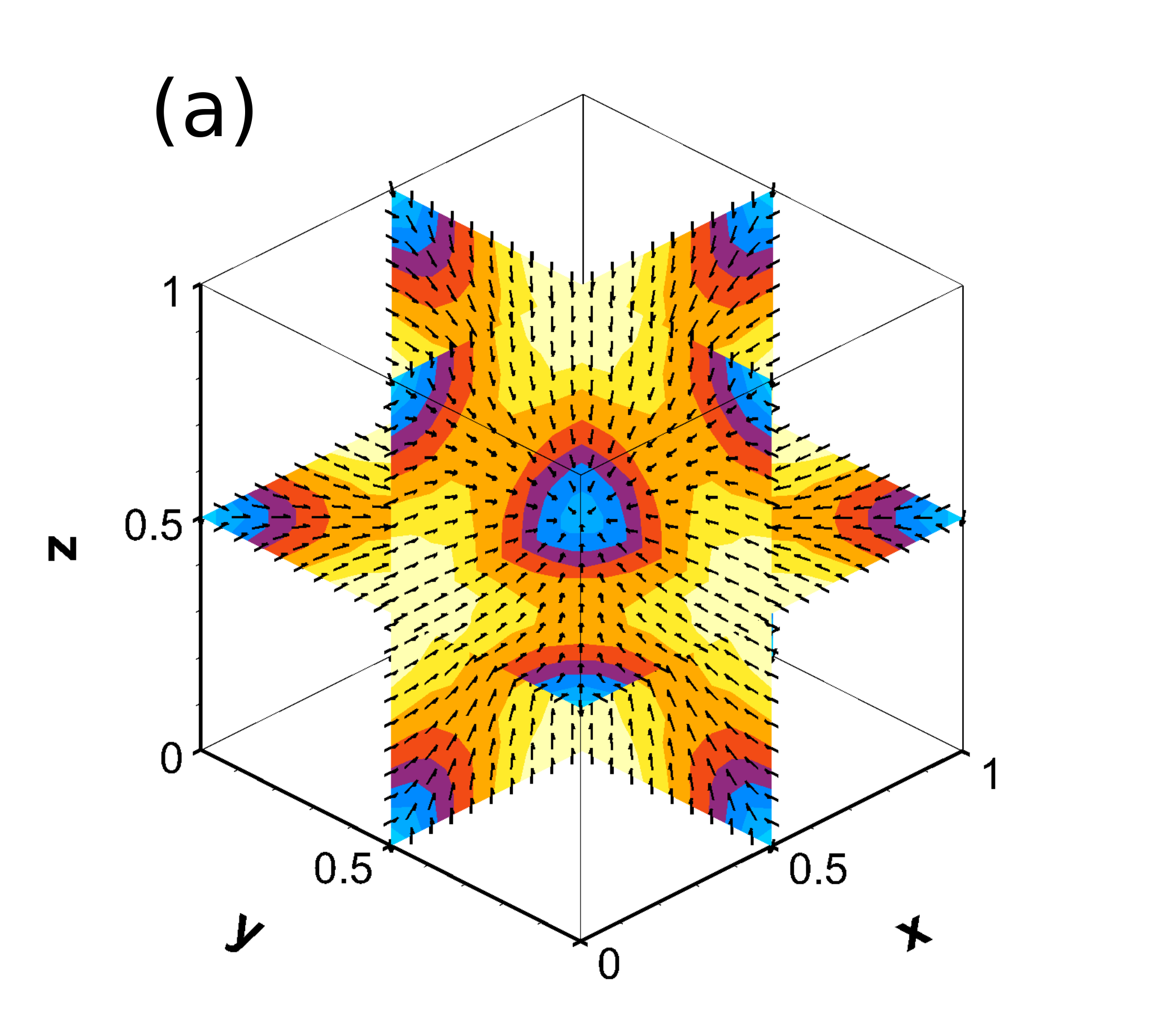}
\includegraphics[width=1.60in]{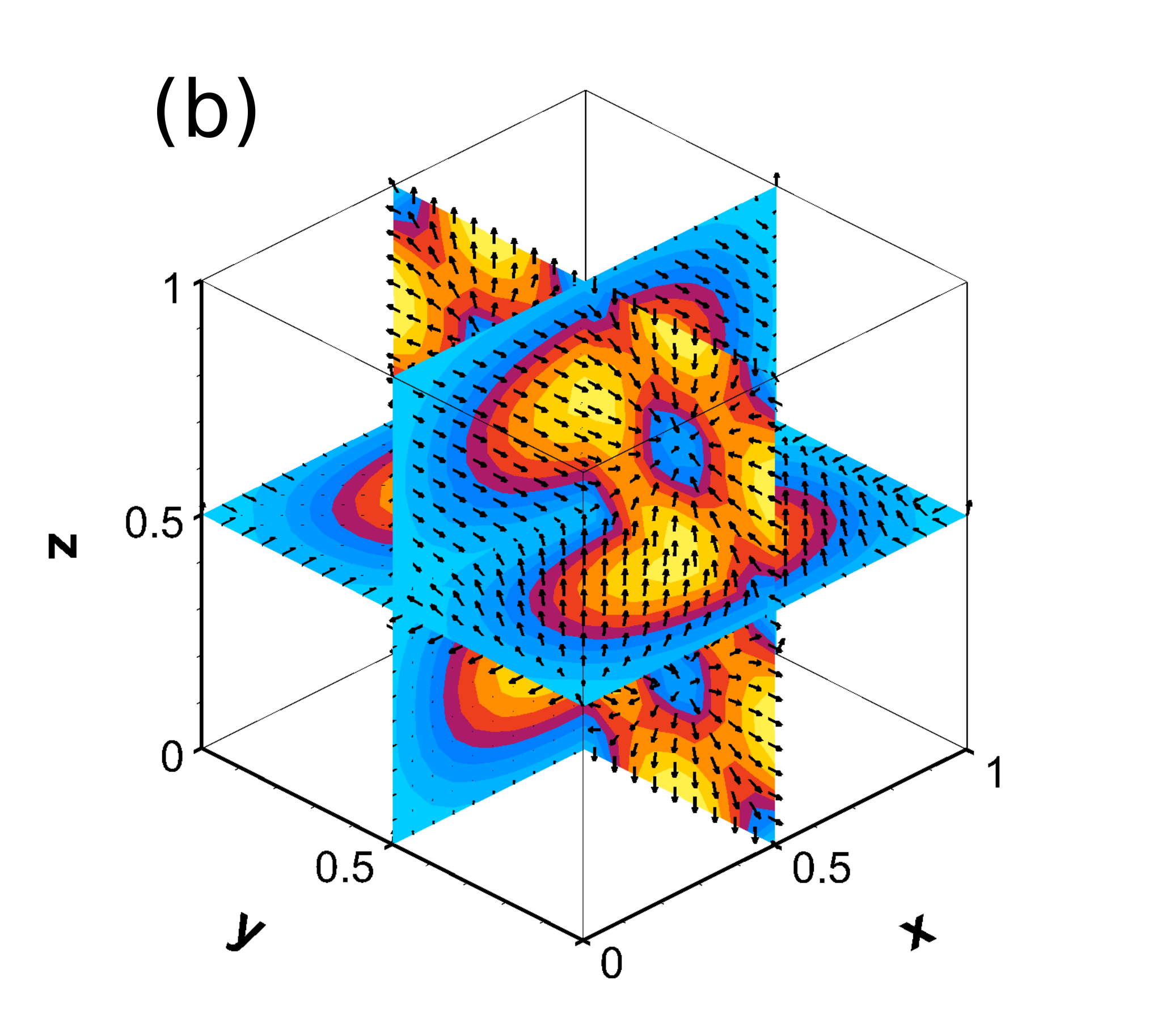}
\  \
\includegraphics[width=1.60in]{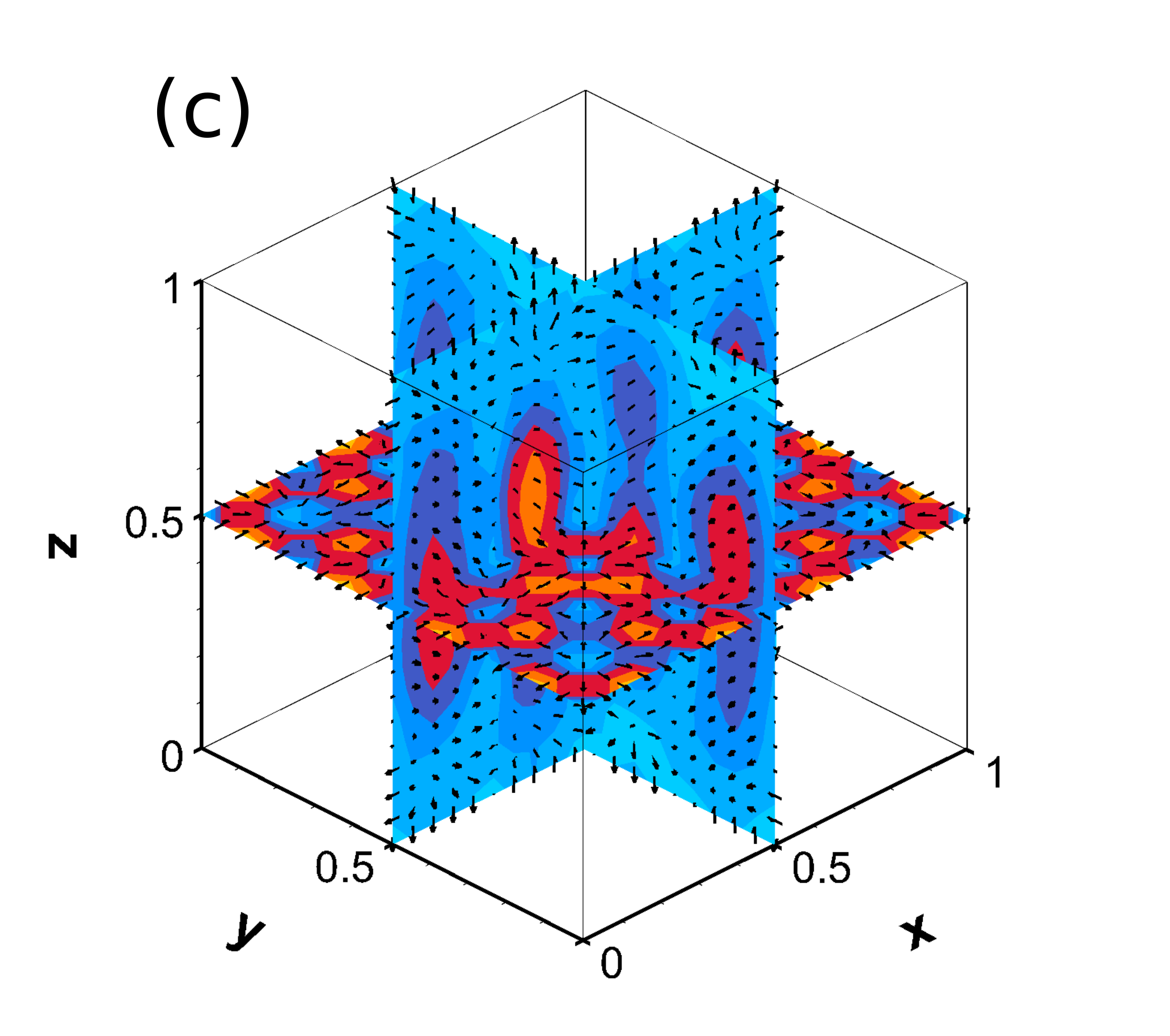} \\
\caption{Spurious  solutions in  the empty  cubic cavity  of
unit  dimensions. Note the field distribution, which is indicative of the presence of
sources,  even though  the cavity  is empty.  Light yellow(gray)
regions correspond to amplitude  antinodes, and blue(darker)
regions correspond to amplitude nodes. (Adapted from Pandey {\it et
  al.}, Ref.~\onlinecite{avs:LRRCavityEM2018}.)}
\label{avsfig:spurious}
\end{figure}

Numerical  solutions   of  Maxwell's   equations  are   polluted  with
non-physical spurious solutions. The divergence of Eq.~\ref{eq:wave1}
leads to
\begin{equation}\label{eq:spurious}
k_0^2 \, \nabla \cdot \epsilon_r \, \mathbf{E} = 0.
\end{equation}
This   condition   is   satisfied   when   either   $k_0   =   0$   or
$\nabla \cdot \epsilon_r \, \mathbf{E}  = 0$. 
In theory, these spurious solutions have zero frequency.  However, due to
discretization, the eigenvalues of the spurious modes are not computed
exactly as zero, and can have numerical  values comparable to those of the
physical solutions.  Consequently, the  spurious eigenvalues cannot be
easily      separated      from     the      desired      eigenvalues.
\cite{avs:Peterson1994,avs:SunCendes1995}
When the divergence condition is not satisfied, it implies the present
of charges inside the cavity. Examples of spurious solutions obtained
when the divergence condition is not imposed are shown in
Fig.~\ref{avsfig:spurious}. Note how the
field  distribution shows  source-like  behavior  within the  cavity,
indicative of non-zero divergence.
For the time-harmonic problem, spurious
solutions  have zero  curl and  a finite  divergence.  This  manifests
numerically  as a  very  large divergence-to-curl  ratio, compared  to
physically  admissible solutions.  In  the following  section we  show
within our HFEM approach we can  increase this ratio with increase in the
mesh density.   We use this  criterion to filter out  spurious solutions
from  the  physical  admissible  solutions.


\subsection{The penalty method and the zero-divergence
 constraint}
\begin{figure}[!tbp] 
\begin{center}
\includegraphics[width=2.9in]{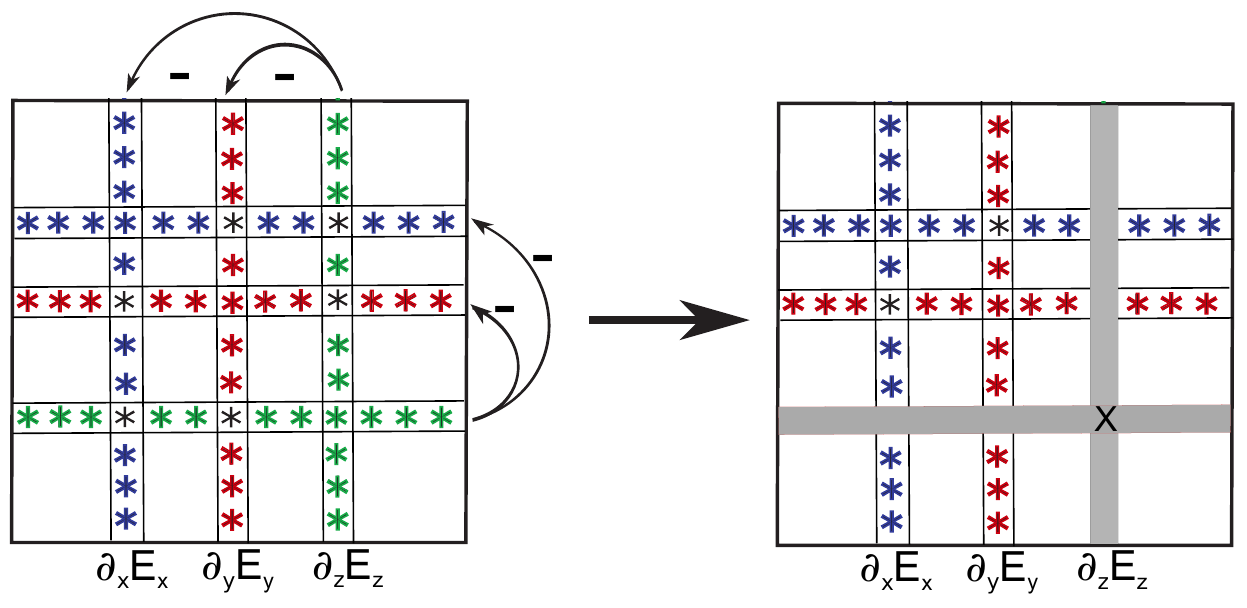}
\end{center}
  \caption{Explicity   imposing  the   divergence-free  condition   by
    performing global  matrix row  and column  operations. One  of the
    derivative DOF is eliminated in  favor of the remaining two.  Here
    $X=1$ for  the left-hand side  matrix, and $X$  is set to  a large
    value in the right-hand side matrix, in the generalized eigenvalue
    problem. This  choice of a  large number pushes the  eigenvalue of
    the redundant  $1\times1$ subspace  out of  the spectral  range of
    interest.}
\label{avsfig:matrix}
\vspace{-0.2in}
\end{figure}
The  penaly method  \cite{avs:RahmanDavies}  has been  proposed to  remove
spurious  solutions  in  nodal finite  element  implementations.   The
penalty term  pushes most  spurious eigenfrequencies outside
the spectral range of interest.   However, a fixed Lagrange multiplier
does not  remove all  spurious modes.  A low  value of the penalty factor
leaves behind some spurious modes, and a large value of the penalty factor
causes errors in the eigenvalues.  One  proposed algorithm is to use a
different multiplier  for each mode.\cite{avs:Webb1988} This,  however, is
an expensive  iterative scheme.   On the other hand, a constant penalty  offers an
inexpensive method for removing most of the zero-frequency spurious modes, and
can be further enhanced.  In our  calculation, we use a fixed Lagrange
multiplier, $\lambda = 1.$

One key feature of spurious  modes is a large divergence-to-curl ratio
$|\nabla\cdot {\bf E}|/|\nabla\times{\bf E}|$.\cite{avs:Webb1988} There is
then  the  possibility  of  identifying  and  removing  any  remaining
spurious solutions based on this ratio, after we determine the eigenfields.
However,  the divergence-to-curl  ratio  for  spurious and  physically
admissible  modes can  become comparable,  as  seen in  the first  and
second columns of Table~\ref{tab:div}. It is clear that the penalty
factor alone does not eliminate all the spurious solutions. There
still are spurious solutions that can intermix with the physically
acceptable solutions. Note that the Hermite shape functions supports
a non-zero divergence value within the brick volume. 
         
We  resolve  this issue  by  explicitly  imposing the  divergence-free
condition   at   each   node,   using  the   derivative   degrees   of
freedom.\cite{avs:Cendes1988,avs:Konrad1989,avs:Pinciuc2014}
At the matrix  level, one of the terms  in the zero-divergence condition,
 $\nabla \cdot  \epsilon_r {\bf E}=0$, is eliminated  in favor of
the other two.
The  procedure is demonstrated in Fig.~\ref{avsfig:matrix}   for   the
simpler   case   of   a   constant $\epsilon_r$.
Applying  this  technique  drives   the  divergence-to-curl  ratio  of
physically  admissible  solutions even  lower,  and  that of  spurious
solutions higher, as can be seen  from the third and fourth columns of
Table~\ref{tab:div}. A comparison of columns 2 and 4 in this table
show the enhancement of the ratio $|\nabla\cdot {\bf
  E}|/|\nabla\times{\bf E}|$ for spurious solutions, and a substantial
reduction of this ratio for the physical solutions. 
        
Additionally, since the divergence condition  is applied at each node,
the total divergence of  the physically admissible solutions decreases
further with  mesh refinement,  whereas it increases  considerably for
spurious solutions (see columns 4 and 6 in Table~\ref{tab:div}).  This
is manifested  as the element  size is reduced, and  the interpolation
from the nodes having the  divergence-free condition into the interior
of the elements is more effective with increasing mesh density.
 
This is in contrast to VFEM,  where the normal
discontinuity of  edge elements leads  to the formation  of artificial
charges at element interfaces, thus increasing the total divergence of
the     solutions;     this      problem     worsens     with     mesh
refinement.\cite{avs:Mur_edge_elements}
Note that  in VFEM, the zero-frequency  spurious solutions are
separated by filtering out the null-space of the curl operator
from        the         spectrum        using        iterative
techniques,\cite{avs:Arbenz,avs:Dziekonski} or  by finding eigenvalues
in the interior  of the spectrum. This is  necessary, since if
the physical solution  space is not normal to  the null space,
the physical solutions will be polluted by null vectors.

We also report the results for HFEM when tetrahedral elements are
used. The eigenvalues, along with the corresponding divergence-to-curl
ratios are listed in Table \ref{table:divtocurl_tet}. The number of
degrees of freedom in this calculation is 42438. As can be seen in the
divergence-to-curl values in Table \ref{table:divtocurl_tet}, using
tetrahedral elements for the breaking of the cavity, imposing the
penalty factor alone does not fully separate the physical solutions
from spurious solutions. Note also from Table
\ref{table:divtocurl_tet} that  both the penalty factor and the
divergence-free condition drastically improves the tagging of spurious
solutions. We observe less number of spurious solutions polluting the
spectrum as compared to the cubic elements. Another feature observed
is that the eigenvalues of the spurious solutions are distinct from
those of the physical solutions, as compared to the cubic element
case, where the spurious solutions are degenerate with the physical
ones.

\begin{figure}[!tbhp] 
\centering
\includegraphics[width=2.3in]{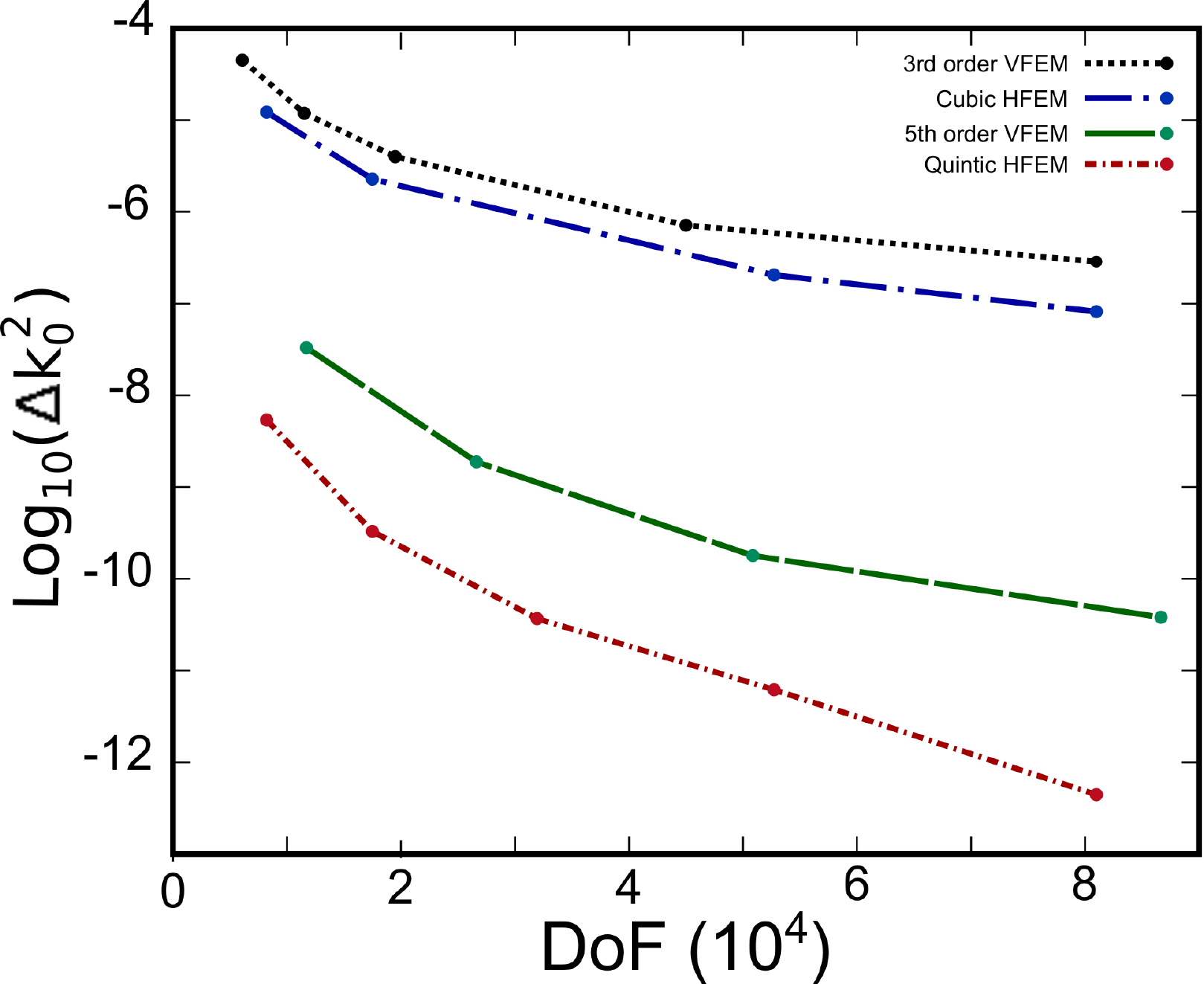} 
\includegraphics[width=2.3in]{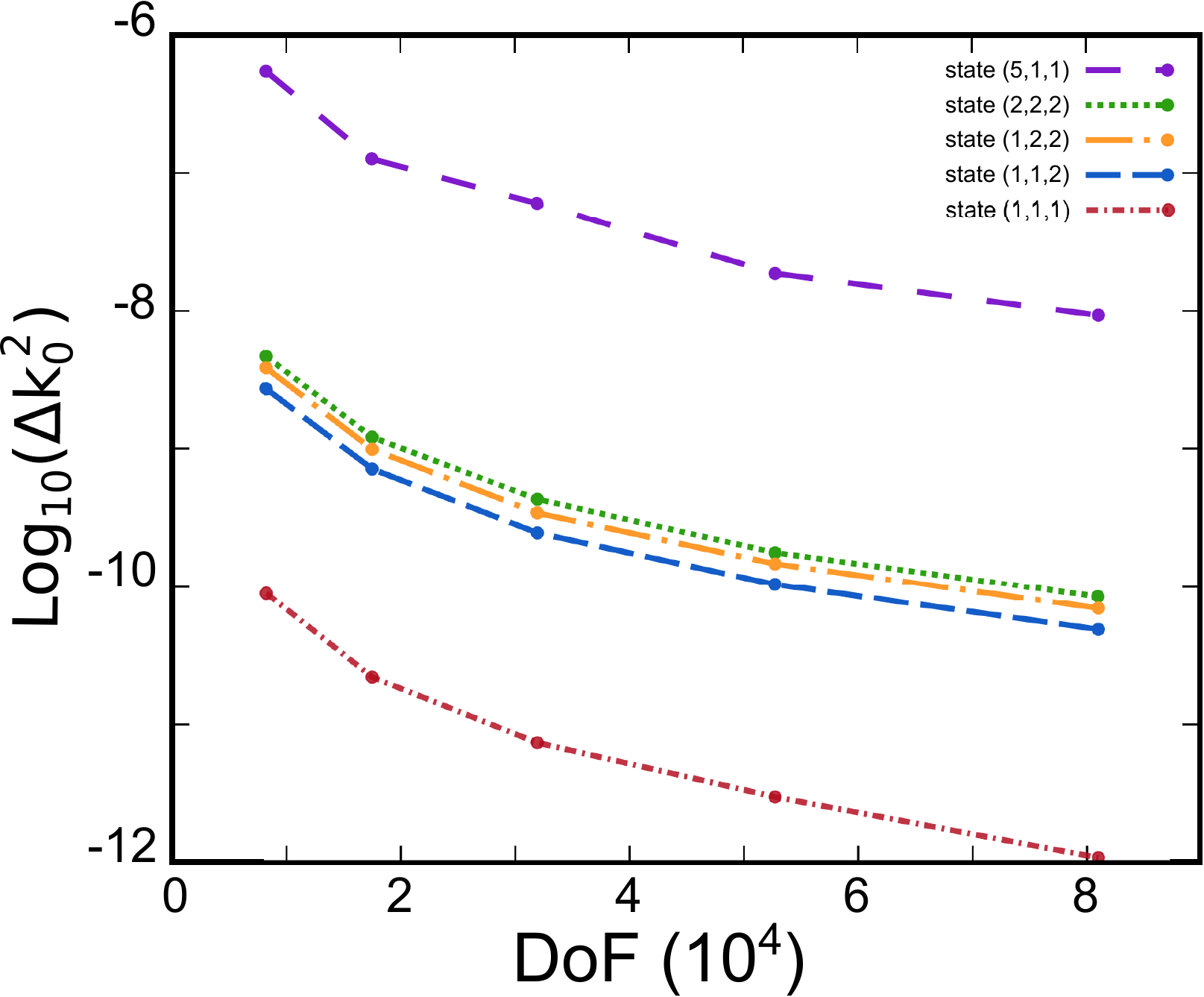}
\caption{\label{avsfig:conv_1} The convergence of errors in eigenvalues of (a) the
first mode for the  3rd  order VFEM,  5th  order VFEM,
Hermite,  and  Hermite interpolation  polynomials,
and (b) the higher frequency modes for HFEM only are shown
for the case of an  empty cubic cavity of unit dimensions.
Using  Hermite  interpolation  polynomials we  can
reduce the  error in  the first  mode upto  $10^{-9}$ with
just 27 elements  and 8232 DOFs, with  further reduction in
error  possible  with  mesh   refinement.  The  total  DOFs
corresponds to the global matrix dimension.}
\end{figure}

\begin{table*}[!htbp]   
  \begin{center}
                \caption{The enhancement of the  global divergence-to-curl ratio, for
                        the lowest few eigenstates, by imposing the
                        divergence-free condition at each node  at the element matrix level
                        are listed. The Hermite elements are defined
                        over a 27-node cubic element with a total of
                        216 DOFs. (From Pandey {\it et al.}, Ref.~\onlinecite{avs:LRRCavityEM2018}.)}
                \vspace{0.1in}
                \label{tab:div} 
                \def\arraystretch{1.3}
                \begin{tabular}{d{3.10}|d{3.9}|d{3.10}|d{3.9}|d{3.10}|d{3.9}|K{0.8in}} \hline \hline
                        \multicolumn{2}{c|}{Penalty factor method only} &
                        \multicolumn{4}{c|}{Penalty factor with $\nabla\!\cdot\!\textbf{E}=0$ condition} & \\
                        \cline{1-6}
                        \multicolumn{2}{c|}{DOF: $8232$} &
 \multicolumn{2}{c|}{DOF: $8232$} & \multicolumn{2}{c|}{DOF: $52728$} & Solution Type\\ 
                        \cline{1-6}
                        \multicolumn{1}{c|}{$k^2_0$} & \multicolumn{1}{c|}{$|\nabla\cdot {\bf
                                        E}|/|\nabla\times{\bf E}|$} & 
                        \multicolumn{1}{c|}{$k^2_0$} & \multicolumn{1}{c|}{$|\nabla\cdot {\bf
                                        E}|/|\nabla\times{\bf E}|$} &  
                        \multicolumn{1}{c|}{$k^2_0$} & \multicolumn{1}{c|}{$|\nabla\cdot {\bf
                                        E}|/|\nabla\times{\bf E}|$} & \\ 
                        \hline
                        19.739334159 & 0.000000000 & 19.739221048 & 0.000000000 & 19.739209007
                        &
                        0.000000000 & Physical \\  
                        19.739334159 & 0.000000000 & 19.739221048 & 0.000000000 & 19.739209007
                        & 0.000000000 & Physical \\ 
                        19.739334159 & 0.000000000 & 19.739221048 & 0.000000000 & 19.739209007
                        & 0.000000000 & Physical \\ 
                        \hline
                        29.609001232 & 0.005891138 & 29.608831573 & 0.000000197 & 29.608813510
                        & 0.000000003 & Physical \\ 
                        29.609001232 & 0.000389894 & 29.608831573 & 0.000000197 &
                        29.608813510 & 0.000000003 & Physical \\ 
                        29.609001232 & 172.872234679 & 31.328579414 & 220.518451137 &
                        31.501166266 & 545.498991584 & Spurious \\ 
                        \hline
                        49.358373749 & 0.000000000 & 49.349260110 & 0.000000000 &
                        49.348046601 & 0.000000000 & Physical \\ 
                        49.358373749 & 0.000000000 & 49.349260110 & 0.000000000 & 49.348046601 & 0.000000000 & Physical \\
                        49.358373749 & 0.000000000 & 49.349260110 & 0.000000000 & 49.348046601 & 0.000000000 & Physical \\
                        49.358373749 & 0.000000000 & 49.349260110 & 0.000000000 & 49.348046601 & 0.000000000 & Physical \\
                        49.358373749 & 0.000000000 & 49.349260110 & 0.000000000 & 49.348046601 & 0.000000000 & Physical \\
                        49.358373749 & 0.000000000 & 49.349260110 & 0.000000000 & 49.348046601 & 0.000000000 & Physical \\
                        \hline
                        59.228040826 & 0.572767636 & 59.218870635 & 0.000006213 & 59.217651104 & 0.000000056 & Physical \\
                        59.228040826 & 0.575823907 & 59.218870635 & 0.000006195 & 59.217651104 & 0.000000080 & Physical \\
                        59.228040826 & 0.137630780 & 59.218870635 & 0.000006005 & 59.217651104 & 0.000000112 & Physical \\
                        59.228040826 & 0.135054548 & 59.218870635 & 0.000006204 & 59.217651104 & 0.000000094 & Physical \\
                        59.228040826 & 0.000065277 & 59.218870635 & 0.000004082 & 59.217651104 & 0.000000043 & Physical \\
                        59.228040826 & 0.277005817 & 59.218870635 & 0.000008226 & 59.217651104 & 0.000000129 & Physical \\
                        59.228040826 & 1.063357739 & 62.432033897 & 167.942572387 & 62.873334761 & 321.742164741 & Spurious \\
                        59.228040826 & 1.083589808 & 62.432033897 & 167.942572387 & 62.873334761 & 321.742164741 & Spurious \\
                        59.228040826 & 3.580885859 & 62.432033897 & 167.942572387 & 62.873334761 & 321.742164741 & Spurious \\
                        \hline \hline 
                \end{tabular}
                \end{center}
\end{table*}
\begin{table*}[ht]
		\centering
		\caption{\label{table:divtocurl_tet} The enhancement
			of the  global divergence-to-curl ratio, for 
			the lowest few eigenstates, by imposing the
			divergence-free condition at each node  at the element matrix level
			are listed. The Hermite polynomials here are defined
			over a tetrahedral element with the 4 vertex
			nodes with 10 DOFs, and 4 face-centered nodes
			with 4  DOFs, resulting in a total of 56 DOFs for each element.
			The eigenmodes are obtained using 42438 DOFs.}
		\renewcommand{\arraystretch}{1.2}
		\begin{tabular}{c|c|c}
			\hline
			\hline
			Penalty factor method only & Penalty factor with $\nabla\!\cdot\!\textbf{E}=0$ condition & Solution \\
			\hspace{0.45in}$k_0^2$ \hspace{0.4in}  $|\nabla\cdot{\textbf E}|/|\nabla\times{\textbf E}|$ &\hspace{0.4in}  $k_0^2$ \hspace{0.4in}  $|\nabla\cdot{\textbf E}|/|\nabla\times{\textbf E}|$ & type\\
			\hline
			19.739208827 \hspace{0.2in} 0.000000000 & 19.739208827 \hspace{0.2in}  0.000000000 & Physical\\
			19.739208829 \hspace{0.2in} 0.000000000 & 19.739208829 \hspace{0.2in}  0.000000000 & Physical\\
			19.739208848 \hspace{0.2in} 0.000000000 & 19.739208849 \hspace{0.2in}  0.000000000 & Physical\\
			\hline
			29.608813476 \hspace{0.2in} 0.486030689 & 29.608813484 \hspace{0.2in}  0.000000002 & Physical\\
			29.608813484 \hspace{0.2in} 0.486018524 & 29.608813497 \hspace{0.2in}  0.000000002 & Physical\\
		    29.608813497 \hspace{0.2in} 0.495455431 & 41.117998502 \hspace{0.2in}  8.090436581 & Spurious\\
		    \hline
		    49.348030093 \hspace{0.2in} 0.000000045 & 49.348030229 \hspace{0.2in}  0.000000041 & Physical\\
		    49.348030963 \hspace{0.2in} 0.000000051 & 49.348031125 \hspace{0.2in}  0.000000045 & Physical\\
		    49.348032032 \hspace{0.2in} 0.000000054 & 49.348032161 \hspace{0.2in}  0.000000046 & Physical\\
		    49.348032074 \hspace{0.2in} 0.000000058 & 49.348032241 \hspace{0.2in}  0.000000049 & Physical\\
		    49.348032134 \hspace{0.2in} 0.000000059 & 49.348032302 \hspace{0.2in}  0.000000051 & Physical\\
		    \hline
		    59.217642131 \hspace{0.2in} 0.274980399 & 59.217643847 \hspace{0.2in}  0.000000072 & Physical\\
		    59.217644056 \hspace{0.2in} 0.242246173 & 59.217645045 \hspace{0.2in}  0.000000080 & Physical\\
		    59.217645004 \hspace{0.2in} 0.230927738 & 59.217654541 \hspace{0.2in}  0.000000124 & Physical\\
		    59.217652841 \hspace{0.2in} 0.276708712 & 59.217656977 \hspace{0.2in}  0.000000135 & Physical\\
		    59.217654456 \hspace{0.2in} 0.287009378 & 59.217657319 \hspace{0.2in}  0.000000137 & Physical\\
		    59.217655419 \hspace{0.2in} 0.276443701 & 59.217657643 \hspace{0.2in}  0.000000135 & Physical\\
		    59.217656808 \hspace{0.2in} 0.635597226 & N/A \hspace{0.7in} N/A & Spurious\\
		    59.217657209 \hspace{0.2in} 0.834678138 & N/A \hspace{0.7in} N/A & Spurious\\
		    59.217657452 \hspace{0.2in} 0.856486293 & N/A \hspace{0.7in} N/A & Spurious\\
			\hline
			\hline
		\end{tabular}
	\end{table*}
        
\subsection{Fields in an empty cubic cavity}
\label{section:emp_cube}
To demonstrate our method, we first model an empty
cubic   cavity   with   conducting   boundaries, with
$\epsilon_r=1$   and   $\mu_r=1$ inside.   Consider    a   cavity   of   unit
dimensions.      
These  calculations are
done using HFEM, with 8232 DOFs, within    a    parallel    computing
environment.\cite{avs:petsc,avs:slepc,avs:mumps1}

From Table~\ref{tab:emp_cube_eigen}, it is  clear that the eigenvalues
of  the empty  cavity  obtained  through our  scheme  have very  small
errors,    when   compared    to    the    analytical   values.     In
Fig.~\ref{avsfig:conv_1},  we  show  the convergence  of  the  calculated
frequencies  to their  analytical  values  in an  empty  cube of  unit
dimensions for both HFEM and VFEM (obtained using the software package
MFEM \cite{avs:mfem}).  As  the global number of DOFs  is increased through
mesh refinement, the accuracy improves steadily. Quintic HFEM delivers
an accuracy of 1 part in $10^9$  with just 8232 DOFs.  The second curve
from bottom (in green) obtained using  5th order VFEM shows about $10$
times  larger  error  for  comparable DOFs.   Even  with  further  mesh
refinement, VFEM has an error higher  than our HFEM scheme. HFEM gives
a higher accuracy  than VFEM, even with half the  number of DOFs.  This
reduction  in  required number  of  DOFs  leads to  improvement  in
computation time.  

The matrix bandwidth  is defined as the sum of  sub- and supradiagonal
arrays together  with the  main diagonal.   For a  total of  8232 DOFs
(52728 DOFs),  the cubic  Hermite polynomials  utilize a  bandwidth of
16175 (105167),  while the  quintic Hermite  interpolation polynomials
occupy a  comparable bandwidth of  16223 (105167). The occupancy  of a
matrix  is  defined  as  the  percentage of  nonzero  entries  in  the
matrix. While  going from  the cubic  to quintic  Hermite polynomials,
there is only  a nominal increase in the matrix  occupancy from 0.04\%
to  0.125\% for  a  matrix of  dimensions  $8232\times8232$.  With  an
increase in DOFs to 52728, the occupancy decreases further to 0.0083\%
for  the  cubic  Hermite,  and  is 0.025\%  for  the  quintic  Hermite
interpolation polynomials.  In the case of cubic Hermite interpolation
polynomials,    with   60    processors,   matrices    of   dimensions
$8232\times8232$  ($52728\times52728$) are  assembled  in 4.9  minutes
(50.9  minutes),  and  diagonalized  in 6.2  minutes  (41.2  minutes),
whereas using the quintic  Hermite interpolation polynomials, with the
same  number  of  processors,  matrices of  the  same  dimensions  are
assembled in  66.4 minutes  (471.4 minutes),  and diagonalized  in 9.6
minutes (78.6 minutes).   The time taken can be  reduced by optimizing
the number of  processors used for the calculation.  We  have used the
Krylov-Schur algorithm as implemented in SLEPc\cite{avs:slepc} for the
calculations.  We also  consider the  error in  eigenvalues of  higher
frequency modes  in Fig.~\ref{avsfig:conv_1}.  The errors  converge at
the same rate as the error in the first mode.

We note that while modeling cavities with curved boundaries, we can discretize the
curved regions with tetrahedral elements. As shown in Table~\ref{table:divtocurl_tet}, we will still be able
to maintain a similar level of accuracy in field calculations. 
        
\begin{table*}[!tbhp]   
  \caption{Eigenvalues for the forty lowest frequency modes of the empty
    cubic cavity, calculated numerically using  Hermite interpolation
    polynomials are listed. The eigenvalues  are compared with  their 
    analytical values and the absolute errors are displayed. (From
    Ref.~\onlinecite{avs:LRRCavityEM2018}.)}
\vspace{0.1in}
\label{tab:emp_cube_eigen} 
\centering
\def\arraystretch{1.3}
\begin{tabular}{c|c|d{3.14}|d{3.14}|l} \hline \hline
 Mode  & Degen-& \multicolumn{1}{c|}{Analytical}
                 & \multicolumn{1}{c|}{Hermite FEM} & Error \\
                 & eracy&&&\\
\hline
$(0,1,1)$ & 3 & 19.7392088021787 & 19.7392088021791 & 4.4\,$\times$10$^{-13}$\\
$(1,1,1)$ & 2 & 29.6088132032680 & 29.6088132032692 & 1.2\,$\times$10$^{-12}$\\
$(0,1,2)$ & 6 & 49.3480220054467 & 49.3480220078586 & 2.4\,$\times$10$^{-9}$\\
$(1,1,2)$ & 6 & 59.2176264065361 & 59.2176264089494 & 2.4\,$\times$10$^{-9}$\\
$(0,2,2)$ & 3 & 78.9568352087148 & 78.9568352135412 & 4.8\,$\times$10$^{-9}$\\
$(1,2,2)$ & 6 & 88.8264396098042 & 88.8264396146369 & 4.8\,$\times$10$^{-9}$\\
$(0,1,3)$ & 6 & 98.6960440108935 & 98.6960442784766 & 2.6\,$\times$10$^{-7}$\\
$(1,1,3)$ & 6 & 108.5656484119829 & 108.5656486795616 & 2.6\,$\times$10$^{-7}$\\
$(2,2,2)$ & 2 & 118.4352528130723 & 118.4352528203570 & 7.2\,$\times$10$^{-9}$\\
\hline \hline 
\end{tabular}
\end{table*}    
\subsection{Accidental degeneracies in EM cavities}
\label{section:group_theory}
Physical properties arising from the symmetry of the system can be
treated efficiently using group representation theory. It has been well
appreciated in quantum mechanics that the degeneracies in the energy
spectrum arise from the symmetry group of the corresponding
Hamiltonian.\cite{avs:tinkham,avs:dresselhaus} The degeneracy of an eigenvalue
is equal to the dimensionality of the corresponding irreducible
representation of the symmetry group. If we have any other additional
degeneracy in the spectrum  which cannot be explained by the  obvious
geometrical symmetry of the system, it is  labeled   as  ``accidental
degeneracy.'' In this section we discuss the presence of such
accidental degeneracy and its removal for EM modes in a
cavity. 
Pedagogical remarks on accidental degeneracy are given in  
Sec.~\ref{section:grouptheory}.

Let  $G$   be  a  group  of   order  $g$  and  $\Gamma^{(i)}$   be  an
$l_i$-dimensional representation of  $G$.  For a group  element $R$ in
$G$, its representation  is given by a $l_i \times  l_i$ square matrix
$\Gamma^{(i)}(R)$.    Then   the   projection   operator\cite{avs:tinkham}
corresponding to $\Gamma^{(i)}$ is given by
\begin{equation}\label{eq:proj}
\mathcal{P}^{(i)}=\frac{l_i}{g}\sum_{R}\chi^{(i)}
\left(R\right)\cdot P_{R},
\end{equation}
where  $\chi^{(i)}(R)$ is  the  character and  $P_R$  is the  operator
corresponding   to  the   element   $R$. The transformation of electric fields
under the operation $P_R$ is defined by 
\begin{equation}
  P_{R}\cdot \mathbf{E}\left(\bm{r}\right) =
  R\cdot\mathbf{E}\left(R^{-1}\cdot\bm{r}\right). 
\end{equation}
The projection  operator $\mathcal{P}^{(i)}$ projects out  the part of
the   field   $\mathbf{E}$   that  belongs   to   the   representation
$\Gamma^{(i)}$. 

The  electric  or  magnetic  fields
corresponding to a degenerate eigenfrequency will form {\it a
set   of   vector   basis-functions  for   the   irreducible
representations of the symmetry  group.} Previously, we have
derived  a coefficient  formula to  recognize the  irreducible
representation  corresponding to  an eigenenergy,  and obtain
the    symmetry     adapted    wavefunctions     in    quantum
dots.\cite{avs:RemAccDegen17}  An analogous  coefficient formula  exists   
even  in the context of electric   (magnetic)  modes   in  EM
cavities.
Let $\left\{\mathbf{E}_i\right\}_{i=1}^{n}$ be the set
of eigenfields for the physical  system under consideration. Then the
coefficient formula\cite{avs:RemAccDegen17} is given by
\begin{equation}\label{eq:coeff_formula}
c^{(i)}_{jk} = \int_{V}d^3r\
\mathbf{E}_j^{\dagger}\cdot\left(\mathcal{P}^{(i)}\mathbf{E}_k\right). 
\end{equation}  
If the  coefficient is  nonzero, then the  field $\mathbf{E}_k$  has a
component  in  the  $i^{th}$-representation and  $\mathbf{E}_j$  is  a
partner.   Using    these   coefficients   we   can    construct   the
symmetry-adapted electric  (magnetic) fields which are  exclusively in
the $i^{th}$-representation.
\begin{table*}[th!] 
\caption{Different possible even and odd combinations of eigenmodes
in an empty cubic cavity, their degeneracy, and corresponding
irreducible representations for the symmetry group $O_h$ are
listed. Here the indices $m,n$ and $\ell$ are non-zero integers. (Adapted from
Ref.~\onlinecite{avs:LRRCavityEM2018}.)}  
\label{tab:basisirrep_oh}
\vspace{0.1in}
\centering
\bgroup
\def\arraystretch{2}
{\begin{tabular}{l l|c|c l}
\hline
&Mode number & Degeneracy &Irreducible representations&\\
\hline\hline
&$(0,2n-1,2n-1)$                & 3 &   $T_{1u}$&\\
&$(0,2n-1,2m)$          & 6 &   $T_{1g} \oplus T_{2g}$&\\
&$(0,2n,2n)$            & 3 &   $A_{2u} \oplus E_u$&\\
&$(0,2n-1,2m-1)$        & 6 &   $T_{1u} \oplus T_{2u}$&\\
&$(0,2n,2m)$                    & 6 &   $A_{1u} \oplus A_{2u} \oplus 2\,E_u$&\\
&$(2n-1,2n-1,2n-1)$     & 2 &   $E_g$&\\
&$(2n, 2n, 2n)$                 & 2 &   $E_u$&\\        
&$(2n-1,2n-1,2m)$       & 6 &   $T_{1u} \oplus T_{2u}$&\\
&$(2n,2n,2m-1)$         & 6 &   $T_{1g} \oplus T_{2g}$&\\
&$(2m-1,2n-1,2n-1)$     & 6 &   $A_{1g} \oplus A_{2g}\oplus 2 E_g$&\\
&$(2m,2n,2n)$           & 6 &   $A_{1u} \oplus A_{2u} \oplus 2 E_u$&\\        
&$(2n-1,2m-1,2\ell)$       & 12 &  $2\,T_{1u} \oplus 2\,T_{2u}$&\\
&$(2n,2m,2\ell-1)$         & 12 &  $2\,T_{1g} \oplus 2\,T_{2g}$&\\
&$(2m,2n,2\ell)$           & 12 &  $2\,A_{1u} \oplus 2\,A_{2u} \oplus 4\,E_{u}$&\\
&$(2m-1,2n-1,2\ell-1)$     & 12 &  $2\,A_{1g} \oplus 2\,A_{2g} \oplus 4\,E_{g}$&\\
\hline\hline
\end{tabular}}
\egroup
\end{table*}

In case of an empty cubic cavity resonator of
length $a$ with conducting 
boundaries. The eigenmodes supported by the cubic resonator have the
eigenvalues 
\begin{align}
\label{eq:eig}
k_0^2 = \left( n_1^2+n_2^2+n_3^2 \right) \, \frac{\pi^2}{a^2},
\end{align}
and the electric field components are given by
\begin{align}\label{eq:cubic_efield}
E_x &= E_{0x}\,\cos\left(\frac{n_1\pi
      x}{a}\right)\sin\left(\frac{n_2\pi
      y}{a}\right)\sin\left(\frac{n_3\pi z}{a}\right); \nonumber \\ 
        E_y &= E_{0y}\,\sin\left(\frac{n_1\pi
              x}{a}\right)\cos\left(\frac{n_2\pi
              y}{a}\right)\sin\left(\frac{n_3\pi z}{a}\right); \\ 
        E_z &= E_{0z}\,\sin\left(\frac{n_1\pi
              x}{a}\right)\sin\left(\frac{n_2\pi
              y}{a}\right)\cos\left(\frac{n_3\pi z}{a}\right),
              \nonumber 
        \end{align}
where $n_1,n_2,n_3$ are non-zero integers, and $E_{0x},E_{0y},E_{0z}$
are the field amplitudes in each direction. 

Three kinds of degeneracies can be identified in the spectrum. The first kind is due to the
permutation of mode numbers.
The second kind is a consequence of the
divergence-free condition $\nabla\cdot\mathbf{E}=0$. On substituting
Eq.~(\ref{eq:cubic_efield}) in the divergence-free condition we obtain
the constraint $n_1\,E_{0x}+n_2\,E_{0y}+n_3\,E_{0z}=0$. Hence, if
$n_1,n_2,n_3\neq0$ we see that there are two independent field
components; hence for a given mode $(n_1,n_2,n_3)$ we will have at
least 2 degenerate field solutions.\cite{avs:nayfeh_book}  The third kind occurs when two disitinct sets
of mode numbers give the same frequency, occurring when the following relation is satisfied:
\begin{equation}
n_1^2+n_2^2+n_3^2 = m_1^2+m_2^2+m_3^2,
\end{equation}
with $n_i\neq m_i$, for $i=1,2,3$.

We know that the cubic cavity has the geometrical symmetry of $O_h$. In
Table~\ref{tab:basisirrep_oh}, we list all different possible
combinations of mode numbers and their corresponding irreducible
representations from the symmetry group $O_h$. Here, we have accounted
for only the first two kinds of degeneracies. For most of the
combinations of mode numbers we observe accidental degeneracy since
they belong to two or more distinct irreducible representations. The
accidental degeneracy associated with permutation of mode numbers can
be rendered normal by identifying the existence
of two dynamical operators ${\boldsymbol \Omega}=(\Omega_1,\Omega_2)$,
given by\cite{avs:LRRCavityEM2018}
\begin{eqnarray}
  \label{eq:NewOperators}
\Omega_1&=&\left(\partial_x^2-\partial_y^2\right) \nonumber\\
\Omega_2&=&\left(2\,\partial_z^2-\partial_x^2-\partial_y^2\right) 
\end{eqnarray}
which 
connect the degenerate field solutions.\cite{avs:fernandez} The full
covering group in this case is ${\mathcal G}=O_h\otimes{\boldsymbol \Omega}$.

As described in the following section, 
the larger  symmetry  group $\mathcal{G}$  of  the cavity  is
reduced to  $O_h$ by  introducing a  concentric  cubic dielectric  inclusion inside  the
cavity.  This  inclusion removes the  accidental degeneracy
due to  the way the fields  occupy the corner regions  exterior to the
cubic dielectric.


\subsection{Fields in dielectrically loaded cubic cavity}
\label{section:load_cube}
\begin{figure}[!tbp] 
\begin{center}
\includegraphics[width=2.3in]{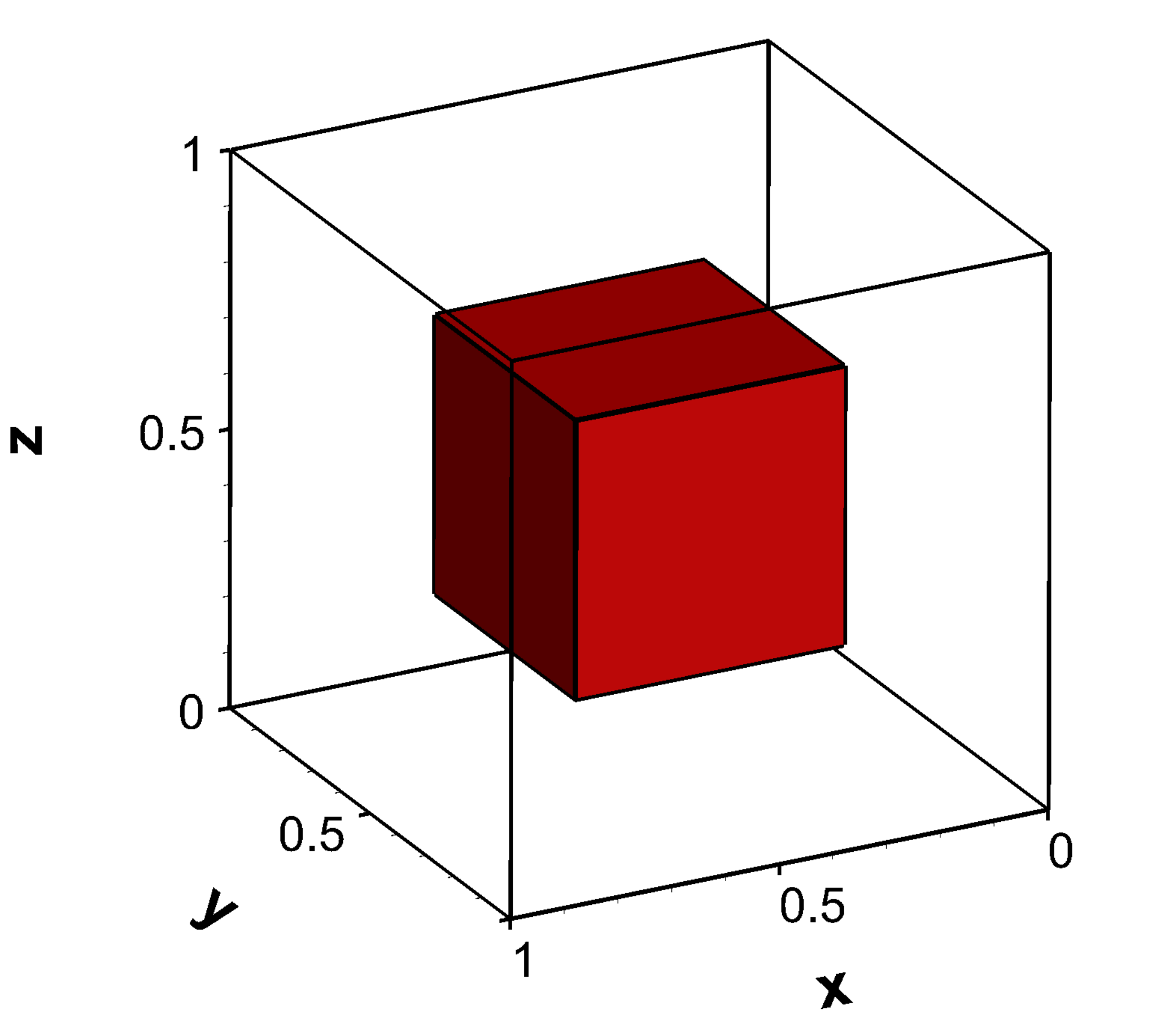}
\end{center}
  \caption{Schematics  of a  dielectrically loaded  cavity with  $O_h$
    symmetry.   The   external   conducting  cavity   has   dimensions
    $1\times1\times1$  mm$^3$;   the  cubic  dielectric   loading  has
    dimensions   $0.5\times0.5\times0.5$   mm$^3$,   with   dielectric
    constant $\epsilon_2$. The  permittivity in the rest  of cavity is
    $\epsilon_1$.}
\label{avsfig:loaded_cavity}
\end{figure}

We consider a cubic  conducting cavity of dimensions $1\times1\times1$
mm$^3$.  This cavity  is  loaded with  a  concentric cubic  dielectric
inclusion,   of   dimensions   $0.5\times0.5\times0.5$   mm$^3$,   and
permittivity $\epsilon_2$,  as shown  in Fig.~\ref{avsfig:loaded_cavity}.
The  permittivity  in   the  rest  of  cavity   is  $\epsilon_1$.  The
eigenvalues of  the first few  modes are tabulated for  the dielectric
ratios         $\epsilon_2/\epsilon_1        =         1.2$        and
$\epsilon_2/\epsilon_1 =  5.0$. The  calculations are done  with 17576
DOFs, to accurately model the dielectric function.

\begin{figure}[!tbp]  
  \centering
\includegraphics[width=1.6in]{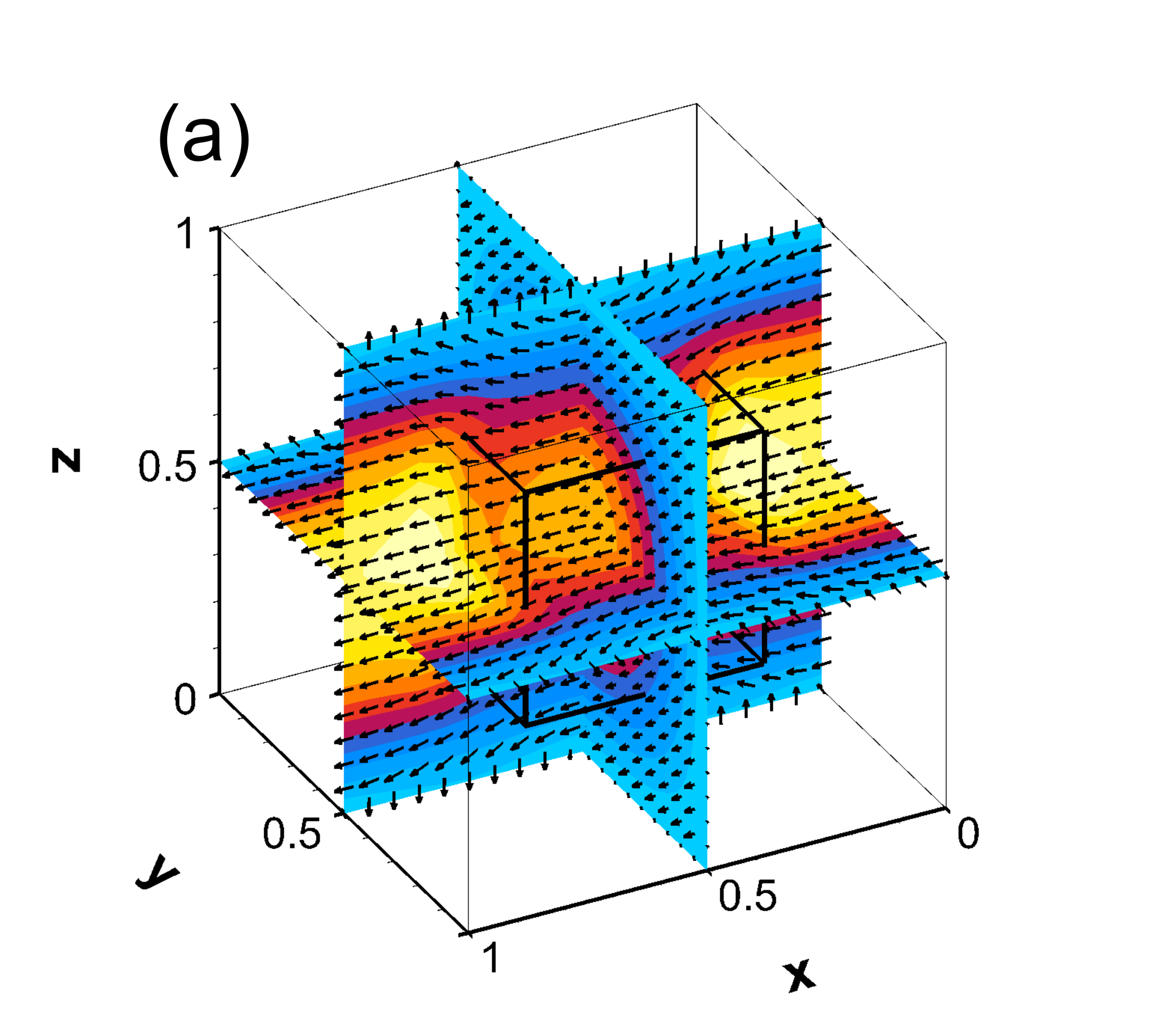} 
\includegraphics[width=1.6in]{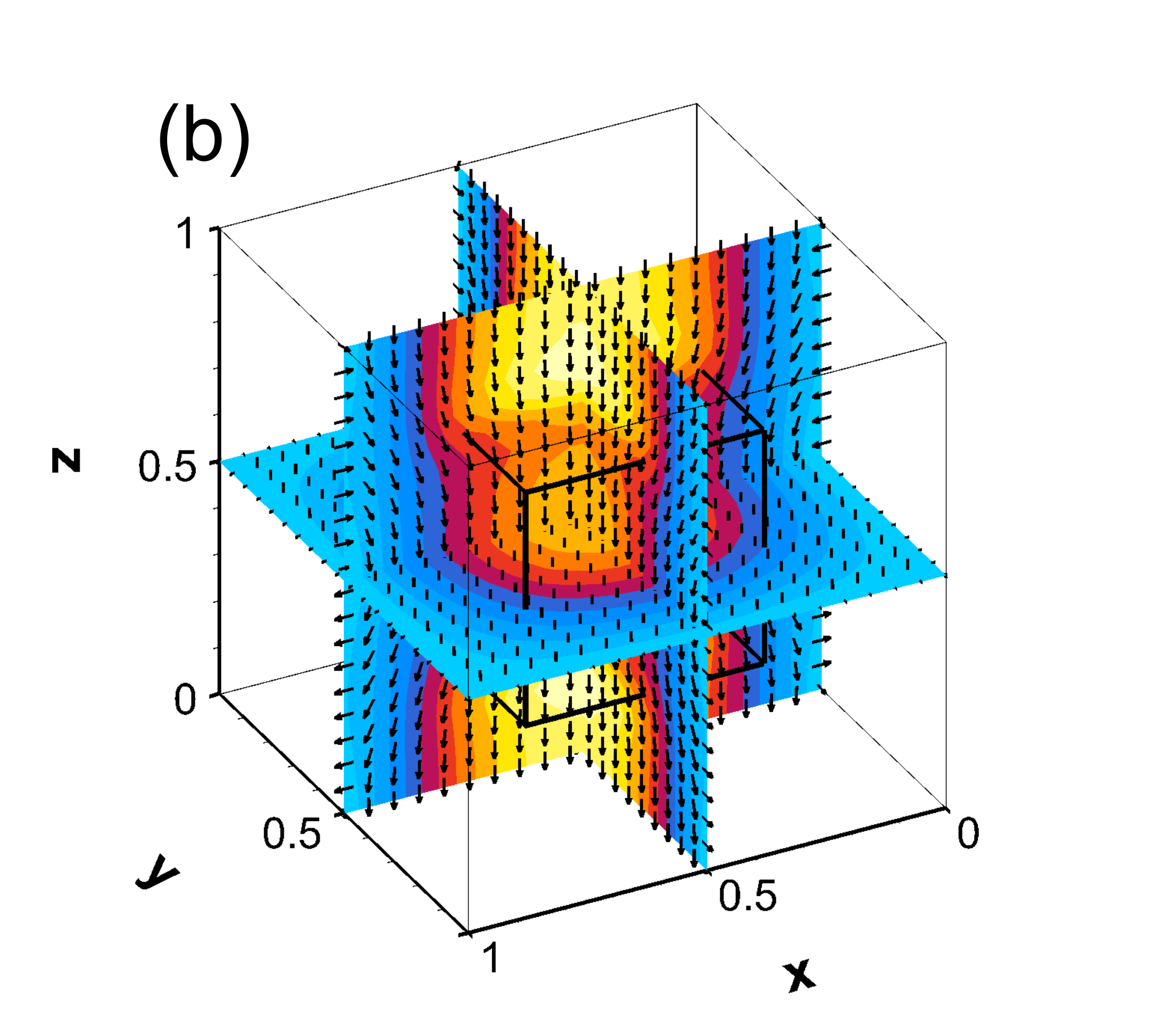} \\
\includegraphics[width=1.6in]{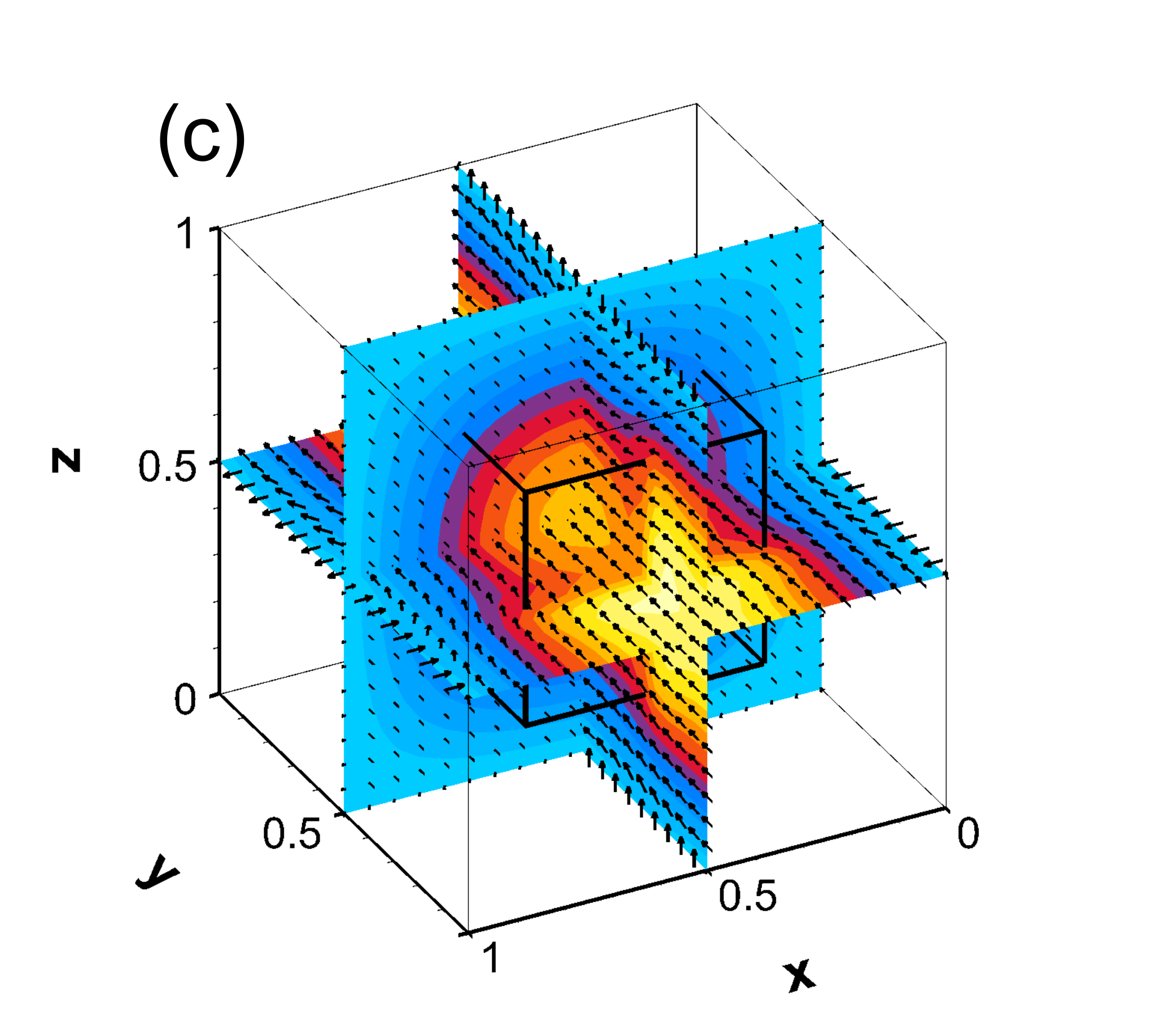}
  \caption{The symmetry-adapted  triplet states  $(0,1,1)$, $(1,0,1)$,
    $(1,1,0)$ of  the dielectrically loaded cavity  with an eigenvalue
    $k_0^2 =  18.5267$ are  shown in (a), (b) and (c), respectively. The  inserted cubic  dielectric has
    dimensions   $0.5\times0.5\times0.5$   mm$^3$,   with   a dielectric
    constant  $\epsilon_2/\epsilon_1  =  1.2$.  Light  yellow (gray)  regions
    correspond  to   amplitude  antinodes,  and  blue (darker)  regions
    correspond to amplitude nodes. (From Pandey {\it et al.}, Ref.~\onlinecite{avs:LRRCavityEM2018}.)}
  \label{avsfig:load_011}
\end{figure}

\begin{figure}[!t]  
  \centering
\includegraphics[width=3.5in]{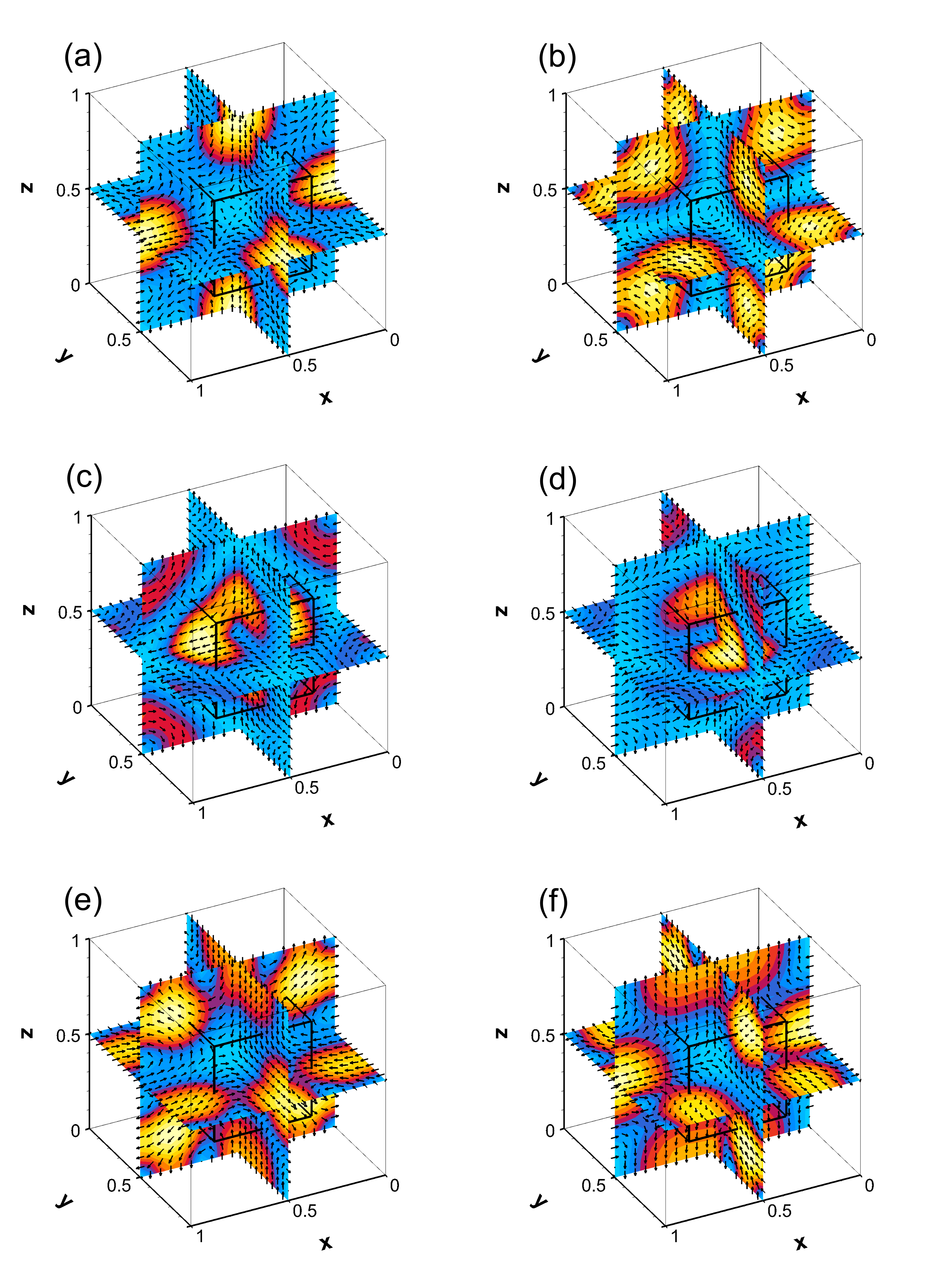}
  \caption{Electric    fields   for    the   symmetry-adapted    modes
    $(1,1,3),(1,3,1),(3,1,1)$,  of the  dielectrically loaded  cavity,
    with eigenvalues (a)  $k_0^2 = 107.43$, (b)  $k_0^2 = 107.59$,
    \mbox{(c, d)} $k_0^2 = 103.73$, and \mbox{(e, f)} $k_0^2 = 107.60$ are shown.  The
    cubic  dielectric has  dimensions $0.5\times0.5\times0.5$  mm$^3$,
    with  a  dielectric  constant  $\epsilon_2/\epsilon_1  =  1.2$.   The
    singlets belong  to the 1D  representations (a) $A_{1g}$,  and (b)
    $A_{2g}$ of the  group $O_h$. Fields in (c, d) and (e, f)  are partners, and each pair 
    form a basis for the 2D representation $E$. Light yellow (gray)  regions correspond to
    amplitude  antinodes,   and  blue (darker)  regions   correspond  to
    amplitude nodes. (From Pandey {\it et al.}, Ref.~\onlinecite{avs:LRRCavityEM2018}.)}
  \label{avsfig:load_cav}
  \vspace{-0.2in}
\end{figure}

In Figs.~\ref{avsfig:load_011}--\ref{avsfig:load_cav} we show electric field
distributions  for the  first  few  modes in  the  loaded cavity  with
$\epsilon_2/\epsilon_1=1.2$. As shown  in Fig.~\ref{avsfig:load_011}, the
first three  modes in  the loaded cavity  remain degenerate,  and they
belong to the  three dimensional representation $T_{1u}$  of the group
$O_h$.
An instance of the removal of accidental degeneracy can be seen in the
$(1,1,3),(1,3,1),(3,1,1)$ modes,  which in the empty  cubic cavity are
part    of   a    degenerate   sextuplet    which   belong    to   the
\mbox{$A_{1g}\oplus    A_{2g}\oplus    2E_g$}   representation    (see
Table~\ref{tab:basisirrep_oh}).  This  sextuplet decomposes into
four       separate       irreducible       representations.        In
Figs.~\ref{avsfig:load_cav}(a,b), we show the singlet modes in the loaded
cavity  belonging  to  the irreducible  representations  $A_{1g}$  and
$A_{2g}$, respectively.  We note that in Fig.~\ref{avsfig:load_cav}(b)
the magnitude of the electric field has complete $O_h$ symmetry, while
the  vectors flip  their  directions under  an inversion.   Similarly,
Figs.~\ref{avsfig:load_cav}(c,d)  and \ref{avsfig:load_cav}(e,f)  show
symmetry-adapted  partners,  which  belong   to  the  two  dimensional
representations $E_g$.
 
\begin{figure}[!th]  
	\centering
	\includegraphics[width=1.6in]{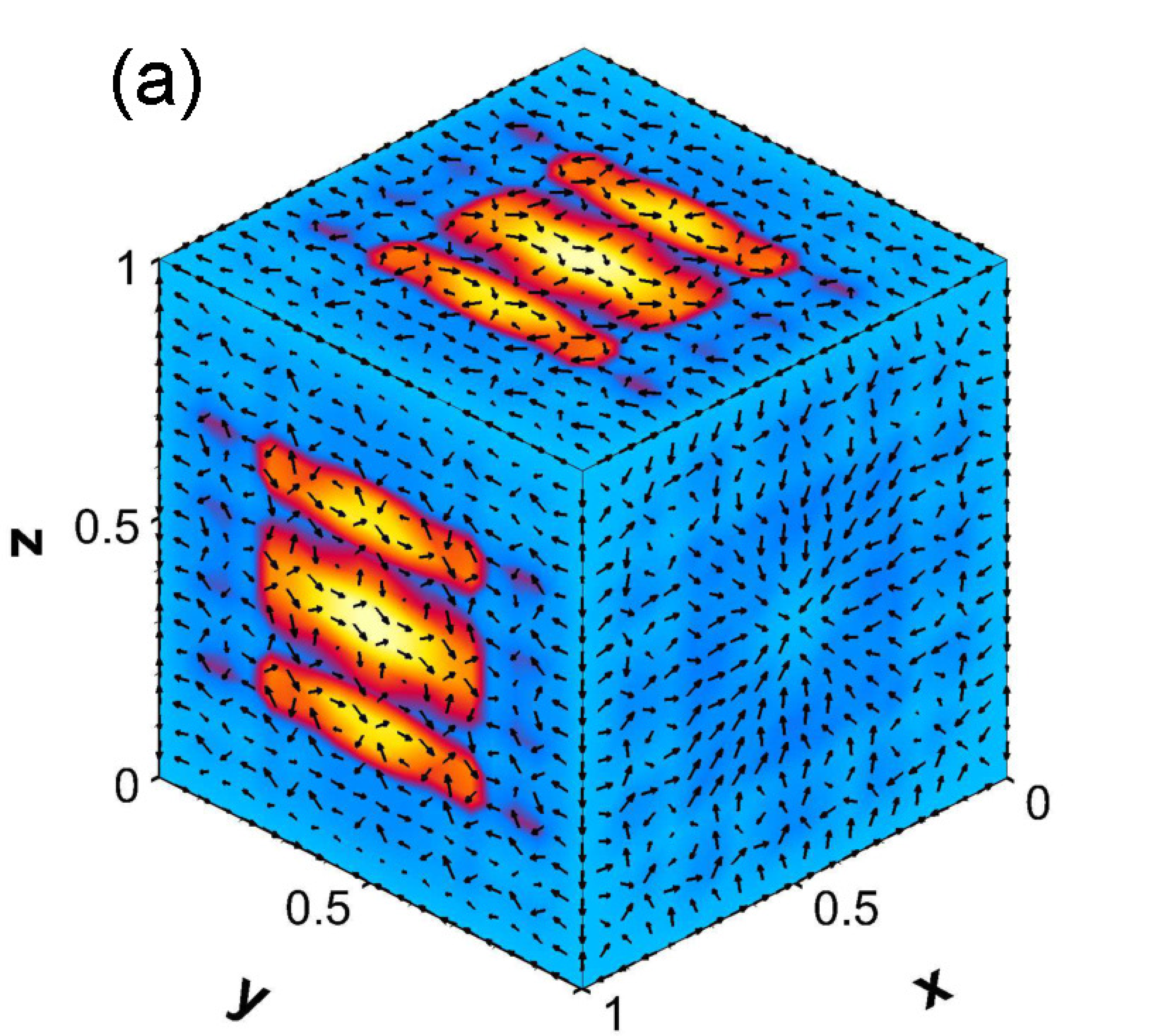}
	\includegraphics[width=1.6in]{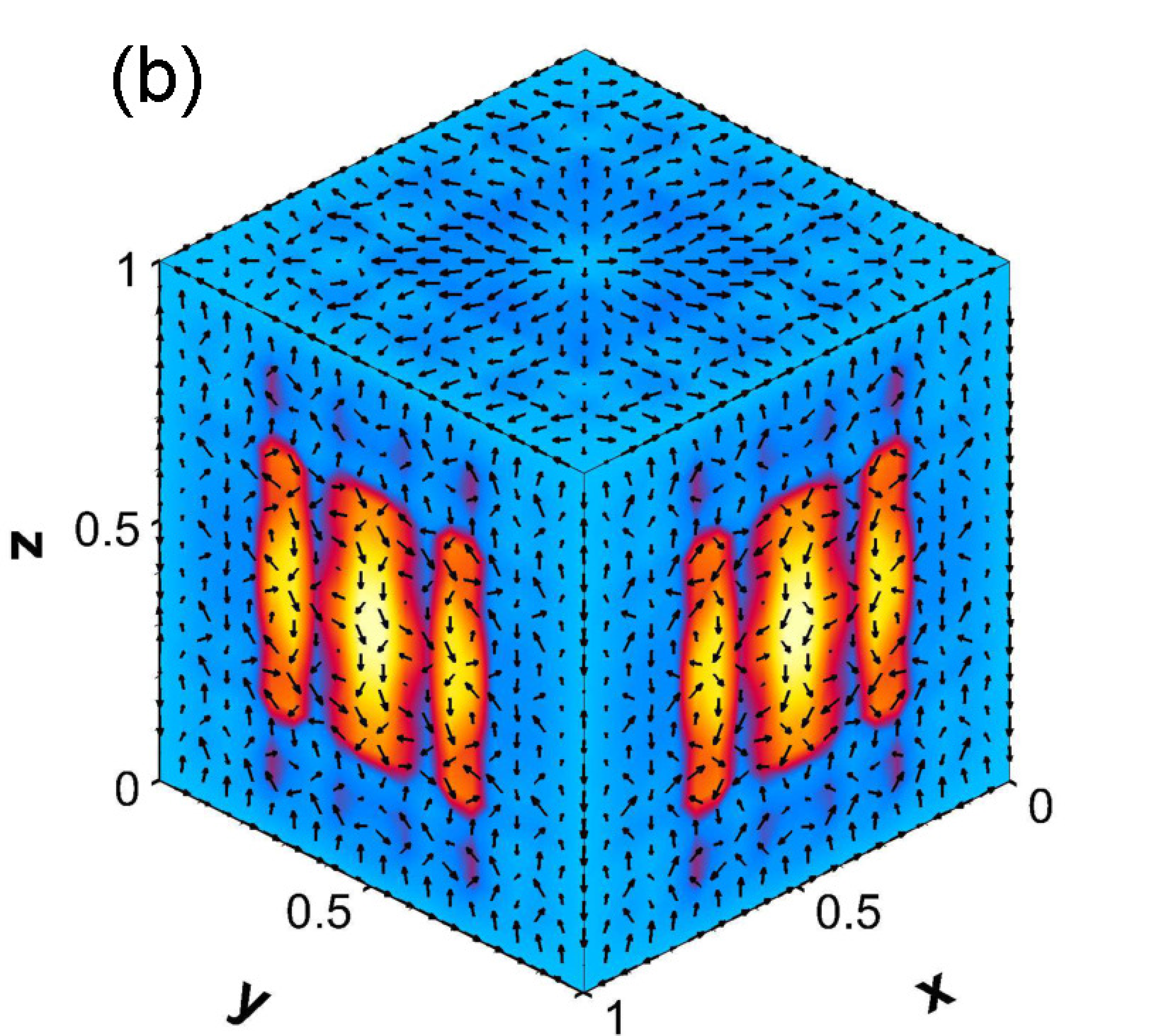}\\
	\includegraphics[width=1.6in]{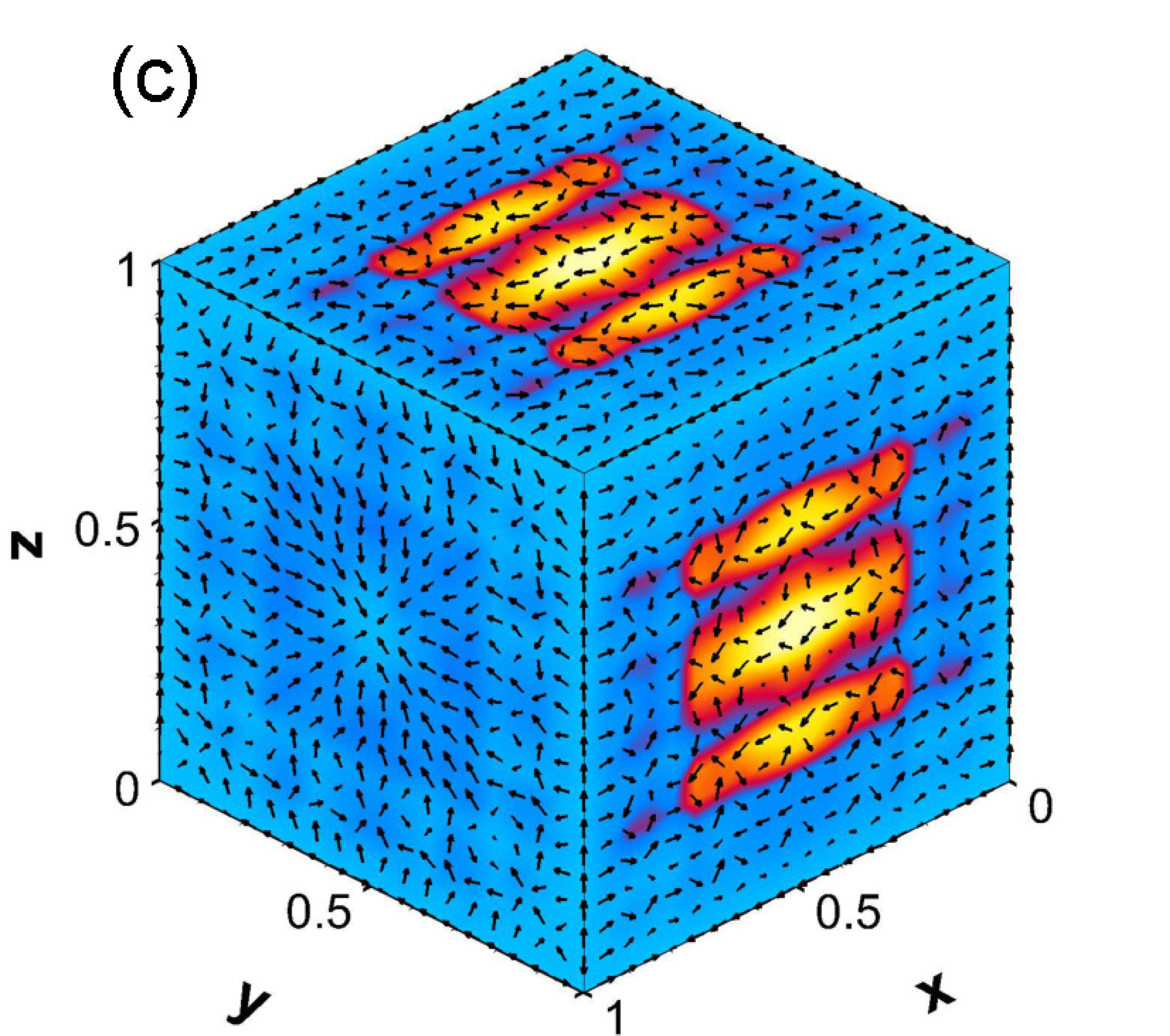}
	\caption{Surface   currents    for   the    symmetry-adapted   modes
		$(0,1,1),(1,0,1),(1,1,0)$,  of the  dielectrically loaded  cavity 
		are shown.  The cubic dielectric   has dimensions $0.5\times0.5\times0.5$
		mm$^3$,  with a dielectric  constant $\epsilon_2/\epsilon_1  = 1.2$.
		Light  yellow (gray)  regions
		correspond  to   amplitude  antinodes,  and  blue (darker)  regions
		correspond to amplitude nodes.}
	\label{avsfig:load_current}
	\vspace{-0.1in}
\end{figure}

\begin{figure}[bh!] 
\begin{center}
\includegraphics[width=2.8in]{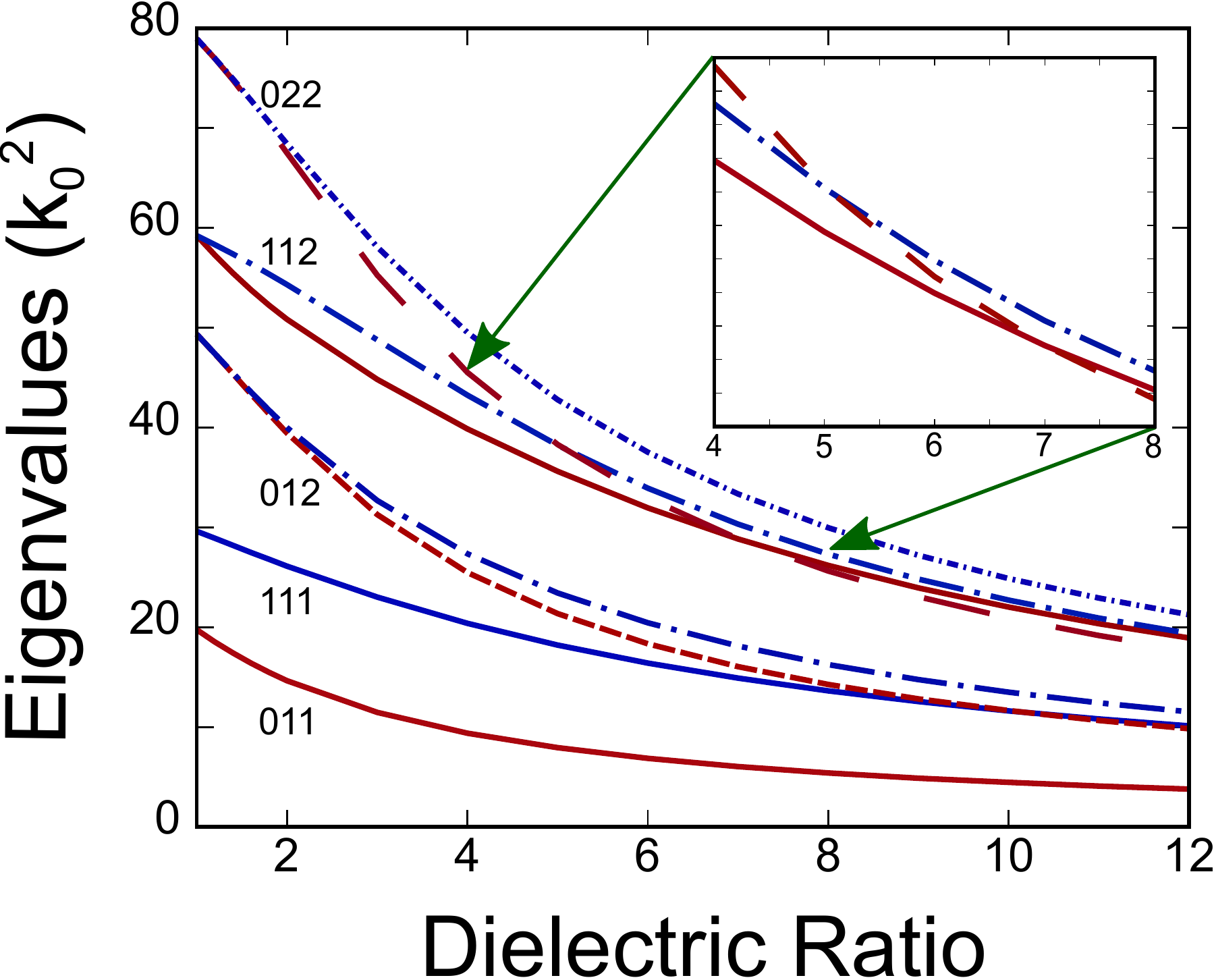}    
\end{center}
\caption{The evolution of eigenvalues as the dielectric constant
in the cavity is varied from 1.0 to 12.0. Sextuplet-level
degenerate modes of the empty cavity are  seen to split into triplets with the inclusion
of cubic dielectric loading. (From Pandey {\it et al.}, Ref.~\onlinecite{avs:LRRCavityEM2018}.)}  
\label{avsfig:crossing}
\end{figure}

In  Fig.~\ref{avsfig:crossing}, the  evolution of  mode frequency
on  varying the  dielectric   constant  $\epsilon_2$  in  the
interior is shown  for the lowest few modes.  We observe level
crossings  akin to  the  case  of bound  states  in a  finite
quantum well as the well depth is varied.\cite{avs:RemAccDegen17}
In Figs.~\ref{avsfig:load_current}, we plot the surface
currents on the conducting cavity. These currents ensure that the
magnetic field outside the cube is identically zero. Note that the
surface currents are symmetry-adapted as well. 
        
\begin{figure}[!htbp]   
	\centering
	\includegraphics[width=1.6in]{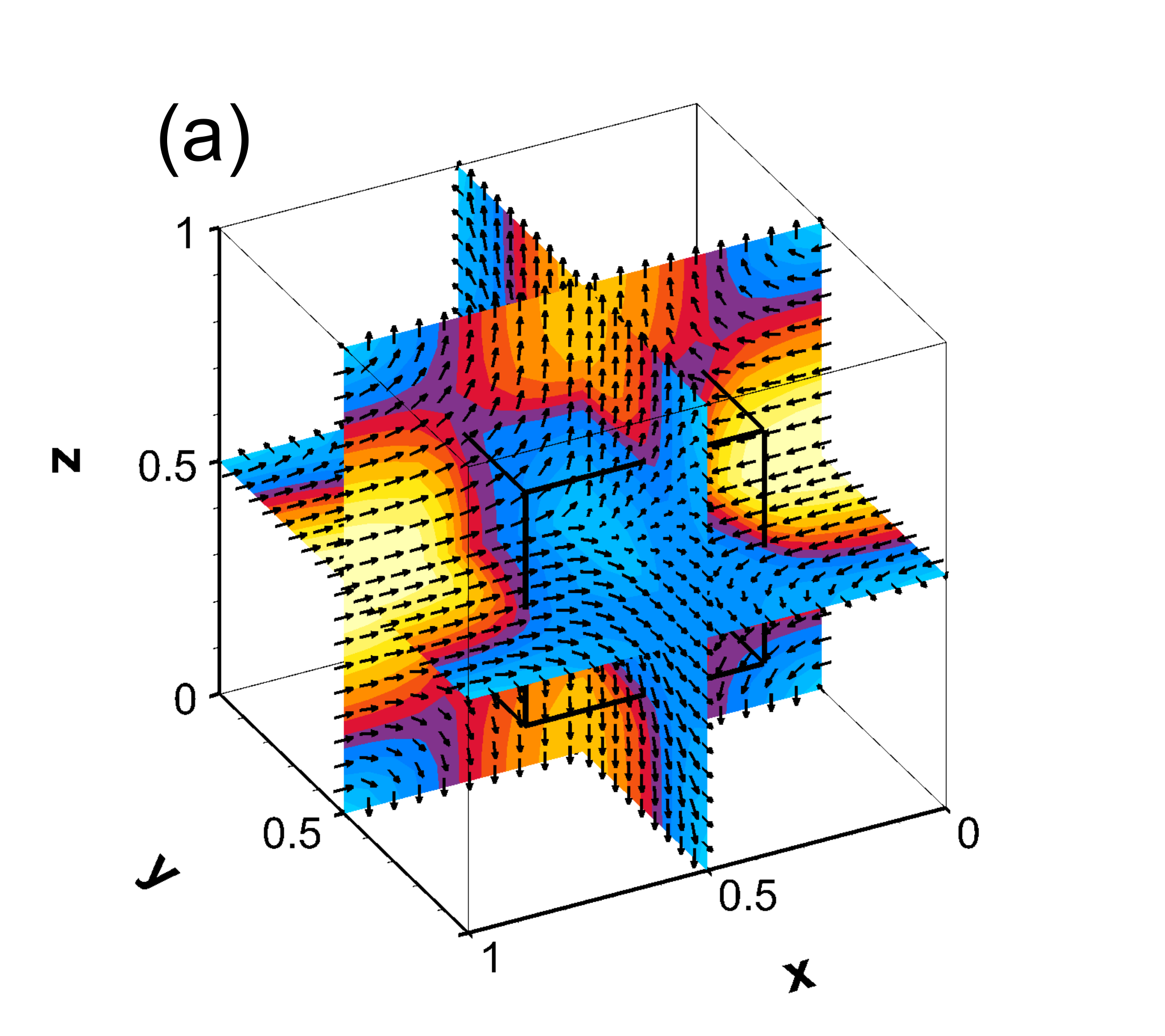}
	\includegraphics[width=1.6in]{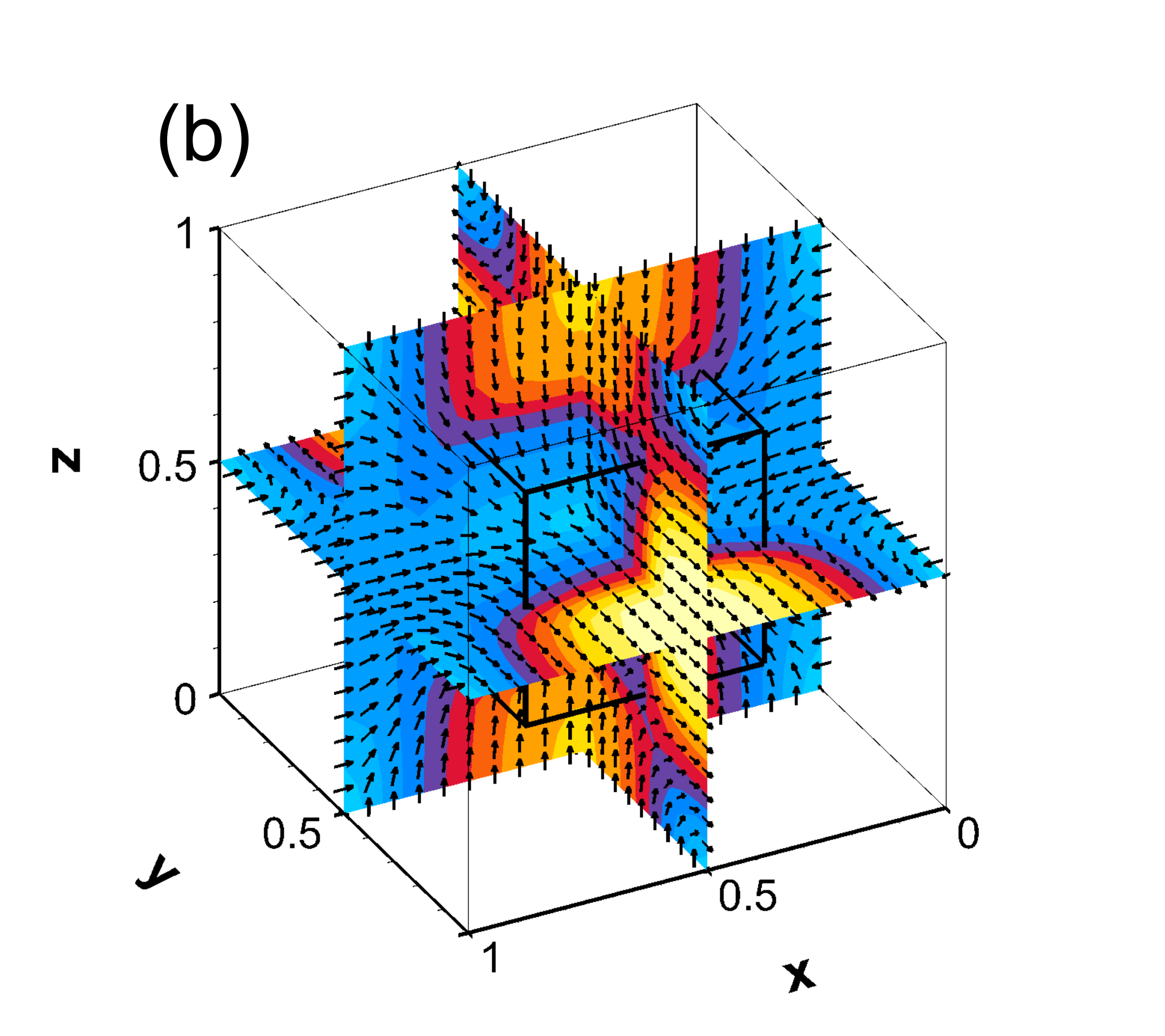}
	\caption{Degenerate modes $(1,1,1)$ of the
		loaded cubic cavity are shown. Light yellow (gray) regions 
		correspond to amplitude antinodes, and blue (darker) regions 
		correspond to amplitude nodes. (From Pandey {\it et
                  al.}, Ref.~\onlinecite{avs:LRRCavityEM2018}.)}  
	\label{avsfig:nonperturb}
\end{figure}

\begin{figure}[!htbp]   
\centering
\includegraphics[width=1.6in]{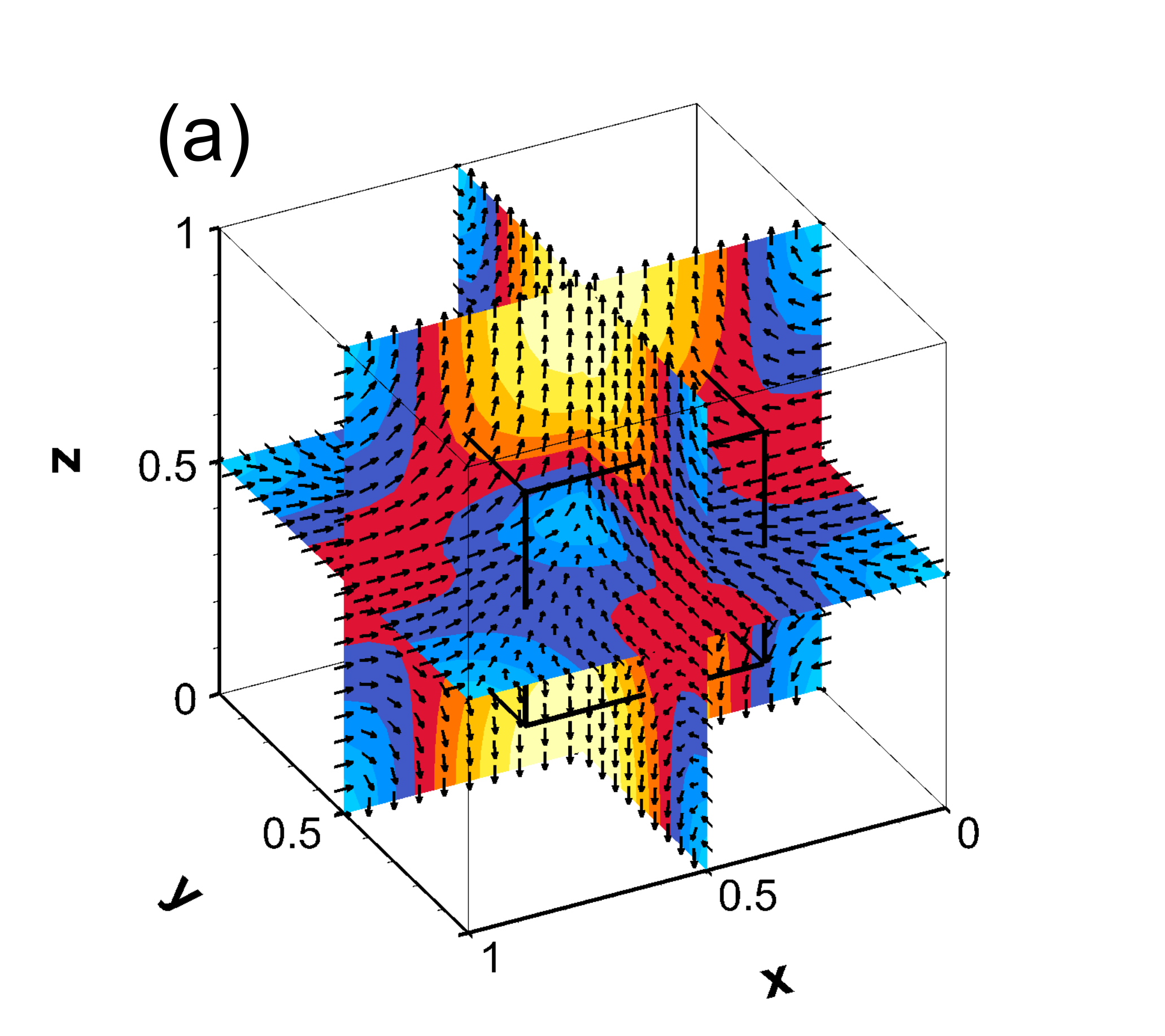}
\includegraphics[width=1.6in]{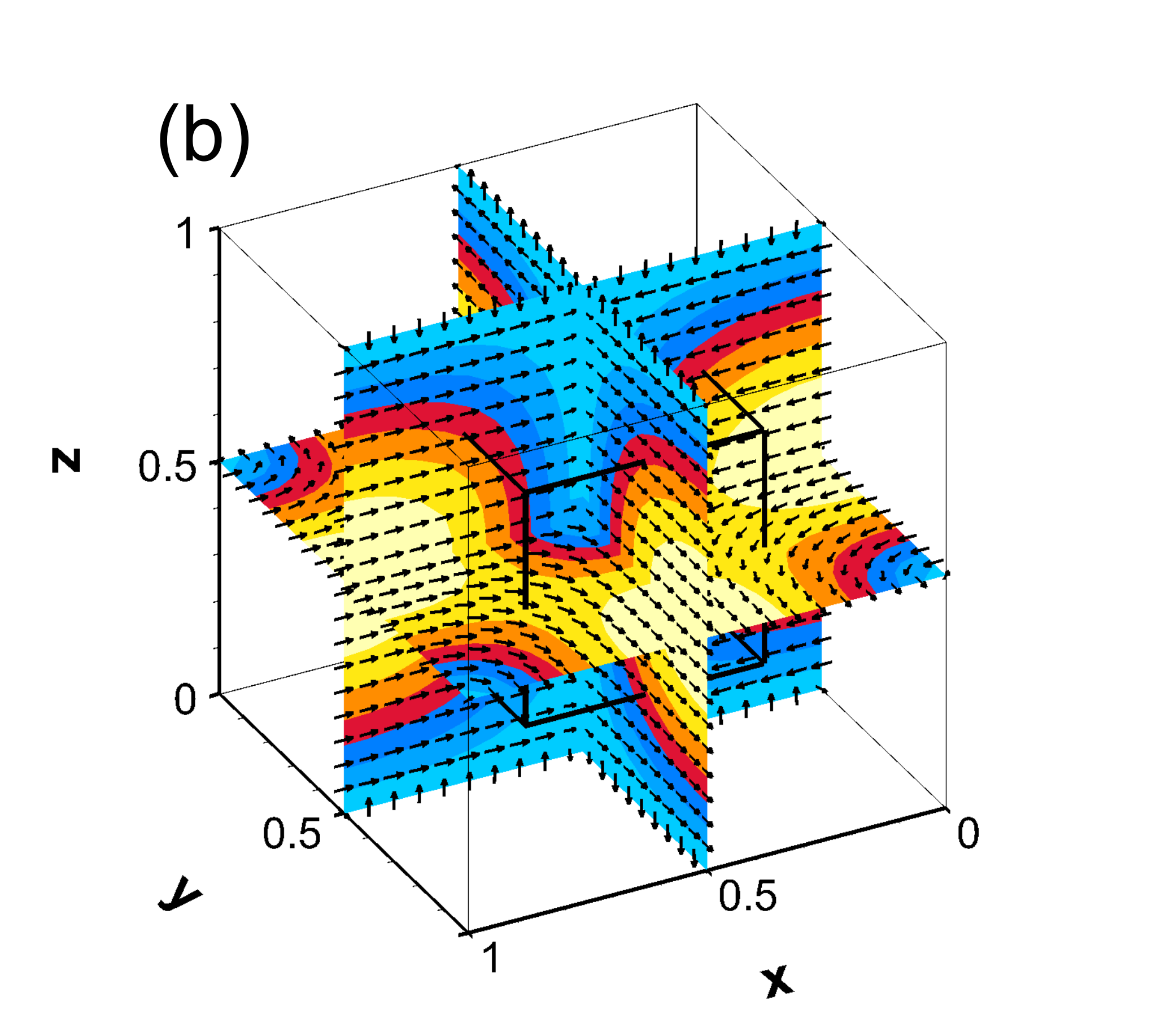}
\caption{Non-degenerate modes $(1,1,1)$ of the perturbed
loaded cavity belonging to the irreducible representations
(a) $A_1$ with $k_0^2 = 26.303562$, and (b) $B_1$ with $k_0^2 =
26.309544$ are shown. Light yellow (gray) regions correspond to amplitude antinodes,
and blue (darker) regions correspond to amplitude nodes. (From  Pandey
{\it et al.}, Ref.~\onlinecite{avs:LRRCavityEM2018}.)}  
\label{avsfig:perturb}
\end{figure}

As  a final  example,  we   consider  a  linear  $z$-dependent
perturbation  to  the  dielectric  function  in  the  interior
dielectric  block.   This  perturbation reduces  the  symmetry
group of the  loaded cavity from  $O_h$ to $C_{4v}$,
resulting in  a further  reduction in mode  degeneracies.  In
Table~\ref{tab:perturb}, we  have listed  the eigenfrequencies
obtained   with  our   method,   and   classified  them   into
corresponding   irreducible  representations   of  the   group
$C_{4v}$. In Fig.~\ref{avsfig:nonperturb} we show  electric field
distributions for the degenerate modes $(1,1,1)$ in the loaded cubic cavity.
 In  Fig.~\ref{avsfig:perturb}, we show  electric field
distributions for the same modes, but for a perturbed cavity. Note how
the modes in the perturbed case are now non-degenerate and 
split in  frequency.  As seen from Fig.~\ref{avsfig:perturb},
the electric field magnitudes have complete $C_{4v}$ symmetry;
the  first mode  belongs to   the representation  $A_1$ of
$C_{4v}$,   while  the   second  mode   belongs to the 
representation $B_1$.

\begin{table*}[th!]  
\caption{Numerically  calculated eigenvalues for the  lowest few
frequency modes of the dielectrically loaded cubic cavity with 17576
total DOFs using quintic Hermite interpolation polynomials. The column
labeled ``irrep'' corresponds to the irreducible representation the
multiplet belongs to. The
conducting cavity has dimensions $1\times1\times1$ mm$^3$; the
cubic dielectric loading has dimensions $0.5\times0.5\times0.5$
mm$^3$. The eigenvalues are
compared against their values for the modes in the empty cavity of
unit dimensions. (From  Pandey {\it et al.}, Ref.~\onlinecite{avs:LRRCavityEM2018}.)} 
\vspace{0.1in}
\label{tab:load_cube_eigen}
\centering
\bgroup
\def\arraystretch{1.3}
{{\begin{tabular}{K{3.5cm}|c|d{3.10}|d{3.10}|d{3.10}}
\hline
Mode number & \multicolumn{1}{p{1.5cm}|}{\centering ``Irrep'' 
} & \multicolumn{3}{c}{$k^2_0$}\\ 
\cline{3-5}
& & \multicolumn{1}{c|}{$\epsilon_2/\epsilon_1 = 5.0$} &
\multicolumn{1}{c|}{$\epsilon_2/\epsilon_1 = 1.2$} &
\multicolumn{1}{c}{$\epsilon_2/\epsilon_1 = 1.0$}\\  
\hline\hline
&  & 7.919791186 & 18.526768246 & 19.739208802\\
$(0,1,1),(1,0,1),(0,1,1)$ & $T_{1u}$ & 7.919791186 & 18.526768246 &
                                                                    19.739208802\\ 
&  & 7.919791186 & 18.526768246 & 19.739208802\\
\hline
$(1,1,1)$ & $E_{g}$ & 18.202154429 & 28.899507681 & 29.608813203\\
& & 18.202154429 & 28.899507681 & 29.608813203\\
\hline
$(0,1,2),(2,1,0)$ &  & 21.354532231 & 47.387744521 & 49.348022007\\
& $T_{1g}$ & 21.354532231 & 47.387744521 & 49.348022007\\
$(1,0,2),(2,0,1)$ &  & 21.354532231 & 47.387744521 & 49.348022007\\
\cline{2-4}
&  & 23.416822776 & 47.404561358 & 49.348022007\\
$(0,2,1),(1,2,0)$ & $T_{2g}$ & 23.416822776 & 47.404561358 & 49.348022007\\
&  & 23.416822776 & 47.404561358 & 49.348022007\\
\hline
$(1,1,2),(2,1,1)$ &  & 35.614536102 & 57.284302371 & 59.217626408\\
& $T_{1u}$ & 35.614536102 & 57.284302371 & 59.217626408\\
$(2,1,1),(1,1,2)$ &  & 35.614536102 & 57.284302371 & 59.217626408\\
\cline{2-4}
&  & 38.219214013 & 58.327254711 & 59.217626408\\
$(1,2,1),(1,2,1)$ & $T_{2u}$ & 38.219214013 & 58.327254711 & 59.217626408\\
&  & 38.219214013 & 58.327254711 & 59.217626408\\
\hline
& \multirow{2}{2.2em}{$E_{u}$} & 38.305292320 & 76.979818578 &
                                                               78.956835213\\ 
$(0,2,2),(2,0,2),(2,2,0)$  &  & 38.305292320 & 76.979818578 & 78.956835213\\
\cline{2-4}
& $A_{2u}$ & 42.790037374 & 77.001769904 & 78.956835213\\
\hline
&  & 55.137384588 & 103.727559345 & 108.565648679\\
& \multirow{2}{2.2em}{$2E_{g}$} &
55.137384588 & 103.727559345 & 108.565648679\\ 
\cline{3-4}
& \multirow{2}{2.2em}{} & 77.568304705 & 107.596740348 & 108.565648679\\
$(1,1,3),(1,3,1),(3,1,1)$  & & 77.568304705 & 107.596740348 & 108.565648679\\
\cline{2-4}
& $A_{1g}$ & 67.549698454 & 107.431515611 & 108.565648679\\
\cline{2-4}
&  $A_{2g}$ & 77.460283215 & 107.591198480 & 108.565648679\\
\hline
\end{tabular}}}
\egroup
\end{table*}

\begin{table}[t!]   
\caption{Lowering of symmetry, and splitting of mode degeneracy in the
  presence of a dielectric inclusion that has a preferential
  z-axis. (From  Pandey {\it et al.}, Ref.~\onlinecite{avs:LRRCavityEM2018}.)} 
\vspace{0.1in}
\label{tab:perturb}
\def\arraystretch{1.3}
\centering
\begin{tabular}{|K{1cm}|K{1cm}|d{3.7}|}
\hline
$O_h$ & $C_{4v}$ & \multicolumn{1}{c|}{$k^2_0$} \\
\hline\hline
\multirow{3}{*}{$T_{1u}$} & $A_{1}$ & 14.840987 \\
\cline{2-3}
& \multirow{2}{*}{$E$} & 14.841426 \\
&  & 14.841426 \\
\hline
\multirow{2}{*}{$E_{g}$} & $A_{1}$ & 26.303562 \\
\cline{2-3}
& $B_{1}$ & 26.309544 \\
\hline
\multirow{3}{*}{$T_{1g}$} & $A_{2}$ & 39.928882 \\
\cline{2-3}
& \multirow{2}{*}{$E$} & 39.937310 \\
&  & 39.937310 \\
\hline
\multirow{3}{*}{$T_{2g}$} & $B_{2}$ & 40.430146 \\
\cline{2-3}
& \multirow{2}{*}{$E$} & 40.431522 \\
&  & 40.431522 \\
\hline
\multirow{3}{*}{$T_{1u}$} & $A_{1}$ & 51.226115 \\
\cline{2-3}
& \multirow{2}{*}{$E$} & 51.202143 \\
&  & 51.202143 \\
\hline
\multirow{3}{*}{$T_{2u}$} & $B_{1}$ & 54.625215 \\
\cline{2-3}
& \multirow{2}{*}{$E$} & 54.631936 \\
&  & 54.631936 \\
\hline
\multirow{2}{*}{$E_{u}$} & $A_{2}$ & 68.152735 \\
\cline{2-3}
& $B_{2}$ & 68.160536 \\
\hline
$A_{2u}$ & $B_{2}$  & 68.993259 \\
\hline
\end{tabular}
\end{table}


{\it Quality factor:} A classic benchmark in computational
electromagnetics is to obtain the $Q$-factor in a  
resonant cavity with lossy walls. Here we calculate the $Q$-factor in
a loaded cavity which has contributions from  
a) the dissipation of energy at the cavity walls, and b) the dielectric loss  when
the permittivity has both real and imaginary parts, \mbox{$\epsilon = \epsilon_r +i\epsilon_i$}. 
For the dielectric losses, the $Q$-factor associated with the resonator\cite{avs:jones_book,avs:Harrington}
is given by  
\begin{align}
  Q_d &=\frac {\DS\iiint_V dV\ \epsilon_r \left|\mathbf{E}\right|^2
        }
        {\DS\iiint_V dV\ \epsilon_i \left|\mathbf{E}\right|^2 }.
\end{align}
\begin{table}
  \caption{We  have tabulated  the numerically  calculated eigenvalues
    and corresponding  $Q$\,-factors for  a loaded cubic cavity with boundary walls
    made out of gold. The 
    conducting  cavity has  dimensions  $1\times1\times1$ mm$^3$;  the
    cubic  dielectric loading  has dimensions  $0.5\times0.5\times0.5$
    mm$^3$,   with   dielectric   ratios    $\epsilon_r   =   2$   and
    $\epsilon_i = 10^{-6}$. The conductivity of gold is taken to be $\sigma=4.46\times10^{7}\,$S/m. (From
    Pandey {\it et al.}, Ref.~\onlinecite{avs:LRRCavityEM2018}.)}
\vspace{0.1in}
\label{tab:img_cube_eigen}
\centering
\bgroup
\def\arraystretch{1.3}
{\begin{tabular}{c|c|c}
		\hline
		$k^2_0$ & $Q_d$\,-factor &$Q_c$\,-factor\\
		\hline\hline
		  14.64474995 $+i$ 0.00000402	 & 3645779.29 &78241.13\\
		 14.64474995 $+i$ 0.00000402	 & 3645779.29 &78241.13\\
		 14.64474995 $+i$ 0.00000402	 & 3645779.29 &78241.13\\
		 \hline
		 26.12581823 $+i$ 0.00000338	 & 7718462.64 &62382.25\\
		 26.12581823 $+i$ 0.00000338	 & 7718462.64 &62382.25\\
		 \hline
		 39.45307840 $+i$ 0.00000942	 & 4186986.51 &156821.95\\
		 39.45307840 $+i$ 0.00000942	 & 4186986.51 &156821.95\\
		 39.45307840 $+i$ 0.00000942	 & 4186986.51 &156821.95\\
		 \hline
		 40.02056958 $+i$ 0.00000864	 & 4633182.94 &110705.70\\
		 40.02056958 $+i$ 0.00000864	 & 4633182.94 &110705.70\\
		 40.02056958 $+i$ 0.00000864	 & 4633182.94 &110705.70\\
		 \hline
		 50.87611080 $+i$ 0.00000720	 & 7061660.73 &78692.17\\
		 50.87611080 $+i$ 0.00000720	 & 7061660.73 &78692.17\\
		 50.87611080 $+i$ 0.00000720	 & 7061660.73 &78692.17\\
		 \hline
		 54.34357813 $+i$ 0.00000540	 & 10065812.56 &86034.59\\
		 54.34357813 $+i$ 0.00000540	 & 10065812.56 &86034.59\\
		 54.34357813 $+i$ 0.00000540	 & 10065812.56 &86034.59\\
		 \hline
		 67.50785266 $+i$ 0.00001276	 & 5290879.93 &211112.95\\
		 67.50785266 $+i$ 0.00001276	 & 5290879.93 &211112.95\\
		 \hline
		 68.46687536 $+i$ 0.00001114	 & 6146166.21 &139534.39\\
		 \hline
		 75.30032691 $+i$ 0.00000986	 & 7636781.09 &77564.09\\
		 75.30032691 $+i$ 0.00000986	 & 7636781.09 &77564.09\\
		 75.30032691 $+i$ 0.00000986	 & 7636781.09 &77564.09\\
		 \hline
\end{tabular}}
\egroup
\end{table}
We note  that when a constant dielectric loading fills the entire
cavity we obtain $Q_d = \epsilon_r/ \epsilon_i$, irrespective of the
eigen-frequency. We have verified this in the limit of full dielectric occupancy.
In Table~\ref{tab:img_cube_eigen}, we have shown the eigenvalues and
their corresponding $Q_d$ values in a partially loaded cubic cavity with
$\epsilon_r = 2$ and $\epsilon_i = 10^{-6}$. The degeneracy spectrum
again follows $O_h$ symmetry.

When the cavity resonator has imperfect conducting walls, we define the corresponding quality factor\cite{avs:jackson_book} as
\begin{align}
  Q_c 
       &= \omega\sigma\delta_s \frac {\DS\iiint_V dV\ \mu \left|\mathbf{H}\right|^2
         }{\DS\oiint\limits_{\substack{\textrm{cavity}\\ \textrm{walls} }} dS\ \left|\mathbf{n\times H}\right|^2},
\end{align}
where  $\sigma$  is  the  conductivity of  the  metallic  surface  and
$\delta_s$ is the  skin depth at the resonant  frequency $\omega$.  In
Table~\ref{tab:img_cube_eigen},  we  list  the  $Q_c$  values  at  the
resonant  frequencies in  a partially  loaded cubic  cavity with  gold
boundary walls.
\subsection{Remarks on the accidental degeneracy}
\label{section:grouptheory}
Let $\tilde{G}$ be an  infinitesimal transformation for the coordinate
system.  Given a Hamiltonian $\hat{H}$, if $[\tilde{G},\hat{H}]=0$, and
$\tilde{G}$  does not  explicitly depend  on  time, then  we say  that
$\tilde{G}$  is  a  constant  of motion.   Such  constants  of  motion
generate symmetries since they transform  one eigenstate to another of
the same eigenvalue.  We expect to find additional constants of motion
whenever we observe accidental degeneracies as explained below. If the
Hamiltonian is separable  in a coordinate system,  then the separation
constants may  be considered  as constants  of motion.\cite{avs:greenberg}
These   are   just  the   generators   of   the  additional   symmetry
operations.  Typically,  accidental  degeneracies  are  then  rendered
normal by identifying the hidden covering group.

The  example  of  the  familiar hydrogen  atom  best  illustrates  the
symmetry argument.  In  the H-atom, with its  Coulomb potential having
geometrical rotational symmetry, the  conservation of the 3 components
of angular  momentum provide us  three constants of  motion associated
with  the  3D  rotation   group  $O(3)$.   Equivalently,  we  consider
the angular momentum components
$\left\{ L_{+},  L_{-}, L_z\right\}$ as  the set of 3  operators which
commute with the Hamiltonian of the H-atom. We know that an eigenstate
$\left|E,\ell,m\right>$ of  the H-atom transforms under  the operation
of ladder operators as
\begin{equation}\label{eq:stepoperators}
L_{\pm}\left|E,\ell,m\right> = \sqrt{(\ell\mp m)(\ell\pm m +1)}
\left|E,\ell,m\pm 1\right>.  
\end{equation}
Hence   angular   momentum  operators   transform   degenerate
eigenstates  of the  same  $\ell$ but  of different  azimuthal
quantum numbers $m$  into one another.  For a  given $\ell$ we
find  that $(2\ell+1)$  states  are  degenerate.  However  the
eigenstates of  different allowed  $\ell$ are  also degenerate
here, leading to a text-book example of accidental degeneracy.
Fock  \cite{avs:Fock1935} identified  the hidden  four-dimensional
rotational symmetry group  $O(4)$ as the true  symmetry of the
H-atom,   which   explains   these   additional   degeneracies
manifesting as the familiar $s,  p, d, f, \ldots$ states being
degenerate  for  a given  principal  quantum  number $n$.   We
expect  to find  additional  operators  (constants of  motion)
which commute with the  Hamiltonian and connect eigenstates of
different $\ell$ quantum numbers. These operators are just the
three components  of the conserved  Runge-Lenz
vector, ,\cite{avs:Pauli1926}  $\bm{A}$,  which  transform  degenerate
eigenstates       of       different      $\ell$       quantum
numbers into one another,\cite{avs:Burkhardt}              analogous             to
Eq.~(\ref{eq:stepoperators}).   The components  of the  angular
momentum $\bm{L}$ and the  Runge-Lenz vector $\bm{A}$ generate
the symmetry  group $O(4)$. 
We  note   that  even  though  the
components  of   $\bm{L}$  and   $\bm{A}$  commute   with  the
Hamiltonian,  they   will  not  mutually  commute   with  each
other. These  components are subject to  kinematic constraints
of  the Casimir  operators  for the  group  $O(4)$.\cite{avs:McIntosh} Hence  the
eigenstates of the Hamiltonian are represented by the complete
set of commuting operators $\left\{H, L^2, L_z\right\}$.
Such an analysis, based  on the symmetries of  a physical system,
is  also  feasible for the EM cavities.    

To further clarify the aspects of degeneracy we briefly consider the example of a
2D empty square cavity of length $a$,
surrounded with metal boundaries. This system has $C_{4v}$ geometrical
symmetry. The character table for different point groups are given in
the texts by Dresselhaus\cite{avs:dresselhaus} and by
Tinkham.\cite{avs:tinkham}  We know that the eigenvalues 
supported in the cavity are given by  
\begin{equation}
k_0^2 = \left(n_x^2+n_y^2\right)\,\frac{\pi^2}{a^2},
\end{equation}
where $n_x$ and $n_y$ are non-zero integers. 
\begin{table*}[th!]  
	\caption{Different possible even and odd combinations of 2D eigenmodes
		and their corresponding     irreducible representations for the
		symmetry group $C_{4v}$ are shown. Here the 
		indices $n,m$ are  non-zero integers, and the column
                labeled ``irrep'' refers to the irreducible
                representations of the multiplet.} 
	\label{tab:basisirrepsc4v}
	\vspace{0.1in}
	\centering
	\bgroup
	\def\arraystretch{2}
	{\begin{tabular}{c|c|c}
			\hline
			 Mode number&
			\multicolumn{1}{p{1.2cm}|}{\centering ``Irrep'' }
			&Basis
			functions\\  
			\hline\hline
			$(2n-1, 2n-1)$ & $B_1$ &
			$\mathbf{E}_{\left(2n-1,2n-1\right)}$\\ 
			\hline
			$(2n, 2n)$ & $A_2$ &
			$\mathbf{E}_{\left(2n,2n\right)}$\\ 
			\hline
			\multirow{3}{*}{\mbox{$(2m-1,2n-1)$}} 
			& $B_1$ &
			$\mathbf{E}_{\left(2m-1,2n-1\right)}+\mathbf{E}_{\left(2n-1,2m-1\right)}$\\ 
			\cline{2-3}
			&$A_1$ &
			$\mathbf{E}_{\left(2m-1,2n-1\right)}-\mathbf{E}_{\left(2n-1,2m-1\right)}$\\ 
			\hline
			\multirow{3}{*}{\mbox{$(2m,2n)$}} 
			& $A_2$ &
			$\mathbf{E}_{\left(2m,2n\right)}+\mathbf{E}_{\left(2n,2m\right)}$\\ 
			\cline{2-3}
			& $B_2$ & $\mathbf{E}_{\left(2m,2n\right)}-\mathbf{E}_{\left(2n,2m\right)}$\\
			\hline
			$(2m-1,2n)$ & $E$ &
			$\mathbf{E}_{\left(2m-1,2n\right)},\
			\mathbf{E}_{\left(2n,2m-1\right)}$\\ 
			\hline\hline
		\end{tabular}}
		\egroup
	\end{table*}
	In Table~\ref{tab:basisirrepsc4v}, we  list all symmetry-adapted basis
	functions and  irreducible representations for  different combinations
	of $(n_x,n_y)$  modes, derived using Eq.~(\ref{eq:coeff_formula}).  
	The  (odd, odd) or  (even, even)  doublet with
	$n_x\neq     n_y$    belongs     to    two     distinct    irreducible
	representations.  Hence,  the  degeneracy  of these  modes  is  not  entirely
	explained by the symmetry  group $C_{4v}$;  therefore  they
	exhibit accidental degeneracy. This is analogous to the situation in
	an  infinite  square quantum  well,  where  the accidental  degeneracy
	occurs for the eigenenergies due  to the separability of the infinite
	well  potential.\cite{avs:LRR_shertzer} Such  an accidental  degeneracy is
	rendered  normal,  in  the  usual parlance,  by  recognizing  that  an
	additional   operator  $\Omega=\left(\partial_x^2-\partial_y^2\right)$
	exists, which connects the basis  functions of $A_1$ ($A_2$) and $B_1$
	($B_2$) representations.   Hence, the  true symmetry  of a  $2D$ empty
	square cavity will be a covering  group, which is a semidirect product
	of  the  geometrical symmetry  group  $C_{4v}$  and a  one-dimensional
	compact    continuous     group    generated    by     the    operator
	$\Omega=\left(\partial_x^2-\partial_y^2\right)$.\cite{avs:Leyvarz} {\it We
		can remove the accidental degeneracy in  a 2D empty square cavity by
		introducing  a   concentric  square  dielectric   inclusion.}   Such
	accidental  degeneracy  and its  removal  occurs  even in  rectangular
	cavities.

\section{Photonic crystals}
The  first   proposals  for  the   design  of  photonic   crystals  by
Yablonovitch\cite{avs:Yablonovitch1987}  and  by John\cite{avs:SajeevJohn1987}
in   1987,  and   further  investigations   by  Ohtaka,   Sakoda,  and
collaborators\cite{avs:OhtakaInoue1982,avs:sakodaPRB1995,avs:OhtakaTanabe1996,avs:sakoda1997,avs:sakoda2001}
and by  Joannopoulos and Johnson\cite{avs:johnson2002} have led  to a full
appreciation of  the physics of  periodic dielectrics. With  the rapid
increases in computing power  and simulation techniques and the design
and  fabrication of  PCs,  a wide  variety  of optoelectronic  devices
including  low-loss  reflecting  surfaces, waveguides,  filters,  flat
lenses, optical  inter-connects and the like, have  made the efficient
prediction of their optical  properties a high priority for physicists
and optical engineers.\cite{avs:johnson2002,avs:pendry1994,avs:Pendry2000,avs:Smith2000,avs:benisty2005,avs:Veselago1964}

By assuming that  the PC contains an arbitrarily  large number of unit
cells, using the Bloch-Floquet Theorem,\cite{avs:Bloch,avs:Floquet,avs:BabyKittel} we
can decompose the magnetic field into two terms,
\begin{equation}
{{\bf H}}\!\left({{\bf r}}\right) = {{\bf U}}\!\left({{\bf r}}\right)
e^{i {{\bf q}} \cdot {{\bf r}}}.
\label{eq:lag9a}
\end{equation}
In the
examples treated in this paper, in which the unit cell is a rectangle
of dimensions $d_x \times d_y$, the vector
{{\bf q}} can be expressed as
\begin{equation}
{{\bf q}}=\dfrac{\pi q_x}{d_x}{{\bf \hat{i}}}
+ \dfrac{\pi q_y}{d_y}{{\bf \hat{j}}} + 0 {{\bf \hat{k}}},
\label{eq:lag9p1a}
\end{equation}
in which the range of $q_x$ and $q_y$ values which comprise
the Brillouin zone \cite{avs:joann2008} are
$-1 < q_x,\, q_y \leq 1,$
\begin{figure}[t!] 
	\begin{center}
		\includegraphics[width=1.10in]{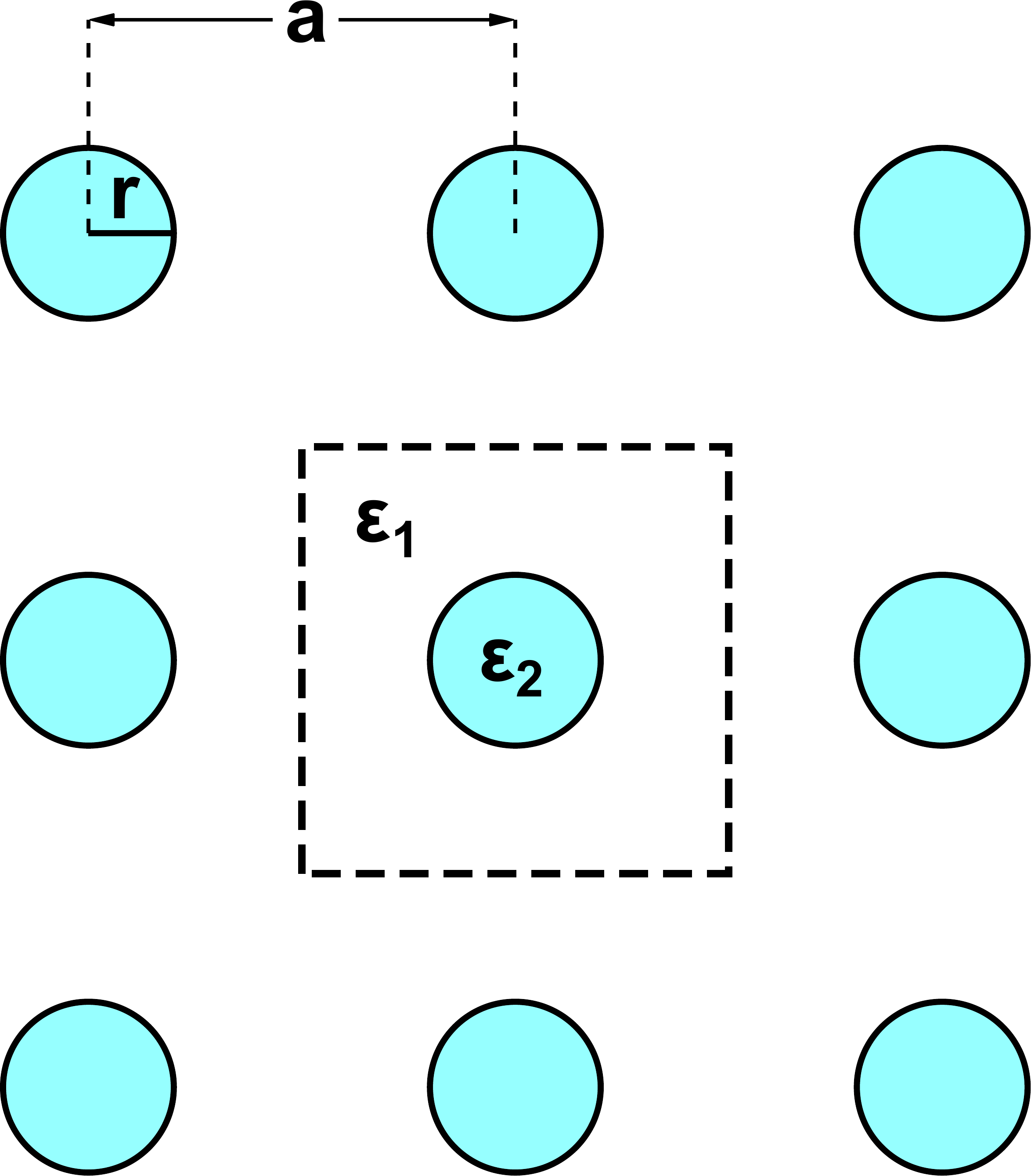} \hspace{0.5in}
		\includegraphics[width=1.3in]{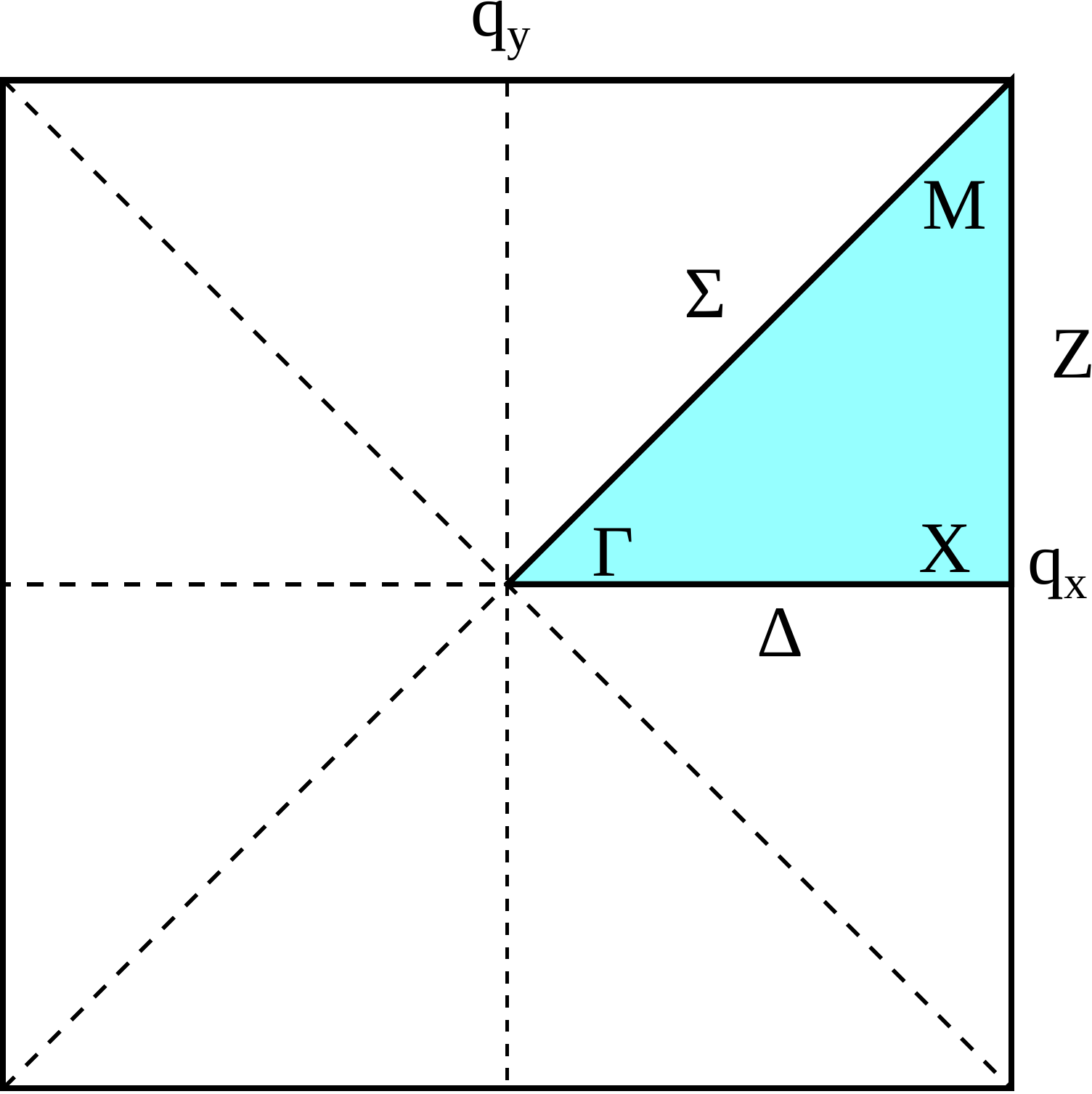}
	\end{center}
	\caption[Geometry of a photonic crystal with dielectric posts]
	{An  array  of  cylindrical  dielectric  posts  of  dielectric
          constant $\epsilon_2$ arranged periodically in a medium with
          a  dielectric  constant  $\epsilon_1$   is  shown.  (b)  The
          Brillouin  zone for  a two-dimensional  photonic crystal  is
          shown, where  {{$-1 < q_x,\,q_y \leq  1$}}.  The irreducible
          component of  the Brillouin zone has  been highlighted.  The
          dashed lines  are symmetry lines within  the first Brillouin
          zone.}
	\label{avsfig:posts}
\end{figure}

It is only necessary  to  consider  the   eigenvalues  over the  first Brillouin zone;
outside     this     zone      the     eigenvalues     will     behave
periodically.  Furthermore,  due   to  reflection  symmetries,  it  is
possible to further reduce the Brillouin zone  to obtain a greater
density of sampling points for the same computational cost. In the first
example considered, a square array of cylindrical dielectric posts, the
Brillouin zone only needs to be treated over the region
marked in Fig.~\ref{avsfig:posts}.

We can decompose the terms in the functional $\bf{L}$ into cell functions and
envelope functions. The second   term    of   the   integrand   of    $\bf{L}$   reduces   to
${{\bf U}}^{*}\cdot k_0^2 \mu_r \cdot{{\bf U}}$.
%
%
The curl  term in the integrand  of $\bf{L}$ can  be simplified using
the   absence    of   propagation   in   the    third   dimension   in
Eq.~(\ref{eq:lag9p1a}) as
\begin{align}\label{eq:lag18}
\hspace{-0.5in}\nabla &\times {{\bf H}} =\\ 
& e^{i{{\bf q}}\cdot{{\bf r}}} 
\left[
\begin{pmatrix}
\partial_y U_z - \partial_z  U_y\\ \partial_z U_x - \partial_x
U_z\\ \partial_x U_y - \partial_y U_x
\end{pmatrix}
\!+ i\pi\begin{pmatrix} q_y U_z / d_y\\ - q_x U_z / d_x\\  q_x
U_y / d_x -  q_y U_x / d_y
\end{pmatrix}
\right]\!.\nonumber
\end{align}
Eq.~(\ref{eq:lag18})  may be  greatly simplified by  classifying  all possible
solutions  into two  distinct  cases: ($i$)  Transverse electric  (TE)
modes  with $E_z =0$, which  from the  imposition  of periodicity
resulting in Eq.~(\ref{eq:lag18})  forces $H_x=H_y=0$, and all
other vector components are nonzero and ($ii$) Transverse magnetic (TM)
modes with $H_z=0$ which forces $E_x =E_y=0$, $H_z=0$,
and  all  other  vector  components are  nonzero.  
It is sufficient to consider TE- and TM-modes only, since any fields  
in the 2D PC can be expressed  as a combination of
these modes. It is well  known that frequencies  absent from the
eigen spectra  of both modes do  not propagate because they  are in the
`band gap'.

The  band gaps of the photonic  crystal may be
determined  by choosing a  large number  of ordered  pairs $(q_x,q_y)$
from  within  the  irreducible   Brillouin  zone  to  determine  which
eigenvalues will  propagate. Then the band structure  in the remainder
of the Brillouin  zone may be determined via  reflection symmetry. The
entire first Brillouin zone may then be translated to adjacent
zones due to the periodicity of the cell function.

Given the  simple modal  decomposition resulting from  2D periodicity,
eigenvalue problems for both TE and  TM modes can be posed in terms of
a  single scalar  quantity,  $U_z$ and  its  spatial derivatives.  For
isotropic media where $\mu_r$ and $\epsilon_r$ are constant, the functional $\bf{L}$
for TM modes simplifies to
\begin{equation}
{\bf  L}  =  \int_\Omega  d^2  r  \left[  U_z^{*}\,  {{\bf A}}\cdot
 \dfrac{1}{\epsilon_r} {{\bf B}}\, U_z - U_z^{*}\, k_0^2 \,\mu_r \,U_z\right],
\label{eq:lag23}
\end{equation}
where
\begin{equation}
{{\bf A}}\!=\!\begin{bmatrix} {\overleftarrow{\partial_y}}\! -\! \dfrac{\pi
  iq_y}{d_y}, -{\overleftarrow{\partial_x}}  +  \dfrac{\pi
  iq_x}{d_x}  \end{bmatrix},  {{\bf B}}\!\!
= \begin{bmatrix} {\overrightarrow{\partial_y}} +\dfrac{\pi iq_y}{d_y}
 \\  {\mbox{\hspace{0.1in}}}  \\-{\overrightarrow{\partial_x}}  -
 \dfrac{\pi iq_x}{d_x}
 \end{bmatrix}
\label{eq:lag22}
\end{equation}
and the arrows over the  derivatives denote the direction in which the
derivatives  operate  on the  quantities  in Eq.~(\ref{eq:lag23}).  To
obtain       ${\bf L}$       for       TE      modes       interchange
$\epsilon_r \leftrightarrow \mu_r  $   in  Eq.~(\ref{eq:lag23})  and
associate $U_z$ with ${\bf E}$.  Within the FEM, a variational solution
is found by the  eigenvalue problem derived from $\delta{\bf{L}}=0$ and yields
the band structure of the 2D PC.

In the following,  we consider 
 three
examples of  photonic crystals. The
first is a square lattice of dielectric posts, for which we obtain the
photonic     band    structure     as    previously     reported    in
Joannopoulos.\cite{avs:joann2008}  We  also  identify  the  symmetries  at
various  points  in  the  dispersion  relations  to  discuss  band
anti-crossing and level degeneracies at special points. The eigenvector
fields at the $\Gamma-$point
 of the frequency dispersion are shown, and frequency
bands over the full Brillouin zone are displayed.

The second example is that of a checker-board lattice of
dielectric regions. Here again we provide the group theoretic
analysis, the band structure, the band surfaces over the Brillouin
zone, and the eigenvector fields.

To demonstrate HFEM's capability in modeling systems of arbitrary shapes,
the third example is chosen to be a photonic crystal structure based on 
M. C. Escher's Horsemen tessellation. The fields, band structure and
band surfaces over the Brillouin zone are calculated.

\subsection{Group Representation Theory and Photonic  Crystals}

\begin{figure}[t!] 
  \begin{center}
\subfloat[][\label{avsfig:TEG1}TE mode 2] 
{\includegraphics[width=1.2in]{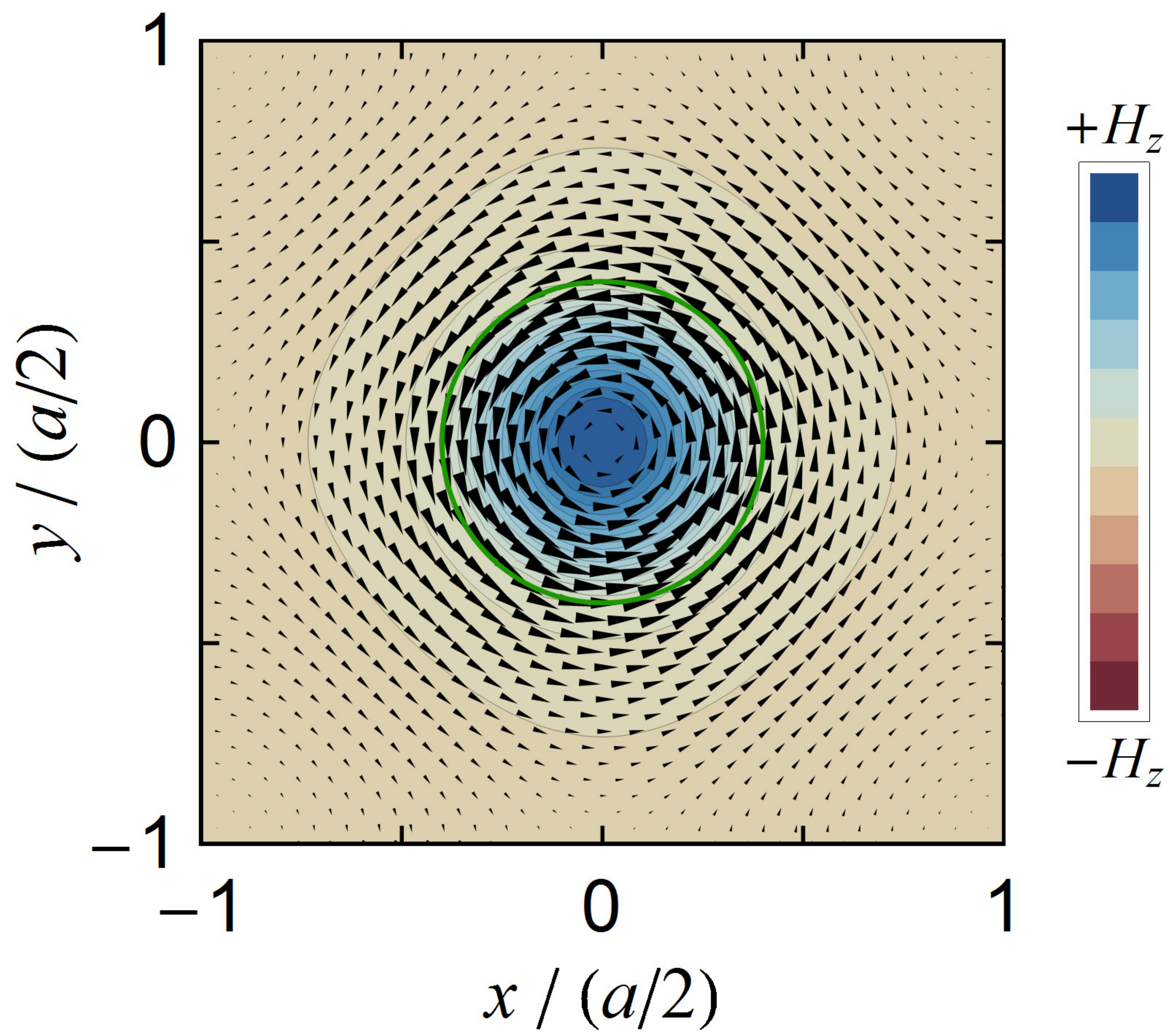}}
\subfloat[][\label{avsfig:TEG2}TE mode 3]
{\includegraphics[width=1.2in]{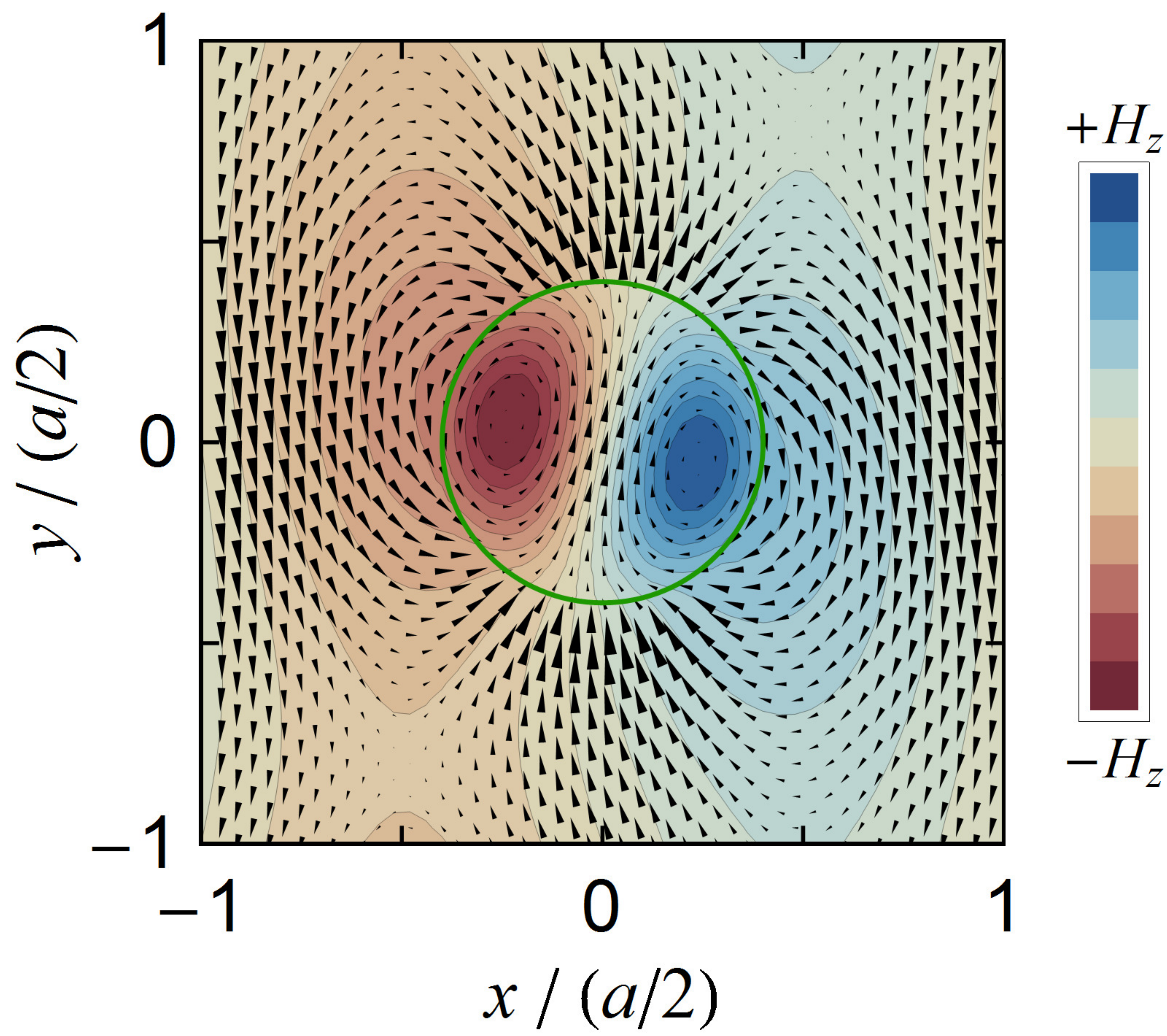}} \\
\subfloat[][\label{avsfig:TMG1}TM mode 2]
{\includegraphics[width=1.2in]{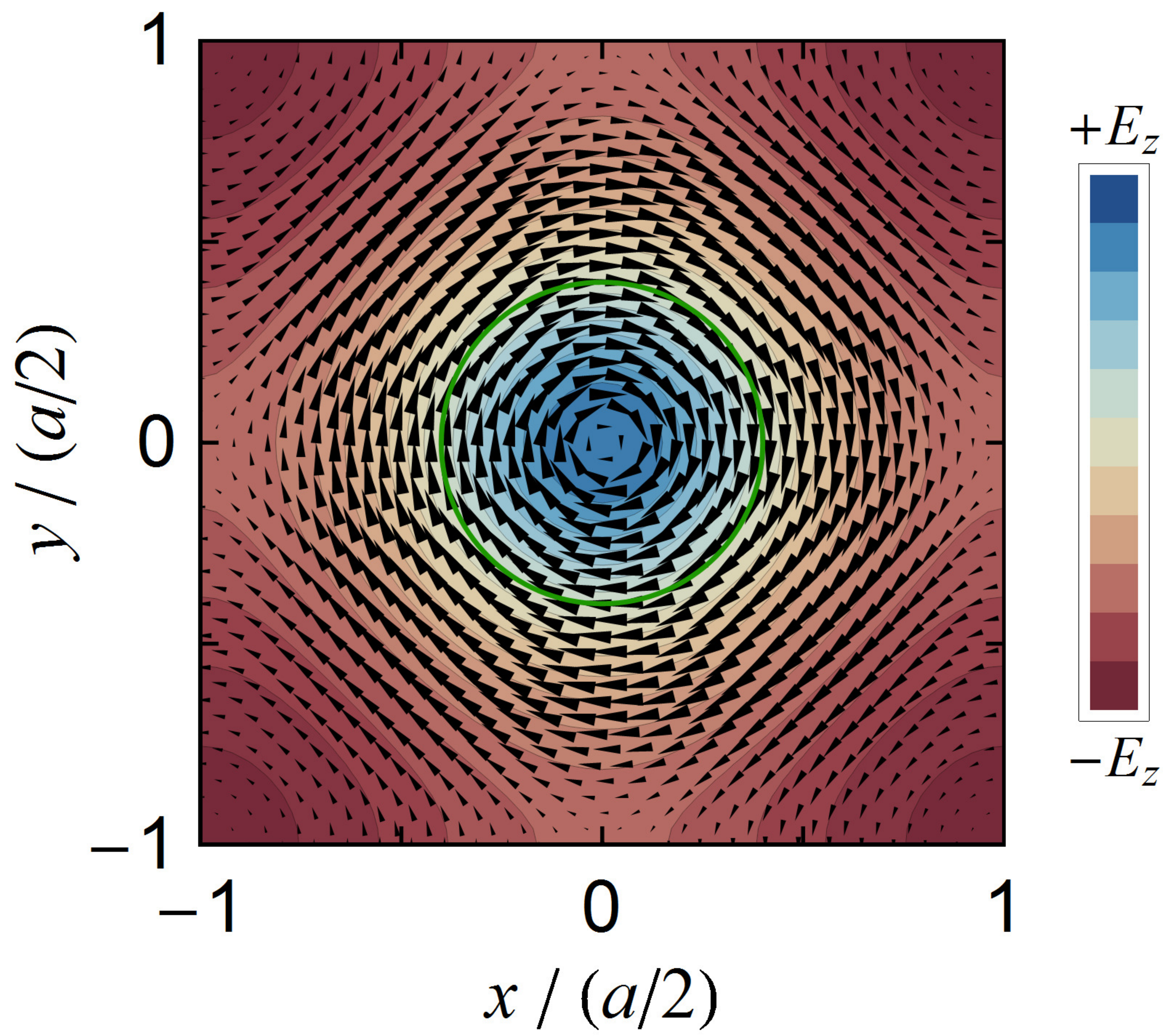}}
\subfloat[][\label{avsfig:TMG2}TM mode 3]
{\includegraphics[width=1.2in]{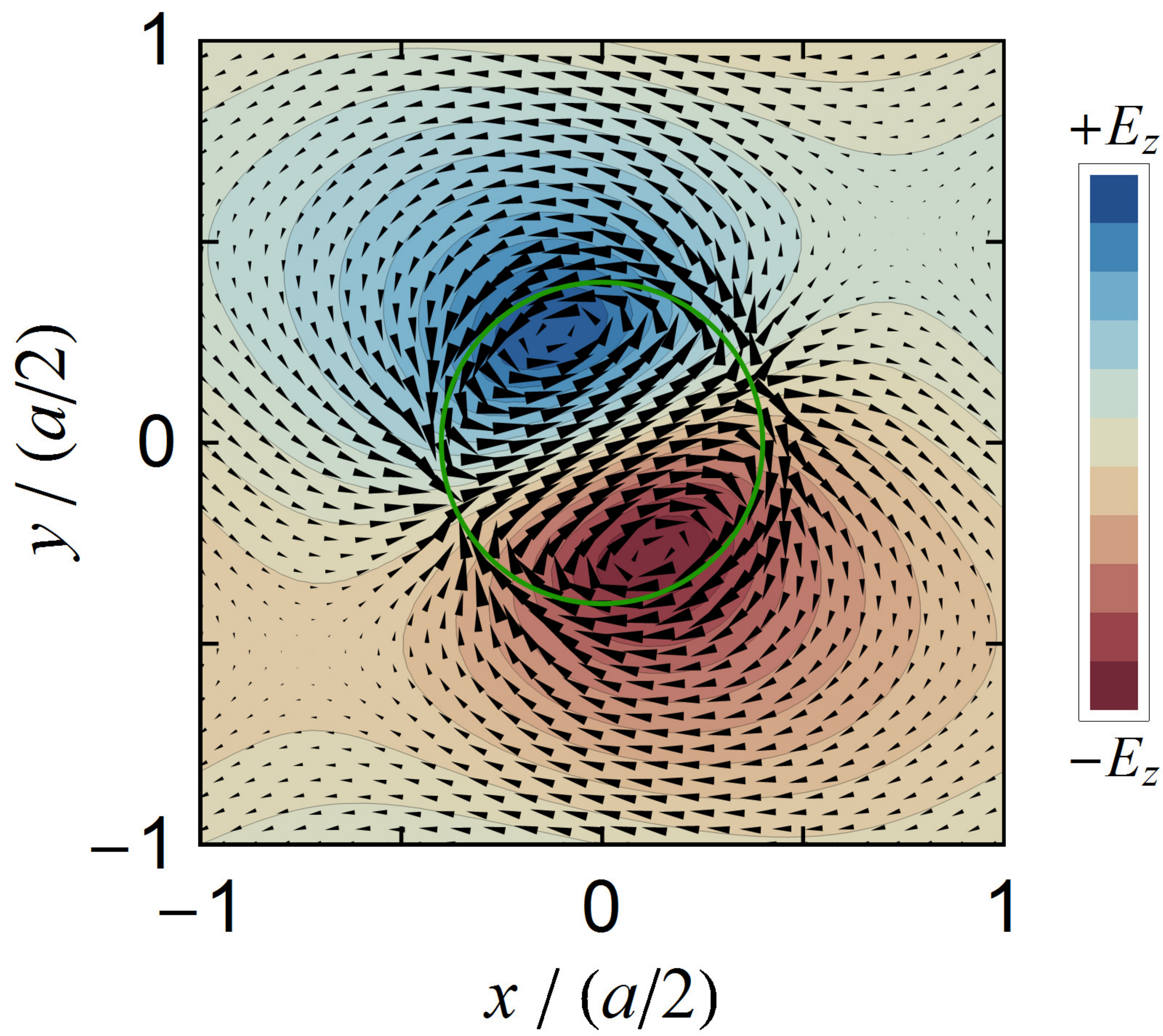}}
\caption[Cylindrical posts: Vector fields, $\Gamma$-point]
{\label{avsfig:Cyl_Gamma}
The electric and magnetic fields of the second and third modes
corresponding to the $\Gamma$-point in the lattice of cylindrical
posts are shown. For TE-modes, the in-plane electric
field is represented by vectors and the out-of-plane
magnetic field is represented by the gradient background.
For TM-modes, the vectors represent the in-plane magnetic
field and the background represents the strength of the
out-of-plane electric field. Note that the first mode
is not shown because the corresponding eigenvalue
is zero, resulting in a trivial solution. }
\end{center}
\end{figure}

 The eigenvector fields can be organized according to
their symmetries with respect to the symmetry group of the crystal and
to the  group   of  the  wavevector. 
In the following, we follow the group-theoretic analysis of
Sakoda.\cite{avs:sakodaPRB1995,avs:sakoda1997,avs:sakoda2001} 
The point group of the cylindrical post unit cell is $C_{4v}$, or the
symmetry of the square. The character table of this group is given in Tinkham.\cite{avs:tinkham}
 The wavevector at the $\Gamma$-point has the
full symmetry of $C_{4v}$. The symmetry of the $\Gamma$-point modes can
be deduced by inspecting the transformation properties of the
eigenvectors that are transverse to the extrusion direction of the
crystal. For a one-dimensional irreducible representation $D_i$, 
operation $R_j$ in class $j$ with character $\chi_i(R_j)$,
and eigenvector field $v$, the eigenvector field will transform
according to
\begin{equation}
	D_i(R_j) \, v = \chi_i(R_j) \, v.
\end{equation}
By inspecting the transformation of $v$ by several $D_i(R_j)$,
the character table can be used to deduce which irreducible
representation the eigenvector field belongs to.
As an example, consider the $\Gamma$-point mode in 
Fig.~\ref{avsfig:TEG1}. 
The transverse vector field satisfies
\begin{equation}
	D_i(C_2)v=v, \quad D_i(2C_4)v=v, \quad D_i(2\sigma_v)v=-v.
\end{equation}
This mode must therefore belong to the irreducible representation
with characters
\begin{equation}
	\chi_i(C_2)=1, \quad \chi_i(2C_4)=1, \quad \chi_i(2\sigma_v) = -1,
\end{equation}
which corresponds to the $\Gamma_2$ representation.
\begin{figure}[t!] 
	\begin{center}
		\includegraphics[width=3.0in]{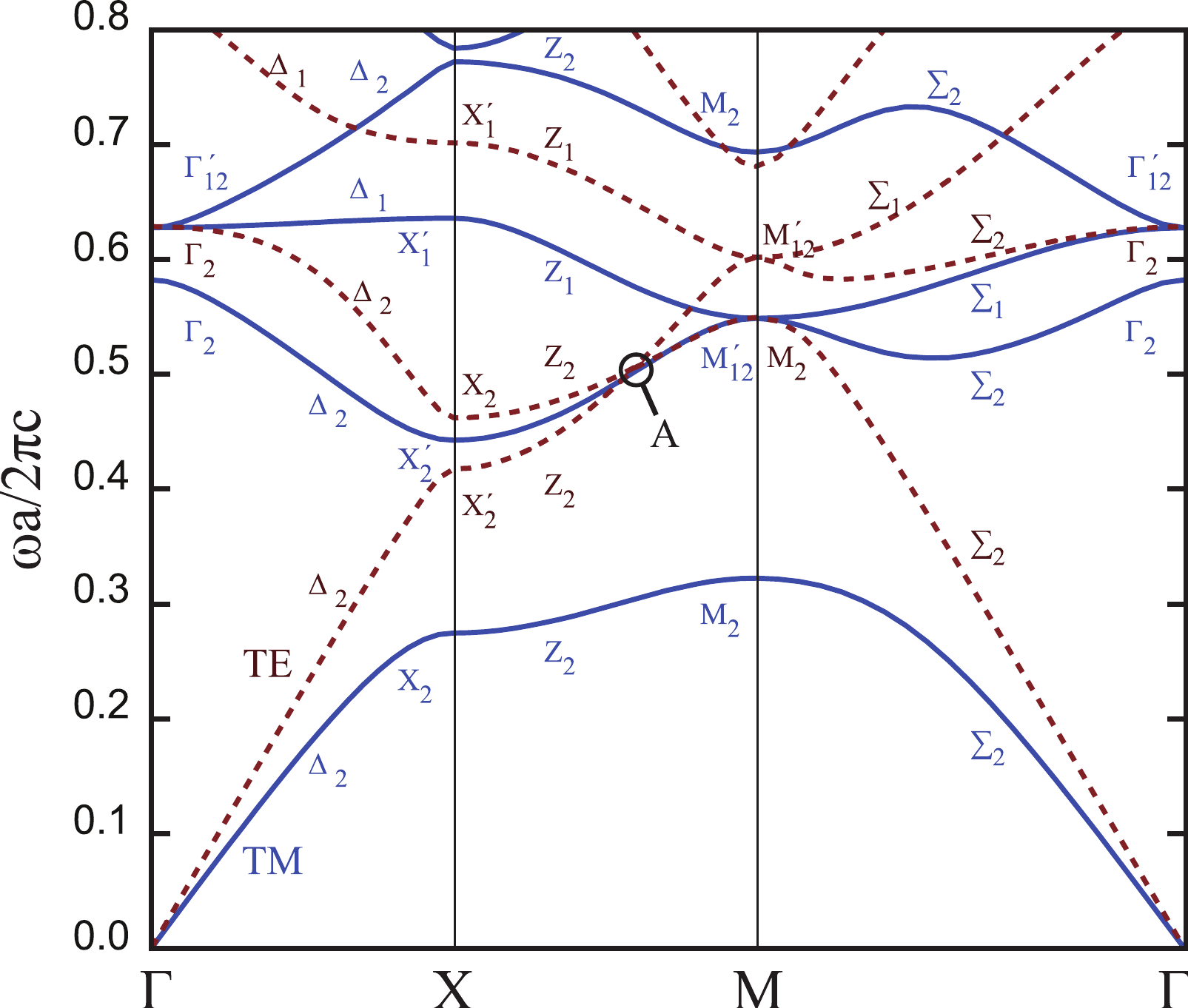}
		\caption[
		Band  structure  of  a  2D  crystal  with  dielectric  posts
		]{ Eigenvalues  for the transverse  electric and  transverse magnetic
			modes as  calculated using the finite element  method with quintic
			Hermite interpolation polynomials.  The  point labeled as A is the
			location  of   an  anticrossing   site  between  the   lowest  and
			second-lowest  TE   modes.  This  location  is   shown  in  higher
			resolution in Fig.~\ref{avsfig:anticross1}.}\label{avsfig:evalsymmetry}
	\end{center}
\end{figure}
\begin{figure}[ht!] 
\begin{center}
\includegraphics[width=3.0in]{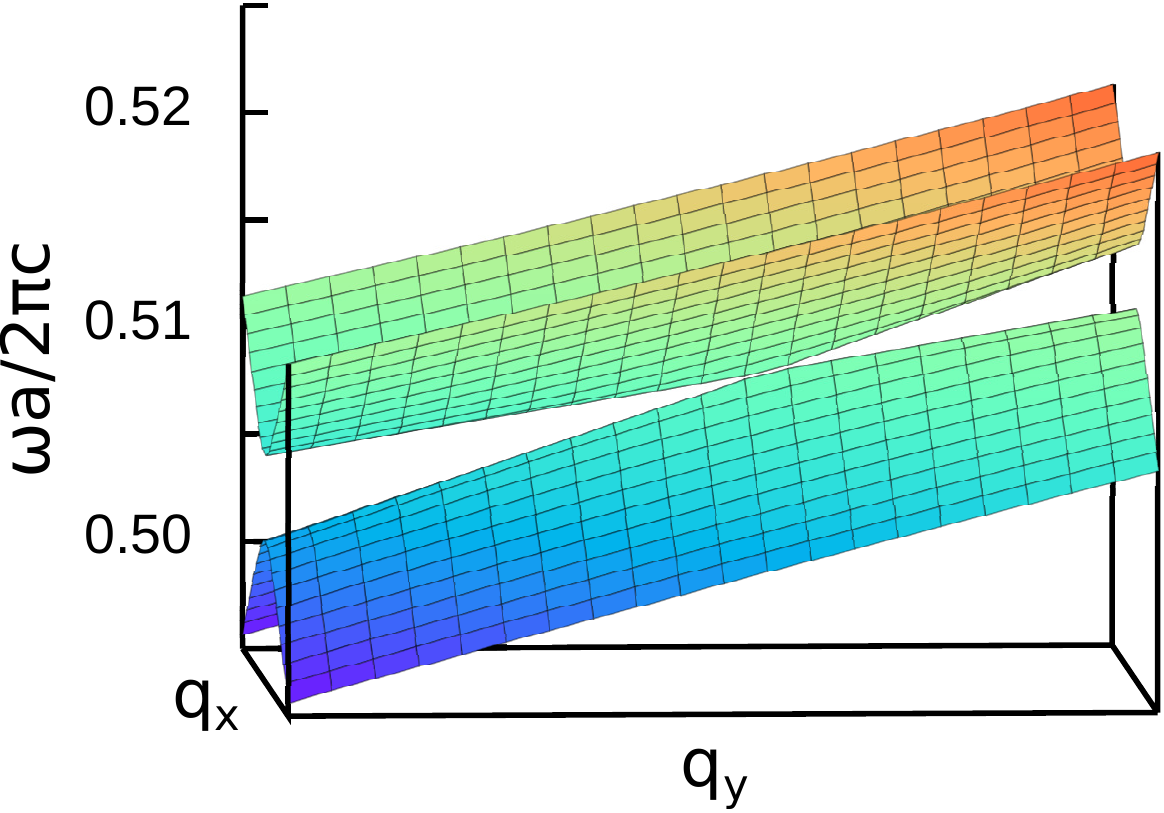}
\end{center}
\caption[Anticrossing  site of  a  2D crystal  with dielectric  posts]
{Close-up  view  of  the  anticrossing   site  shown  at  point  A  in
  Fig.~\ref{avsfig:evalsymmetry}. (Adapted from Boucher {\it et al.},
  Ref.~\onlinecite{avs:CRB_PhCTs}.)}
\label{avsfig:anticross1}
\end{figure}
For modes with  wavevector away from the  $\Gamma$-point, the symmetry
of  the  wavevector itself  must  also  be  taken into  account.   The
$M$-point has  the full  symmetry of $C_{4v}$.  The $X$-point  has the
reduced symmetry group  $C_{2v}$ (the symmetry of  the rectangle) with
the character  table given in Tinkham.\cite{avs:tinkham}  Points along
$\Delta$, $Z$, and $\Sigma$ have the still further reduced symmetry of
$C_{1h}$ (bilateral symmetry) with the  character table given in books
on   group  representation   theory.\cite{avs:tinkham,avs:dresselhaus}
Points along $Z$ have $C_{1h}$ symmetry  due to the fact that a mirror
through  the line  orthogonal to  the  $q_x$ direction  brings $Z$  to
$Z+Q$, where $Q$ is a reciprocal lattice translation vector.

The dispersion relations for the lowest few modes of the cylindrical post 
labeled by their irreducible representations are shown in 
Fig.\ref{avsfig:evalsymmetry}. Notice that in Fig.~\ref{avsfig:anticross1}
there is an anticrossing site 
in the TE modes along $Z$. Since the irreducible representations
form an orthogonal basis, anitcrossings can only occur between
modes within the same irreducible representation.
Indeed, this is the case here, as the two anticrossing modes are
in the $Z_2$ irreducible representation.

\begin{figure}[b!] 
\begin{center}
\subfloat[][\label{avsfig:TEX0}TE mode $1$]
{\includegraphics[width=1in]{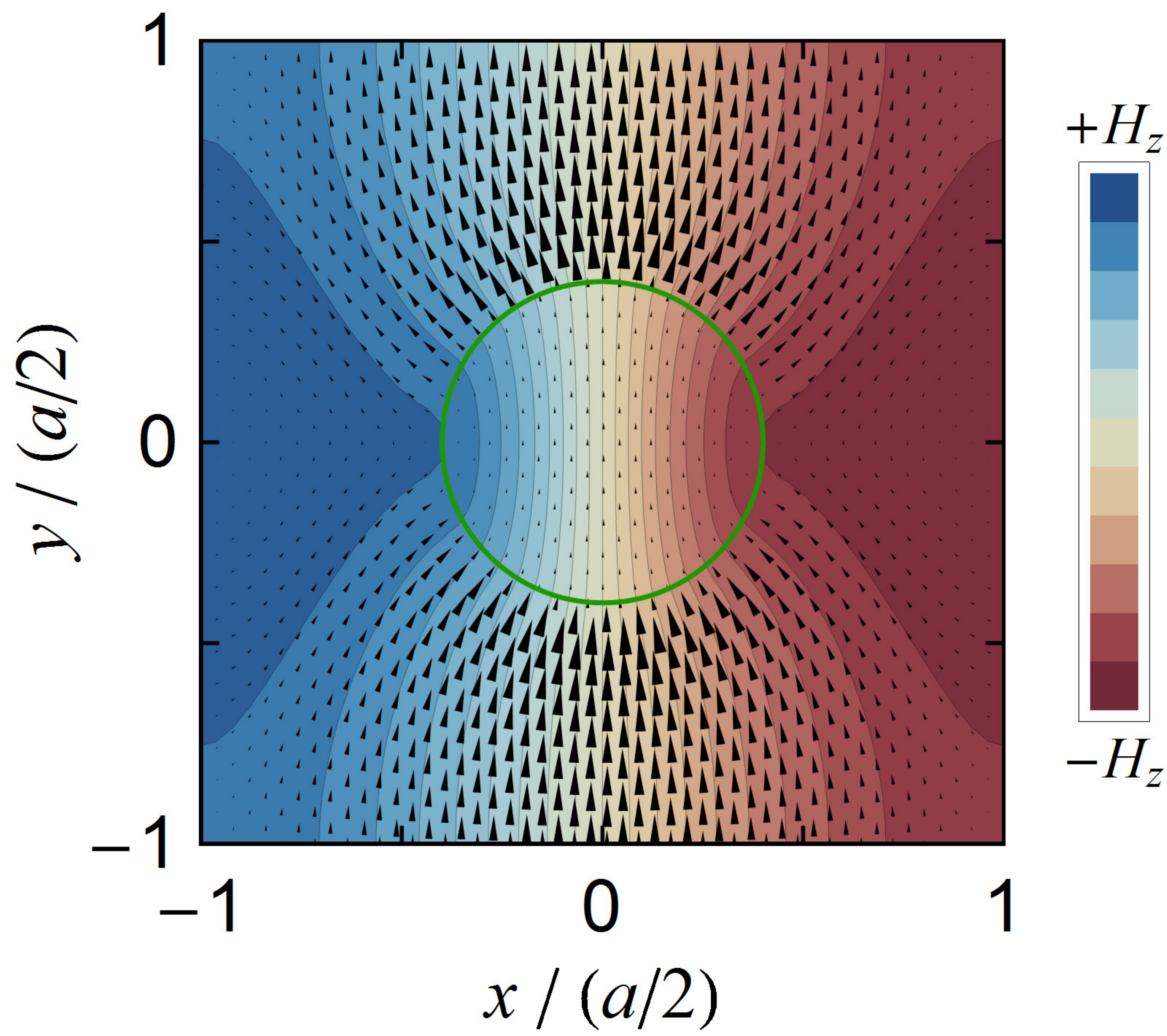}}
\subfloat[][\label{avsfig:TEX1}TE mode $2$]
{\includegraphics[width=1in]{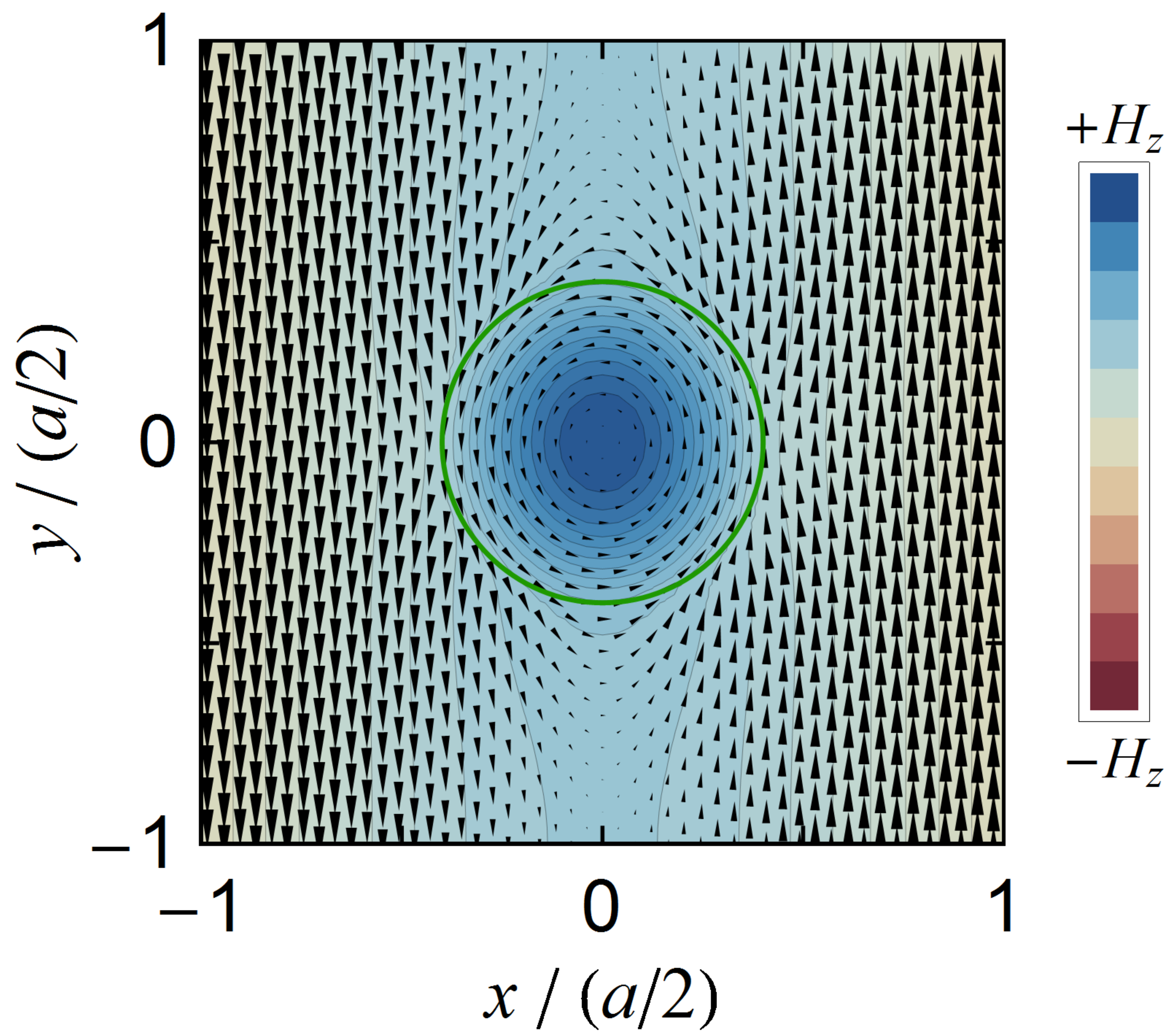}}
\subfloat[][\label{avsfig:TEX2}TE mode $3$]
{\includegraphics[width=1in]{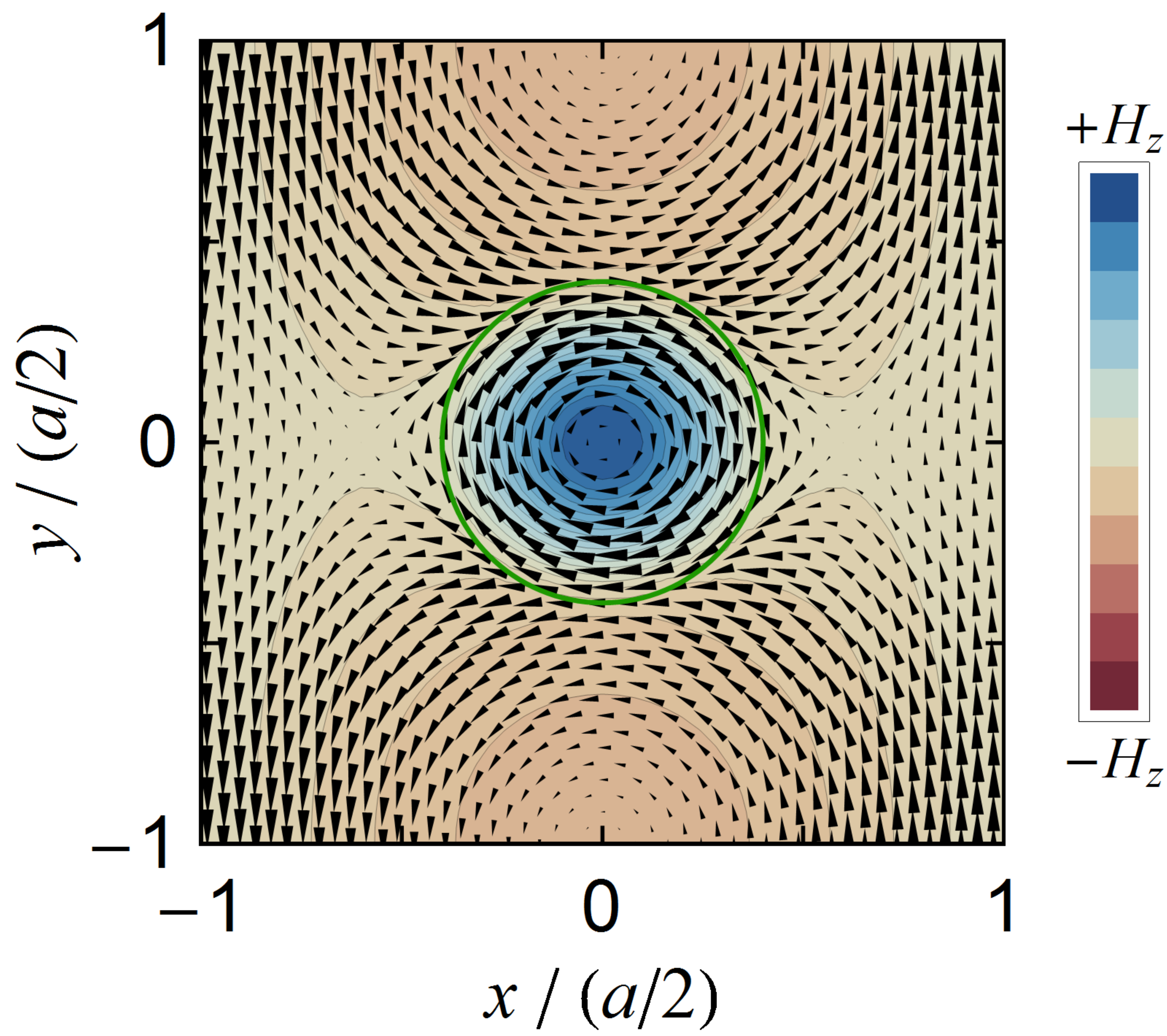}}
\\
\subfloat[][\label{avsfig:TMX0}TM mode $1$]
{\includegraphics[width=1in]{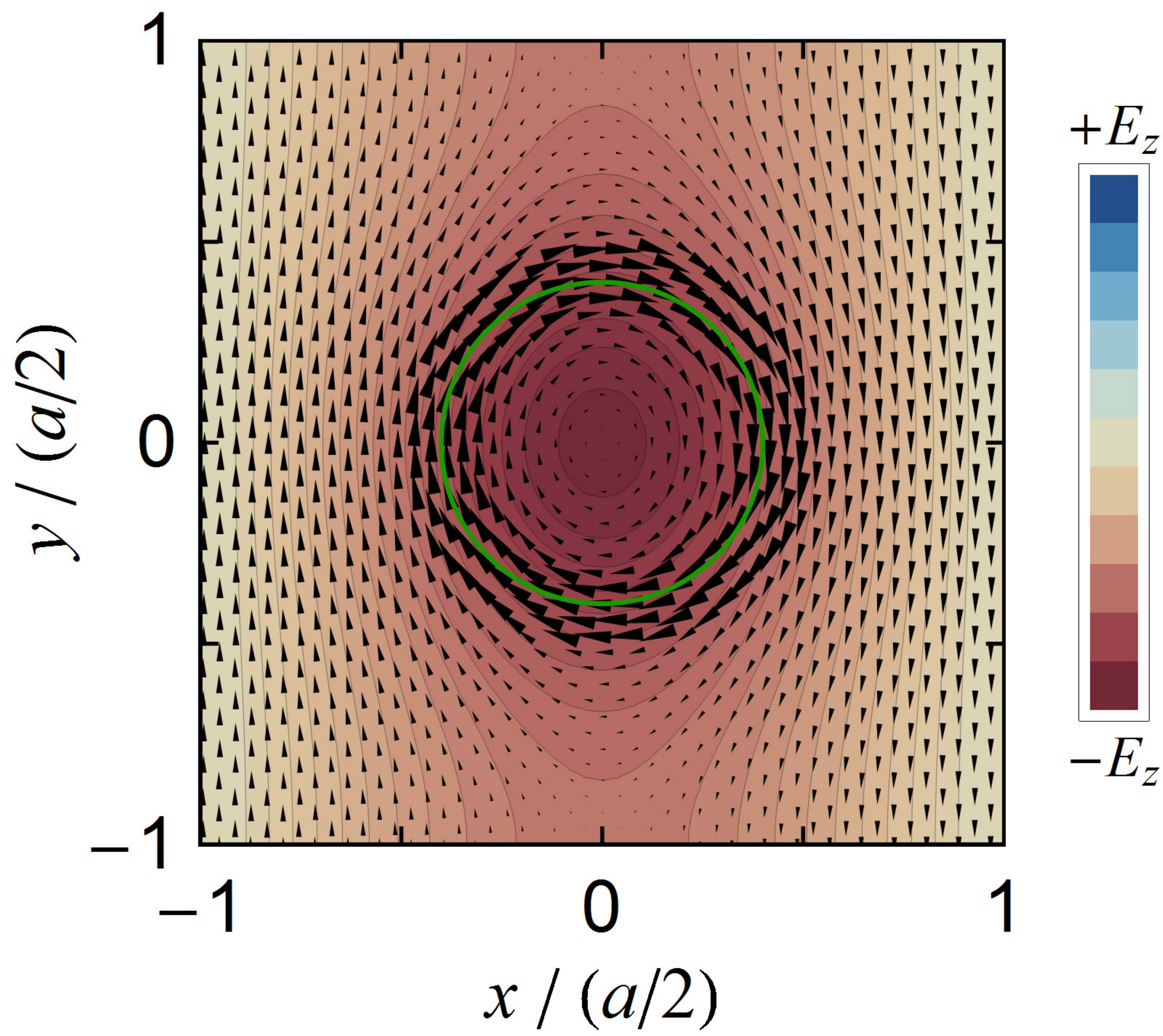}}
\subfloat[][\label{avsfig:TMX1}TM mode $2$]
{\includegraphics[width=1in]{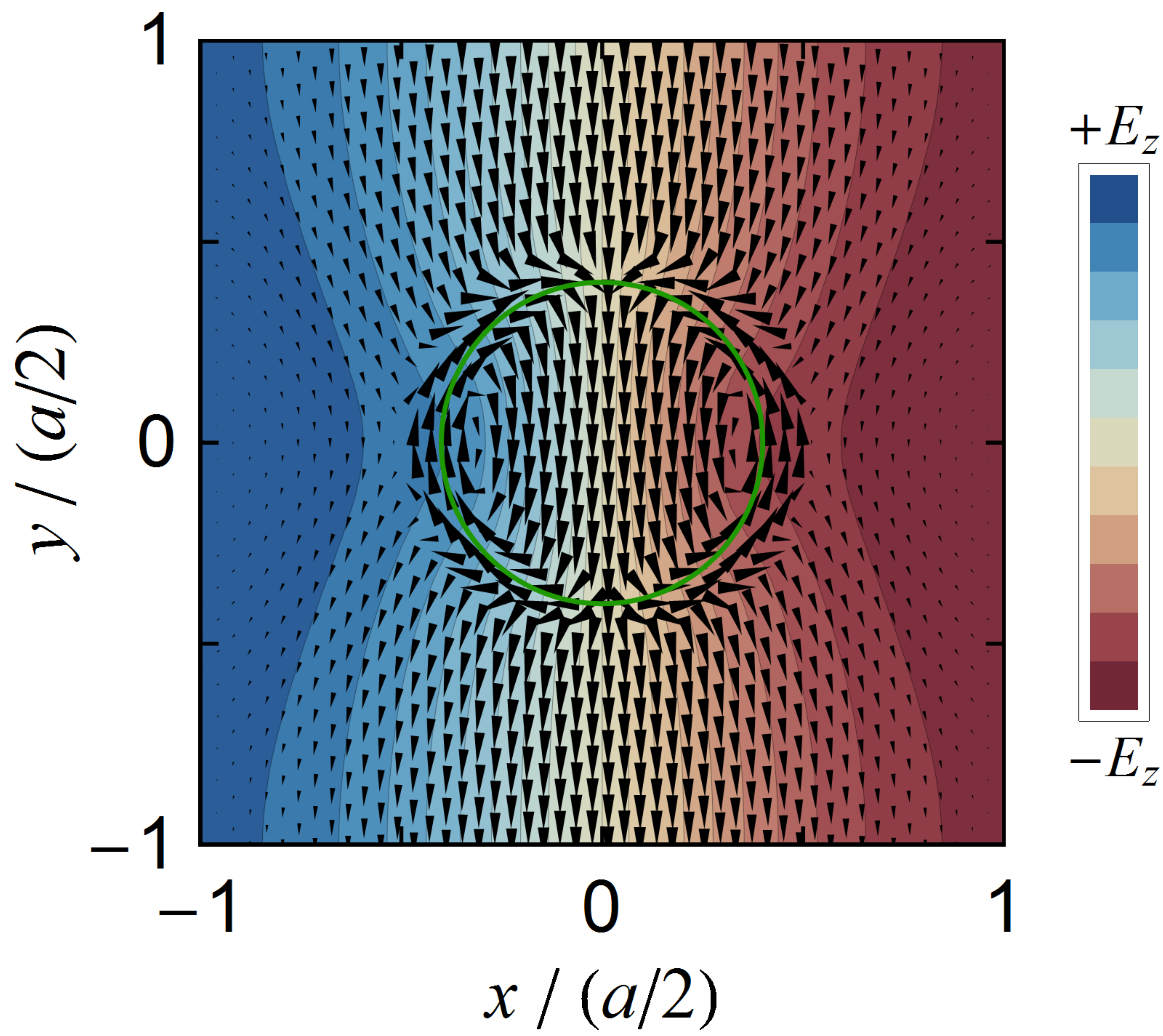}}
\subfloat[][\label{avsfig:TMX2}TM mode $3$]
{\includegraphics[width=1in]{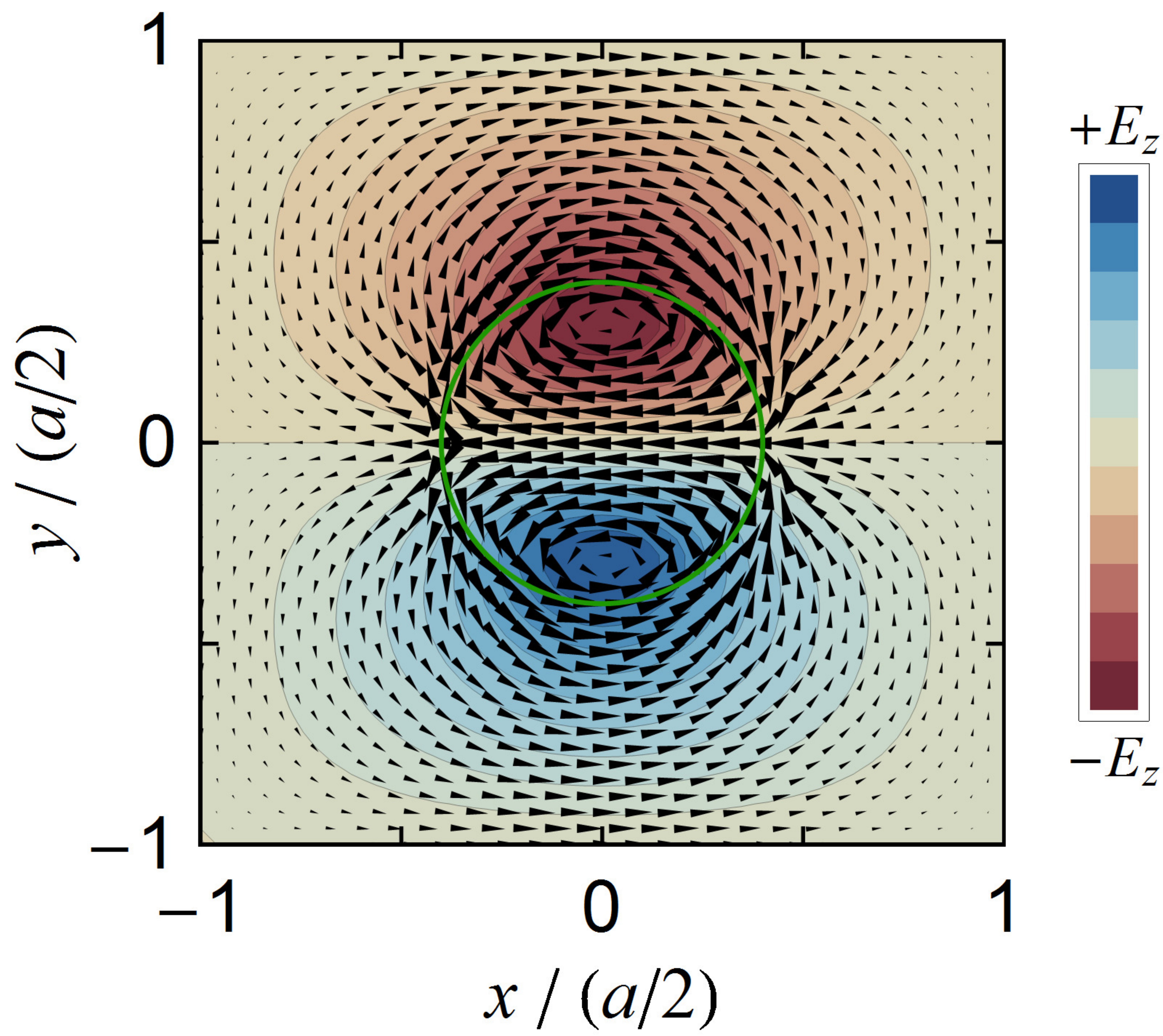}}
\caption[Cylindrical posts: Vector fields, $X$-point]
{\label{avsfig:Cyl_X}
The electric and magnetic fields of the first three modes
corresponding to the $X$-point in the lattice of cylindrical
posts are shown. For TE-modes, the in-plane electric
field is represented by vectors and the out-of-plane
magnetic field is represented by the gradient background.
For TM-modes, the vectors represent the in-plane magnetic
field and the background represents the strength of the
out-of-plane electric field.}
\end{center}
\end{figure}
\subsection{Eigenstates for periodic dielectric posts}

\begin{figure}[b!] 
\begin{center}
\subfloat[][\label{avsfig:TEPostDisp}TE mode]
{\includegraphics[width=1.5in,height=1.8in]{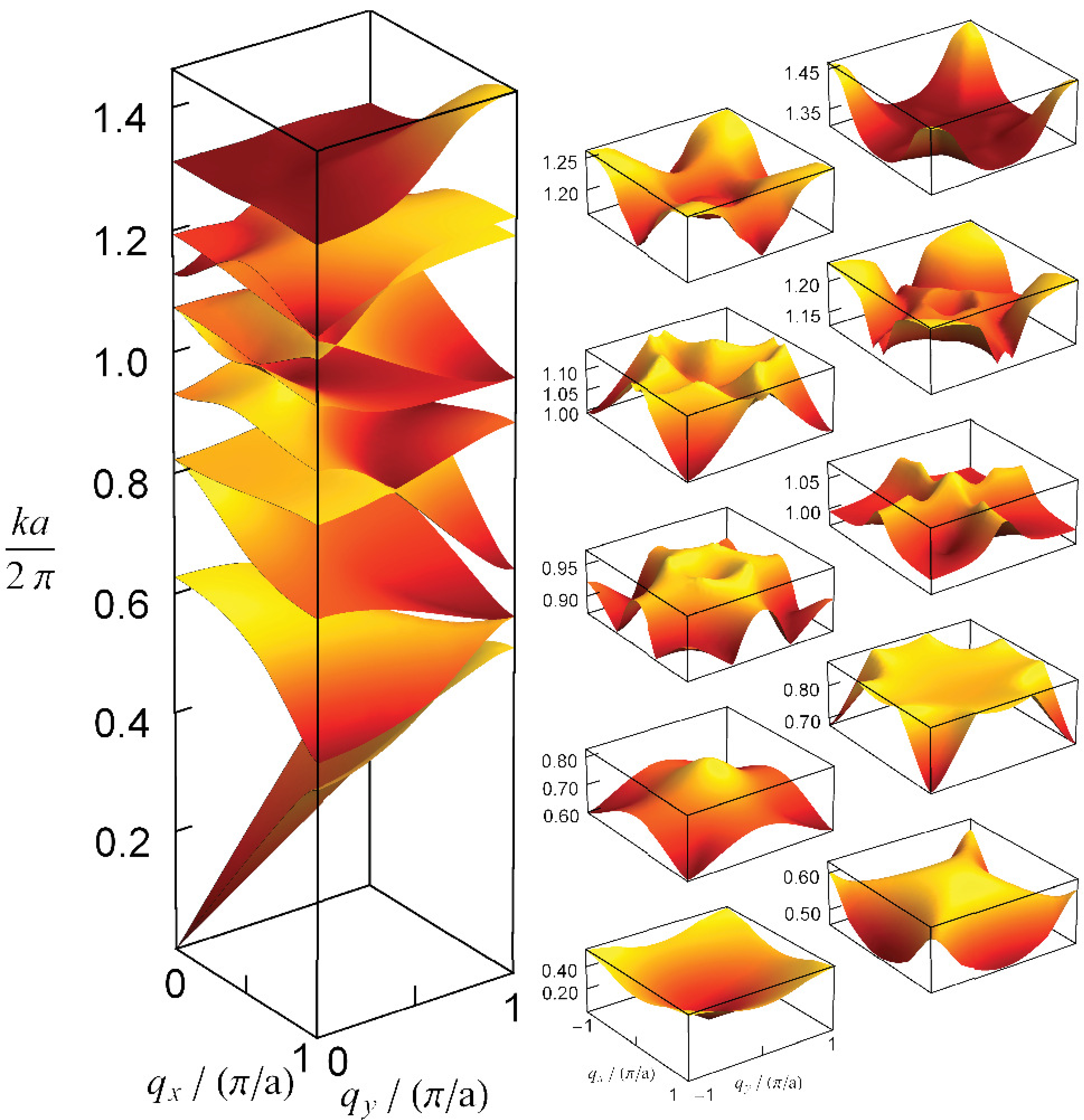}}\hspace{0.2in}
\subfloat[][\label{avsfig:TMPostDisp}TM mode]
{\includegraphics[width=1.5in, height=1.8in]{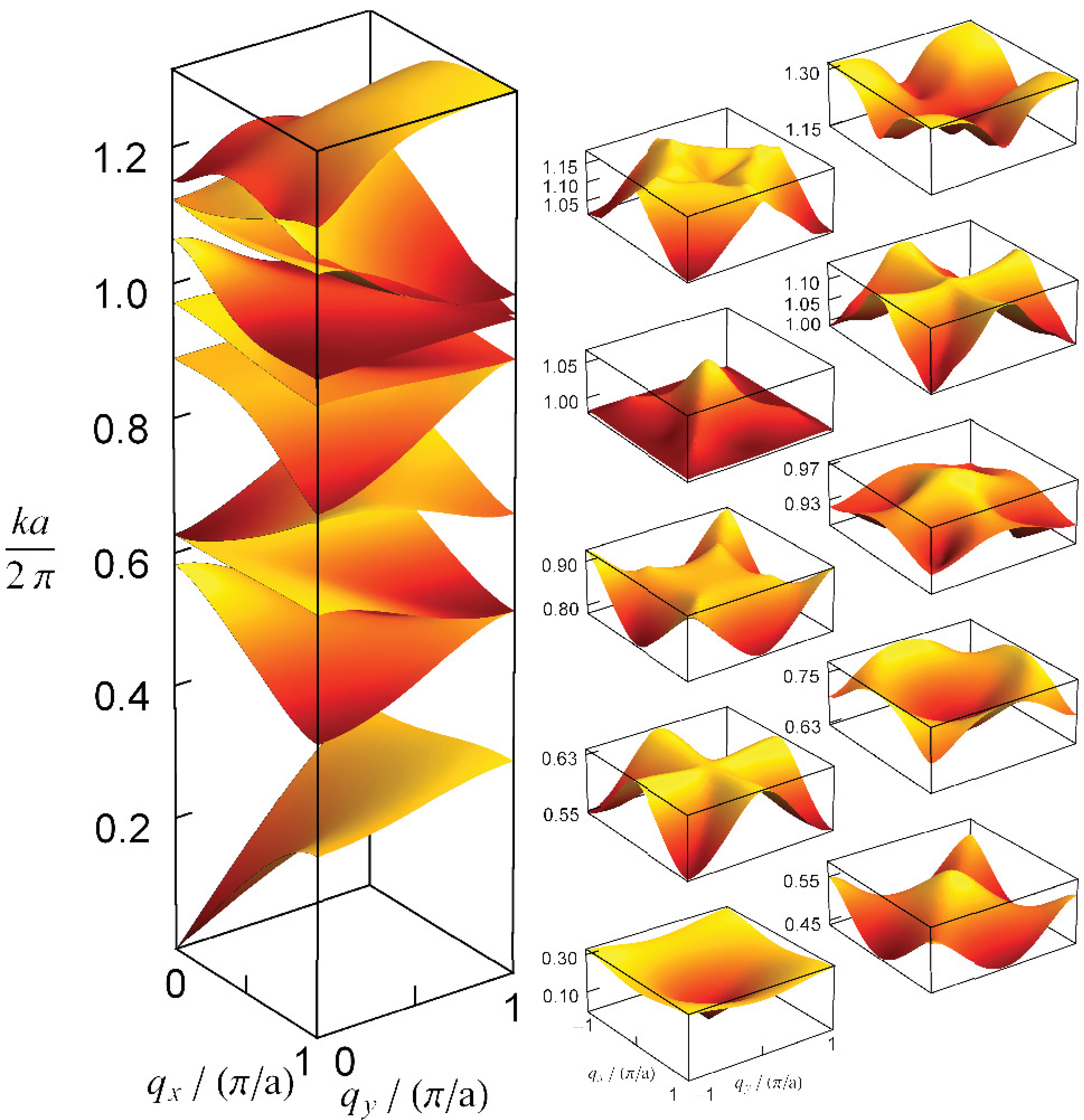}}
\caption{The eigenvalues of the transverse
(a) electric modes and (b) magnetic modes of the periodic lattice
of dielectric posts are plotted as surfaces in the first
Brillouin zone. On the left side of (a) an (b), the first ten eigenvalues
are shown in the irreducible part of the Brillouin zone for TE and TM modes,
respectively. On the right side, each eigenvalue has been separated from
the rest and extended to the full Brillouin zone through
symmetry operations.}
\end{center}
\end{figure}


  The   band  structure  for   the  lattice  of
dielectric posts was computed using a mesh of $4420$ nodes, yielding a
matrix size of $26520\times 26520$.  The mesh was refined in the region
surrounding  the  edge of  the  cylindrical  post.  The
curves  shown  in  Fig.~\ref{avsfig:evalsymmetry}  give  the  behavior  of  the
propagating frequencies of radiation  at various points along the edge
of the  irreducible part  of the Brillouin  zone.  The  finite element
method reproduces a  band gap in the TM modes  which is also predicted
by the planewave method.  Using  finite elements, it  is also possible
to increase  the resolution close  to the anticrossing site  marked in
Fig.~\ref{avsfig:evalsymmetry}.    This  is  a   location  at   which  multiple
eigenvalues of the same polarization  (i.e. both TM or both TE) appear
to  touch.  The close-up  view  of  this  point  on  the  edge of  the
Brillouin zone
is given in Fig.~\ref{avsfig:anticross1}.\\

The  eigenfunctions  for the  arrangement  of
cylindrical       dielectric       posts       are      shown       in
Figs.~\ref{avsfig:Cyl_Gamma}-\ref{avsfig:Cyl_X}. Note that the point symmetries
of each mode at the high symmetry points of $\Gamma, X$ and $M$ can
be used, along with their character tables,
 to verify the symmetry groups
shown in Fig.~\ref{avsfig:evalsymmetry} by direct observation of the
eigenvector fields.

The dispersion relations are calculated for the irreducible Brillouin
zone, which is only one eighth of the full Brillouin zone as shown in
Fig.~\ref{avsfig:posts}, and then their full
reconstruction over the entire zone is performed. This can reduce
computation time by a factor of 8. The lowest few
TE and TM dispersion relations
are shown in Figs.~\ref{avsfig:TEPostDisp}-\ref{avsfig:TMPostDisp}. These
three-dimensional dispersion surfaces also provide another means
of visualizing band gaps in the TE and TM modes, which are of great
interest in photonic crystal applications. 

\begin{figure}[ht!] 
  \begin{center}
\subfloat[][\label{avsfig:checkers}]
{\includegraphics[width=1.1in]{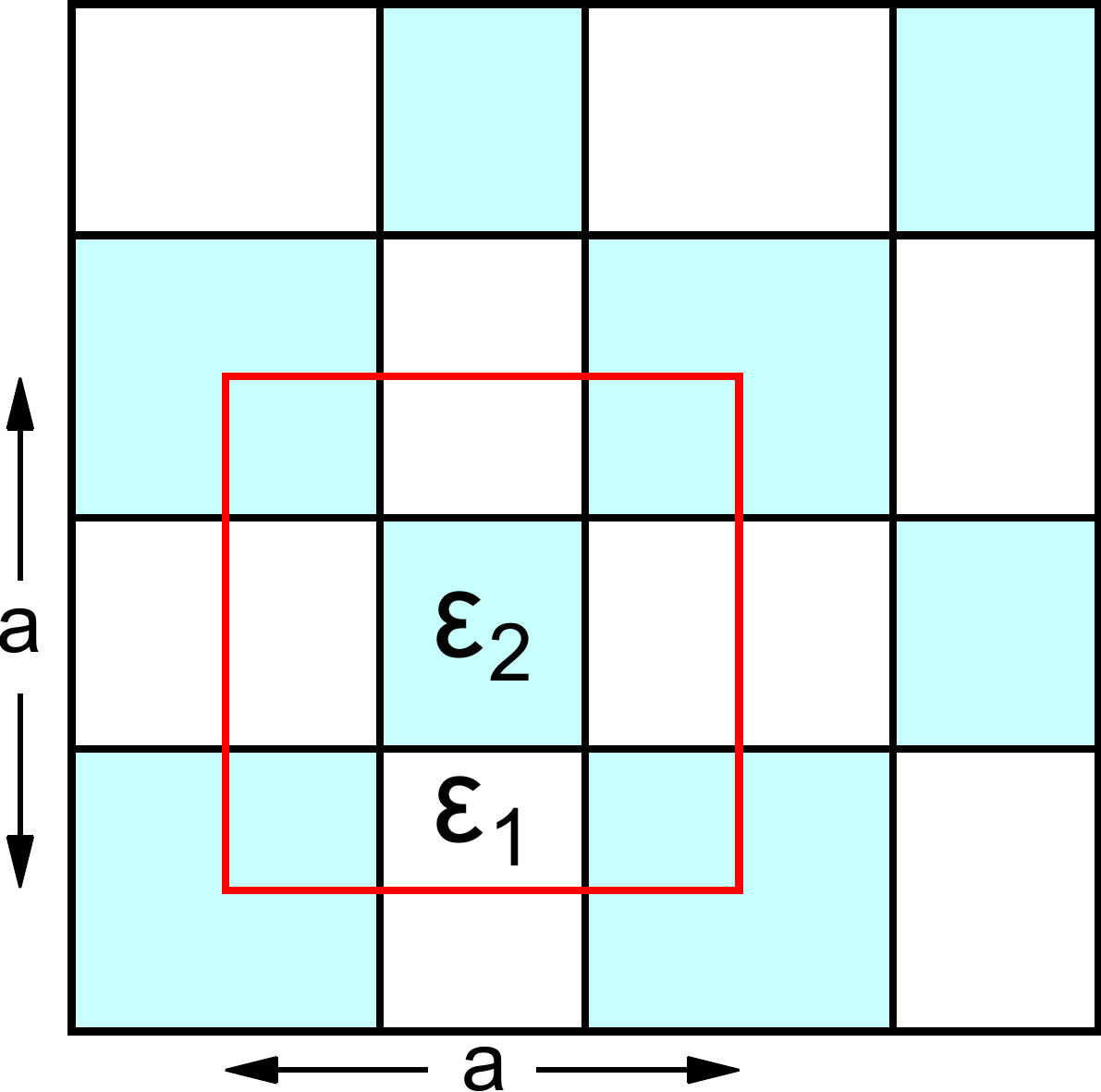}}\hspace{0.1in}
\subfloat[][\label{avsfig:checkermesh}]
{\includegraphics[width=1.1in]{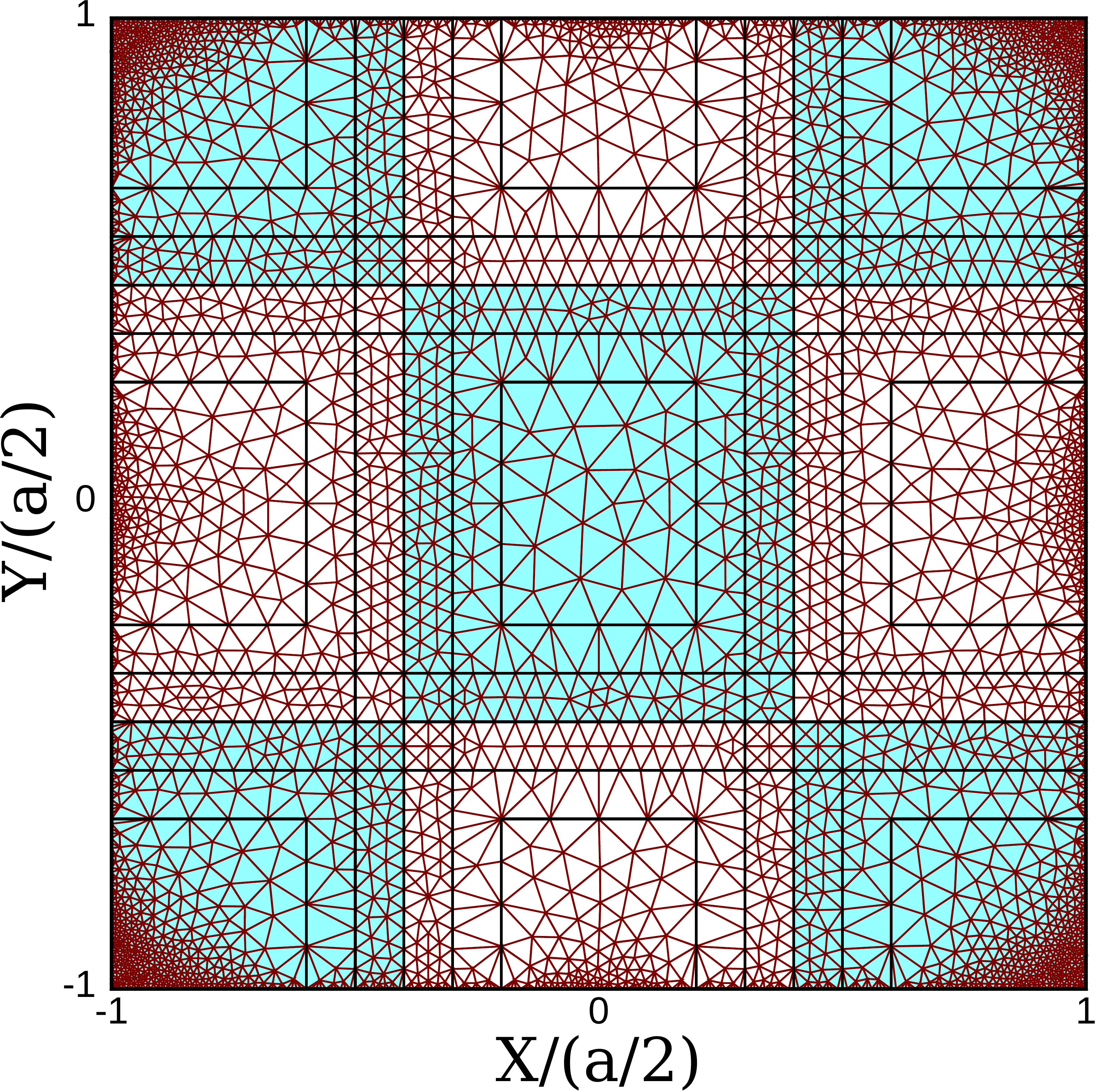}}    
\end{center}
\caption[Geometry of  a photonic crystal with  a checkerboard pattern]
{(a) A  periodic  checkerboard  pattern  with two  alternating  dielectric
  materials is shown. Note that the sizes  of adjacent checkers within a single
  unit cell do not necessarily match. (b) A sample finite element mesh is given for the unit cell of a
  checkerboard lattice. Mesh refinement occurs at all of the checker
  boundaries.}
\end{figure}

\begin{figure}[ht!] 
\begin{center}
\includegraphics[width=2.0in]{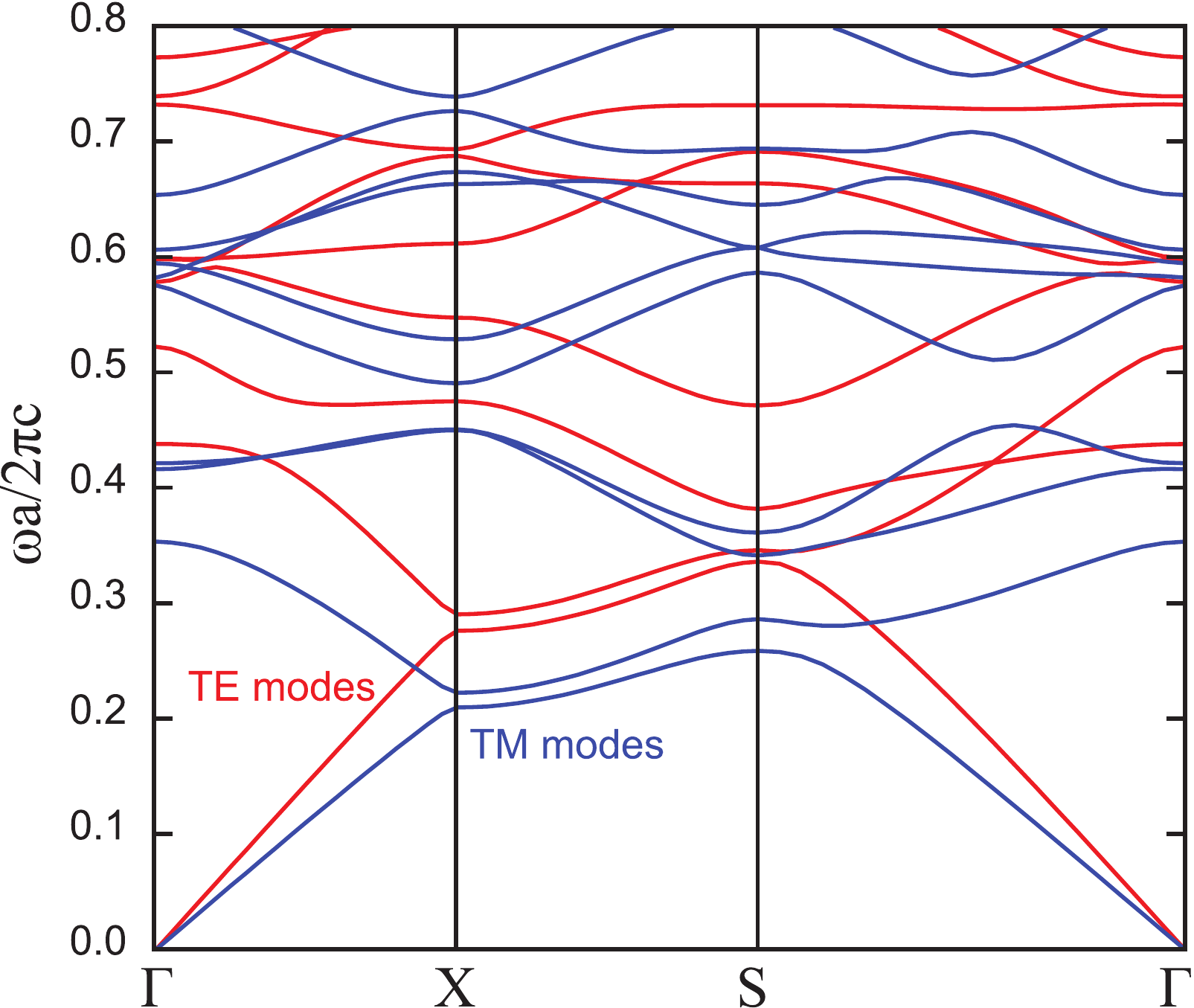}
\end{center}
\caption[Band structure of a 2D crystal with the checkerboard pattern]
{\label{avsfig:checkevals}
  Eigenvalues for the transverse electric and transverse magnetic modes
  for the checkerboard arrangement.}
\end{figure}
\begin{figure*}[t!] 
\begin{center}
\subfloat[][\label{avsfig:TEG12}TE mode 2 at $\Gamma$]
{\includegraphics[width=1.3in]{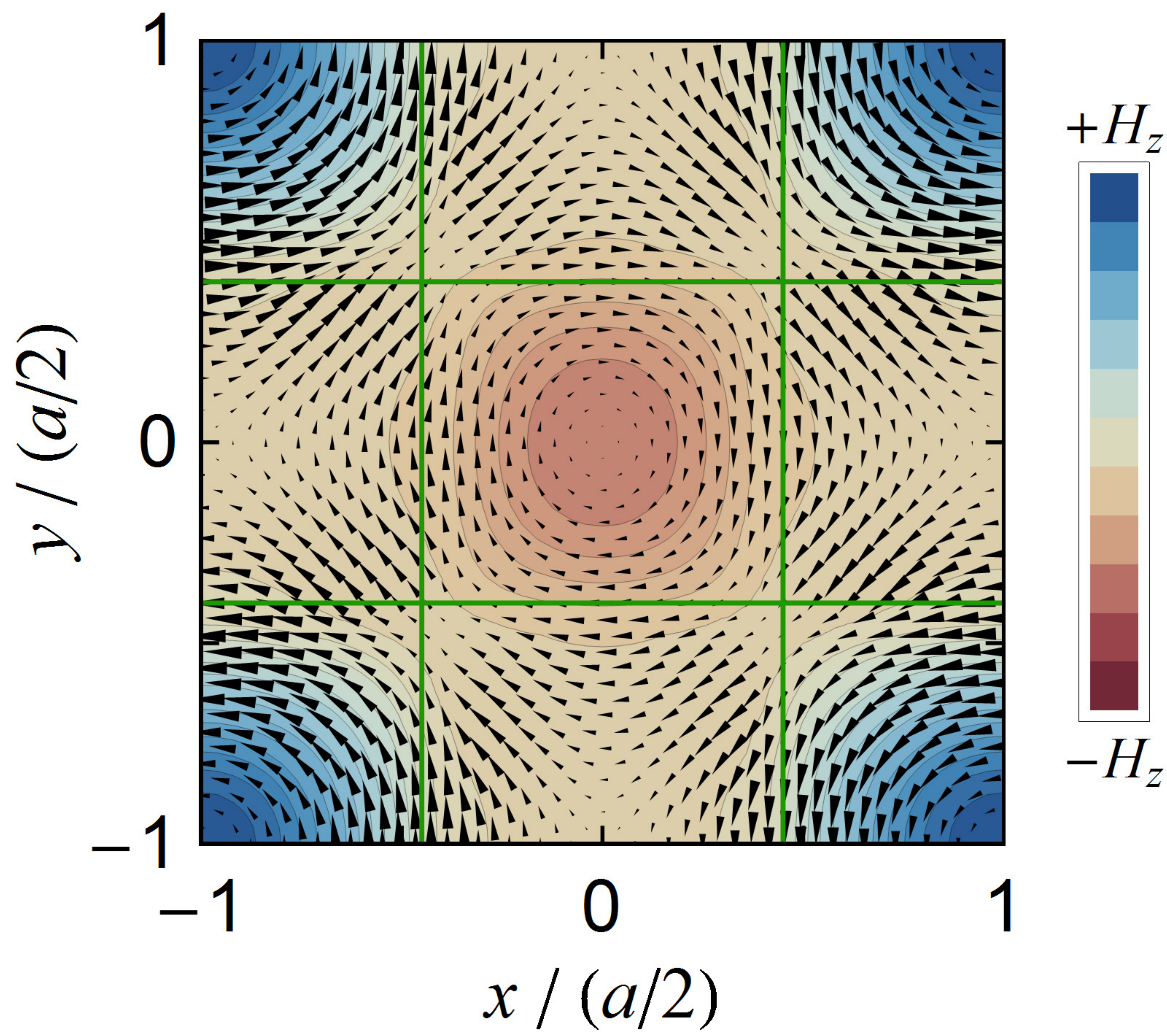}}
\subfloat[][\label{avsfig:TEG22}TE mode 3 at $\Gamma$]
{\includegraphics[width=1.3in]{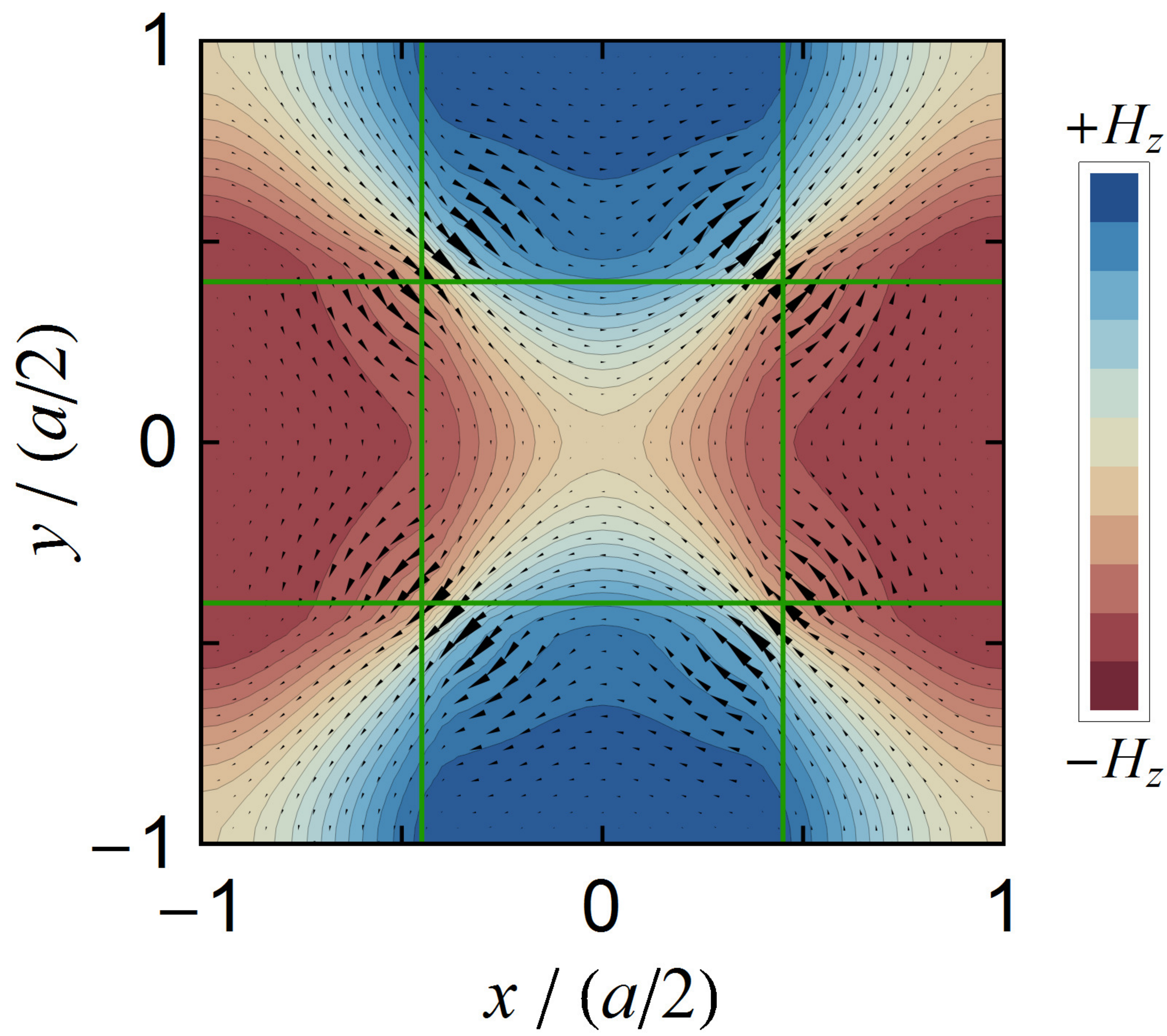}}
\subfloat[][\label{avsfig:TMG12}TM mode 2 at $\Gamma$]
{\includegraphics[width=1.3in]{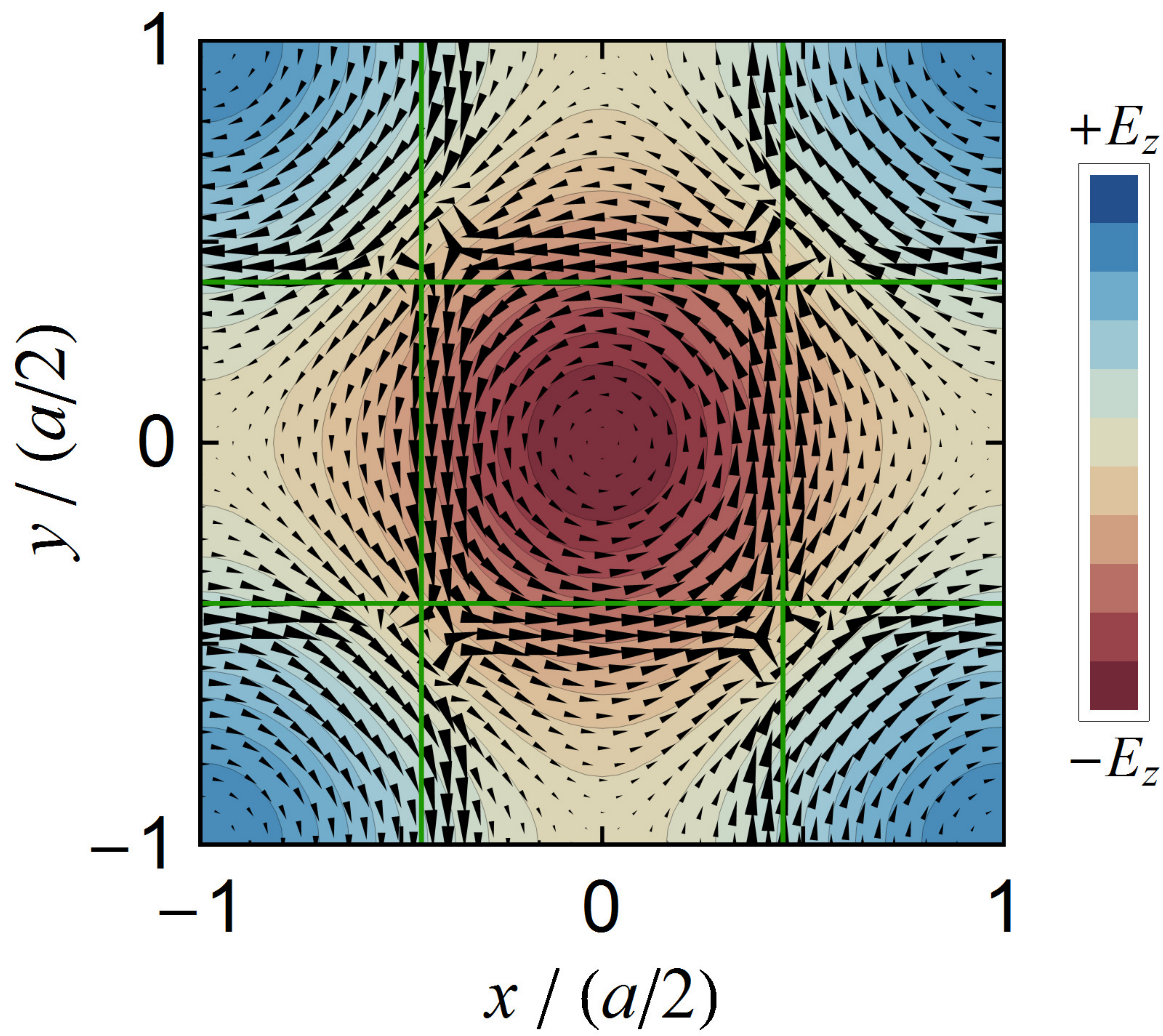}}
\subfloat[][\label{avsfig:TMG22}TM mode 3 at $\Gamma$]
{\includegraphics[width=1.3in]{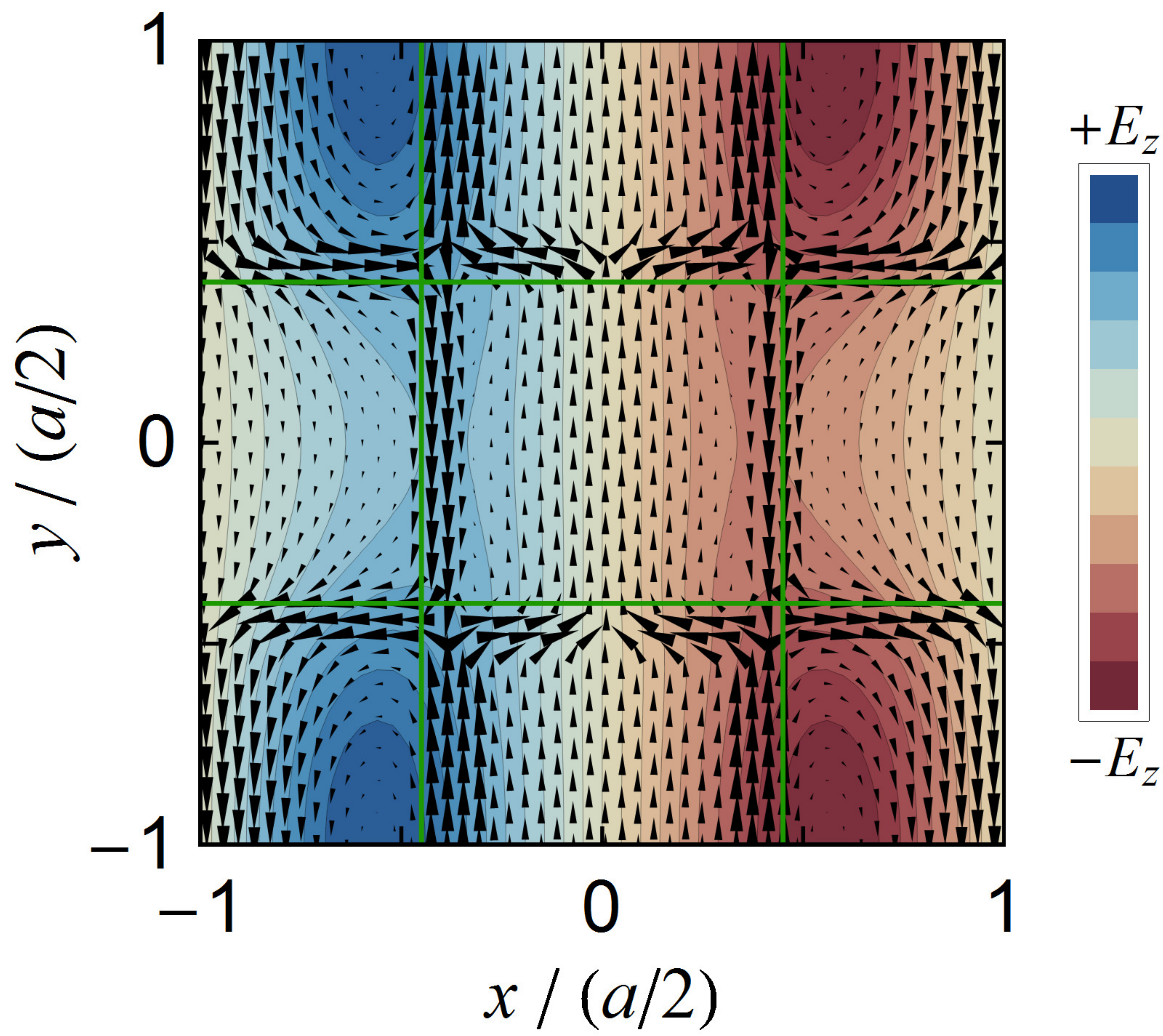}}
\\
\subfloat[][\label{avsfig:TEX02}TE mode $1$ at $X$]
{\includegraphics[width=1.3in]{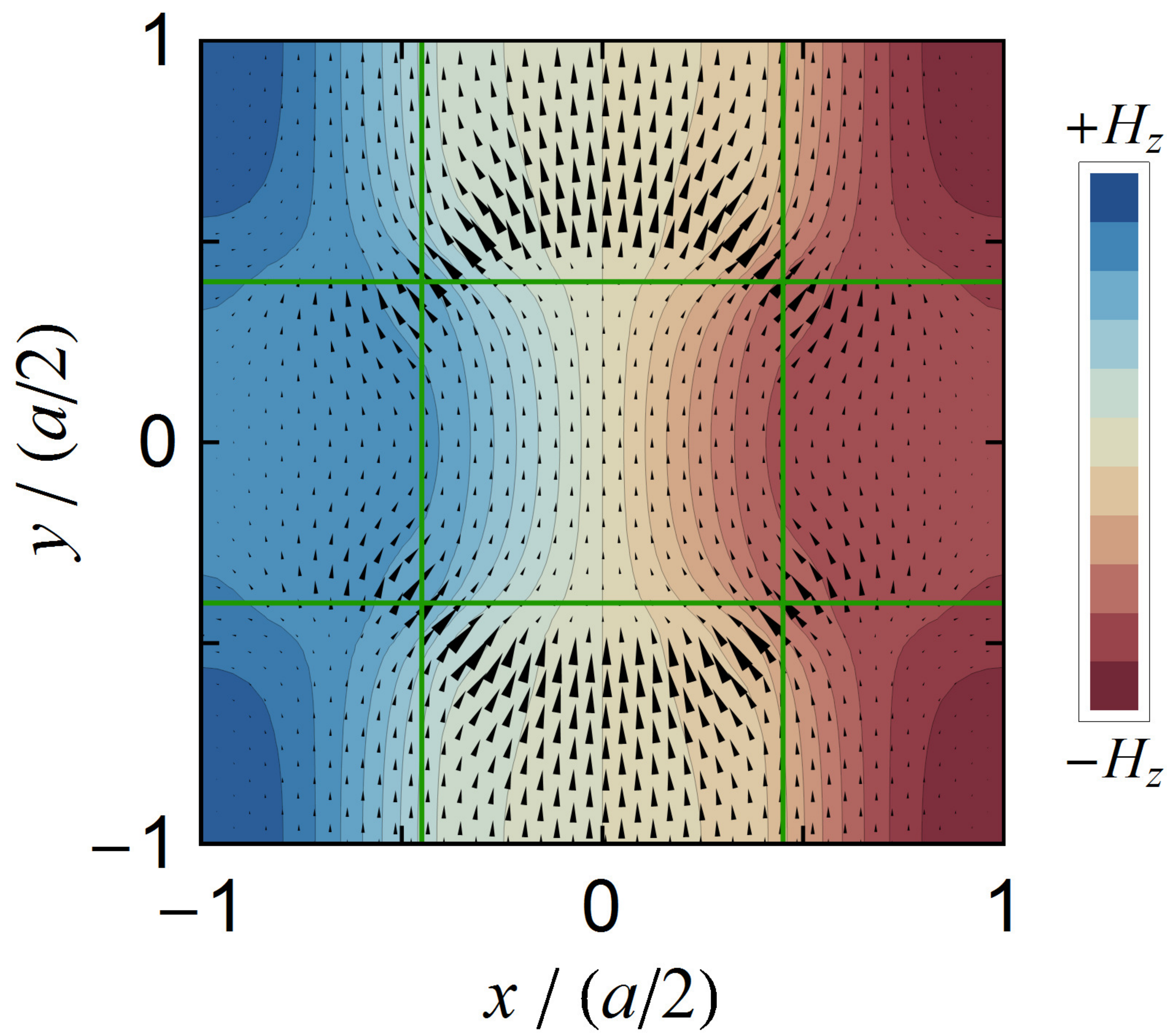}}
\subfloat[][\label{avsfig:TEX12}TE mode $2$ at $X$]
{\includegraphics[width=1.3in]{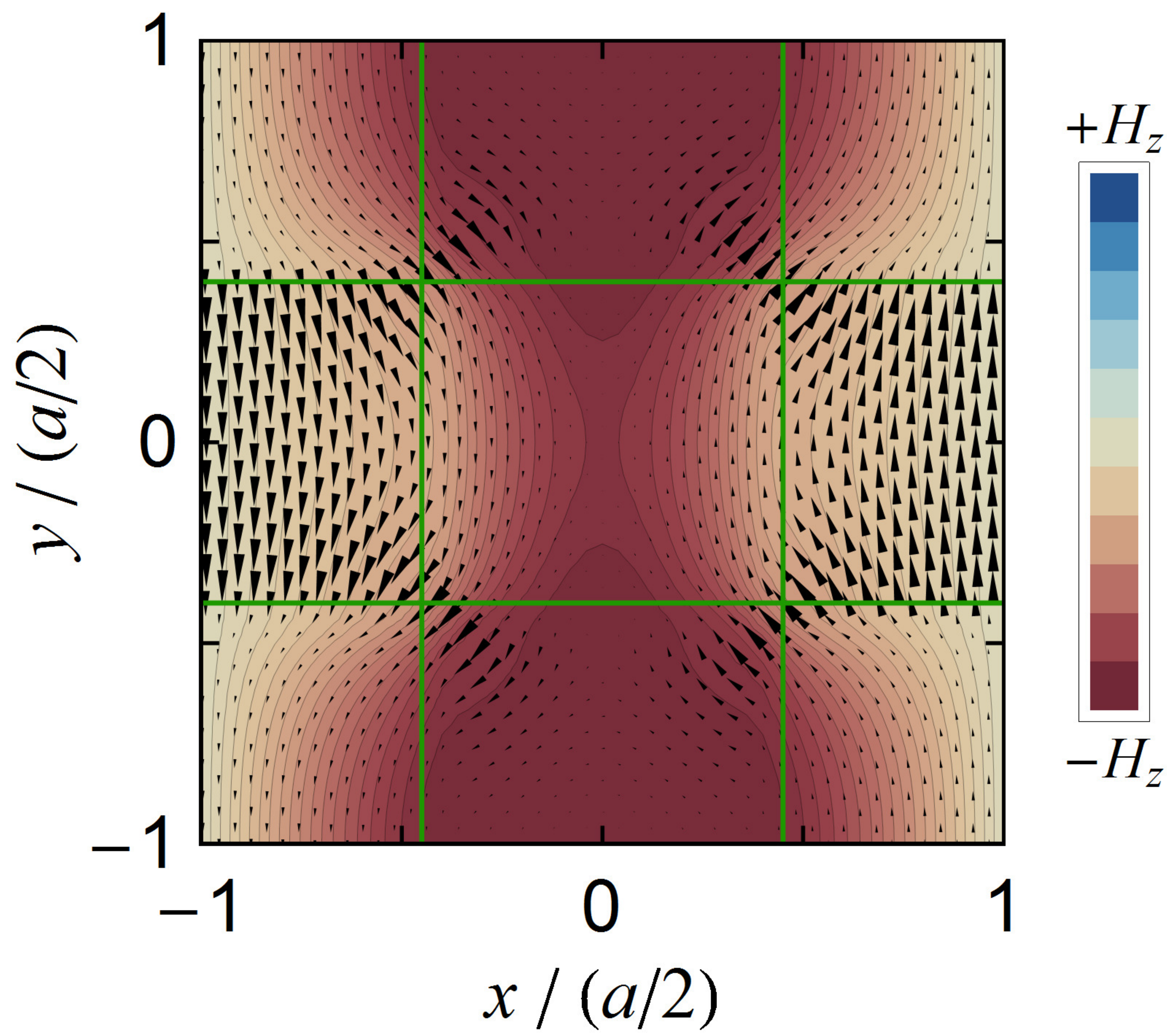}}
\subfloat[][\label{avsfig:TEX22}TE mode $3$ at $X$]
{\includegraphics[width=1.3in]{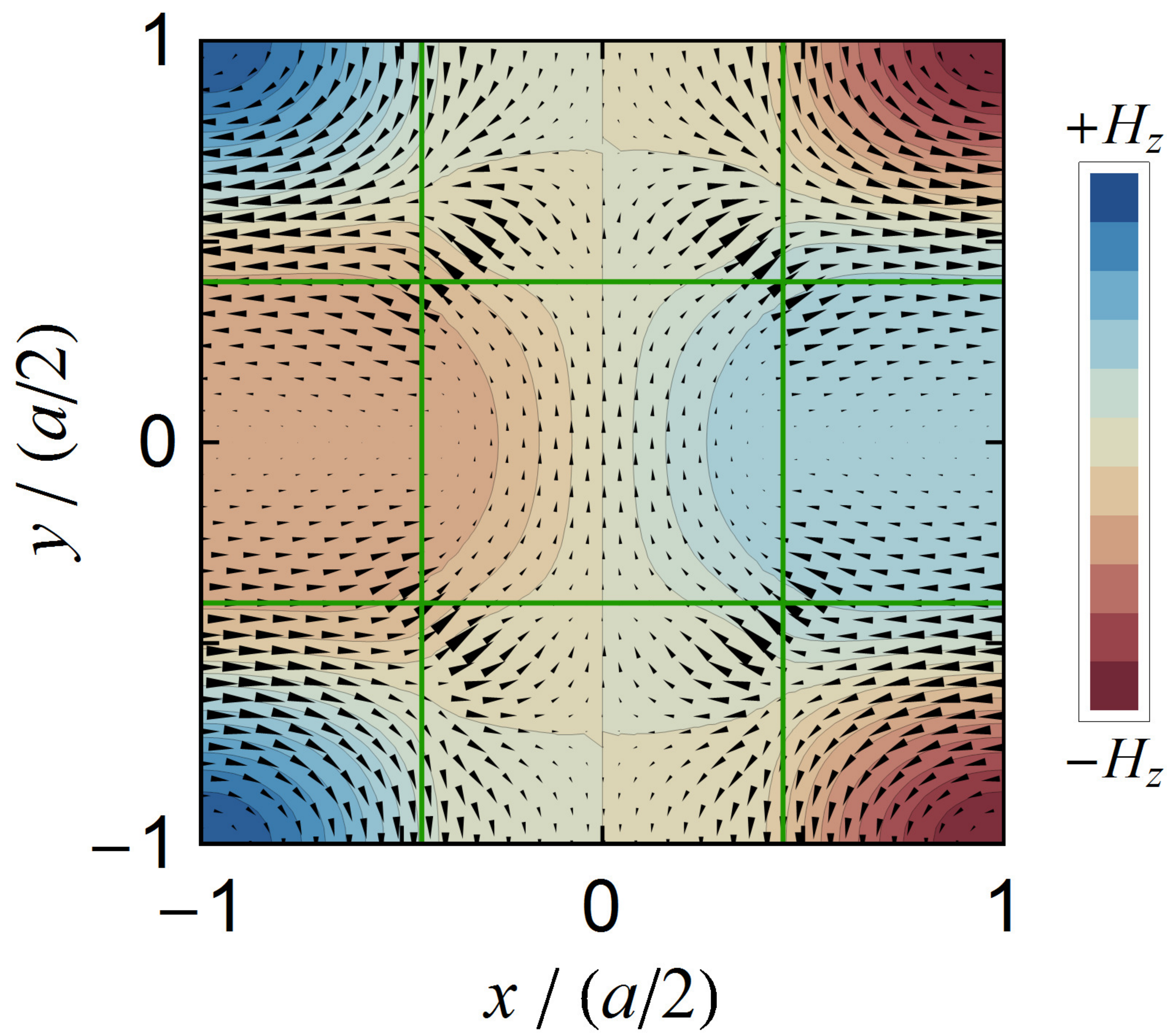}}
\\
\subfloat[][\label{avsfig:TMX02}TM mode $1$ at $X$]
{\includegraphics[width=1.3in]{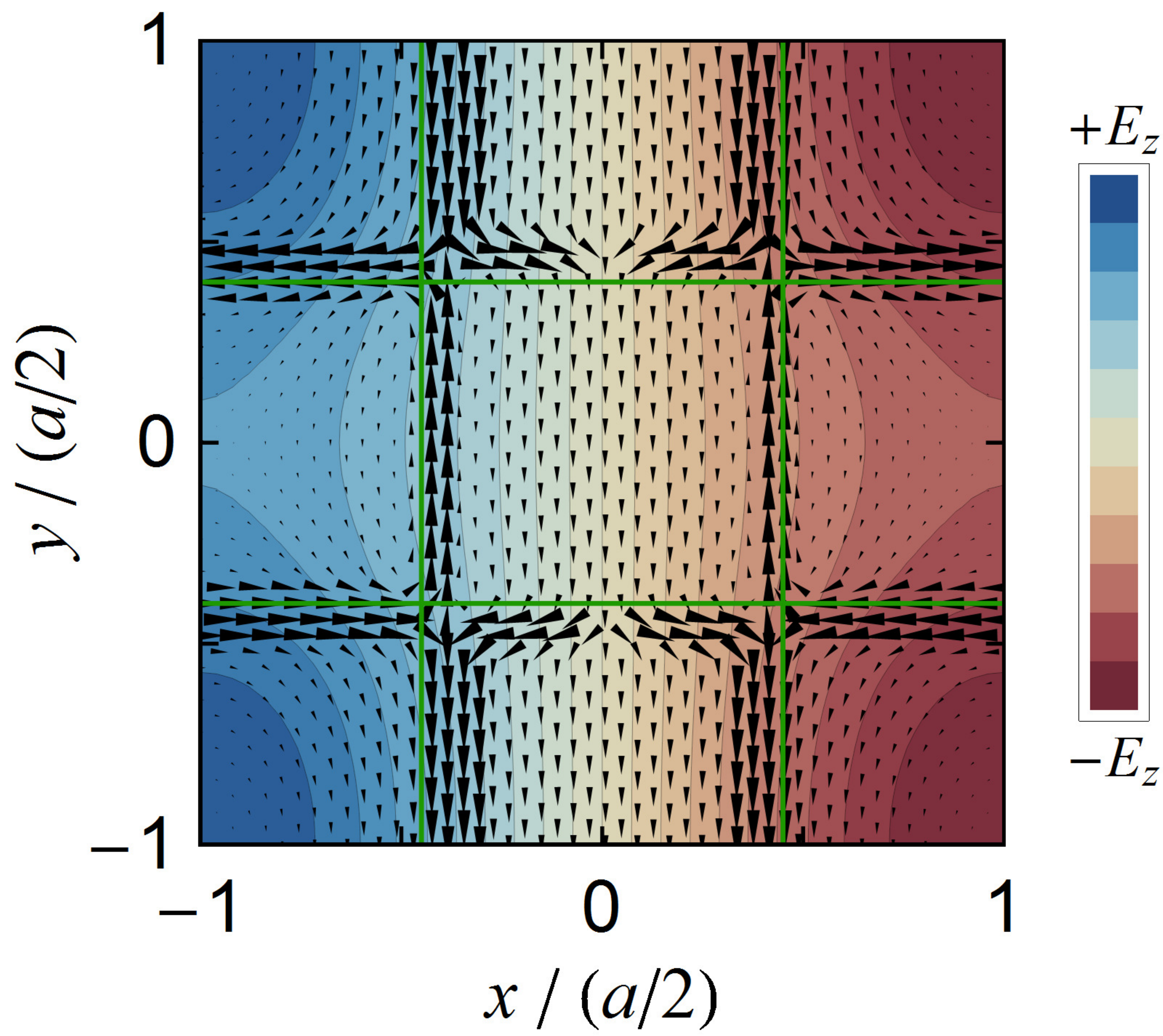}}
\subfloat[][\label{avsfig:TMX12}TM mode $2$ at $X$]
{\includegraphics[width=1.3in]{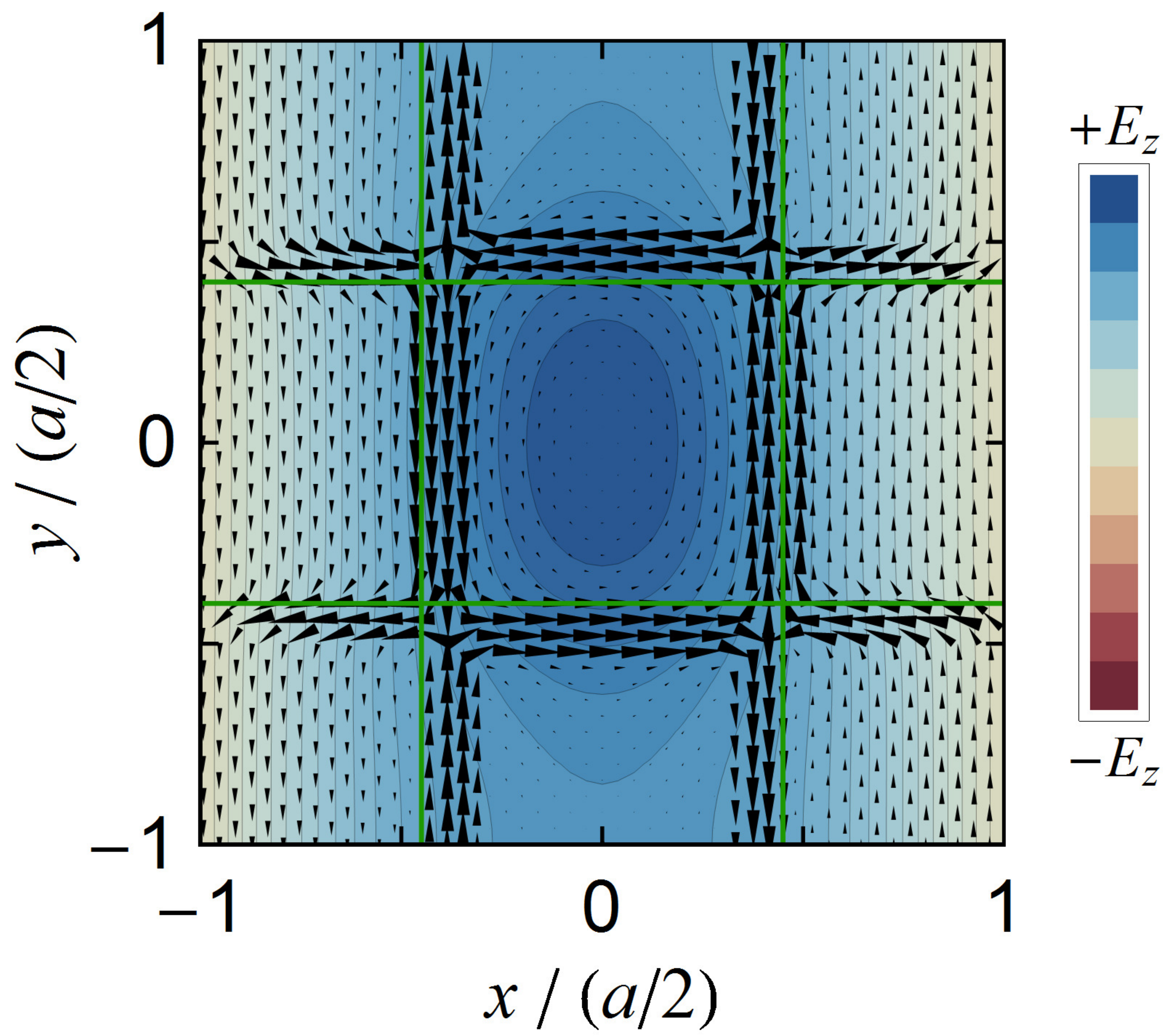}}
\subfloat[][\label{avsfig:TMX22}TM mode $3$ at $X$]
{\includegraphics[width=1.3in]{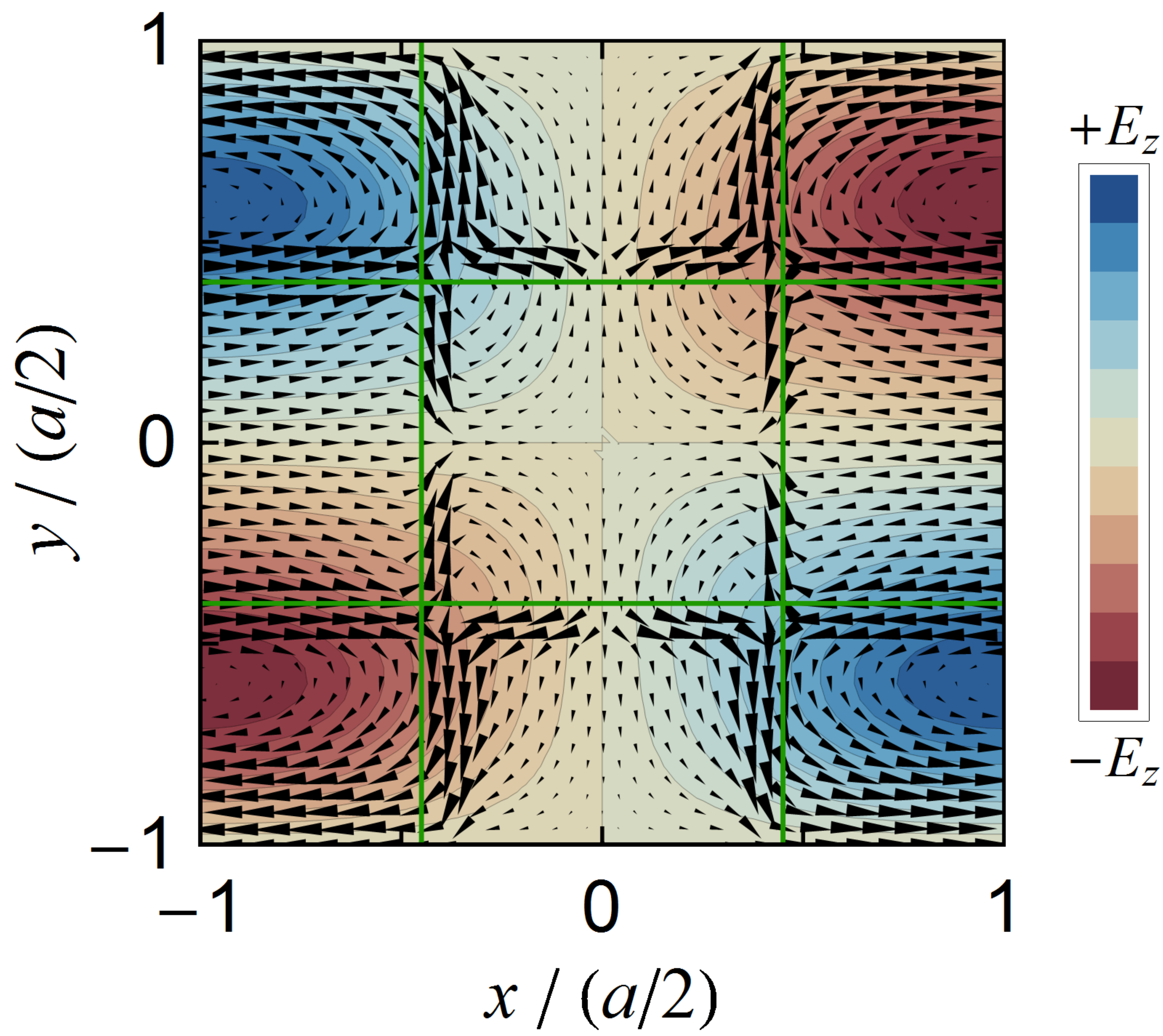}}
\caption[Checkerboard: Vector fields, $\Gamma$-point]
{The electric and magnetic fields of the second and third modes
corresponding to the $\Gamma$-point in the checkerboard
lattice are shown from (a) to (d), while the fields of the first three
modes corresponding to the $X$-point are shown from (e) to (j).
For TE-modes, the in-plane electric
field is represented by vectors and the out-of-plane
magnetic field is represented by the gradient background.
For TM-modes, the vectors represent the in-plane magnetic
field and the background represents the strength of the
out-of-plane electric field. Note that the first mode
is not shown because the corresponding eigenvalue
is zero, resulting in a trivial solution.}
\end{center}
\end{figure*}

\subsection{Eigenstates for a checkerboard lattice}

\begin{figure}[b!] 
\begin{center}
\subfloat[][\label{avsfig:TECheckDisp}TE mode]
{\includegraphics[width=1.5in,height=1.8in]{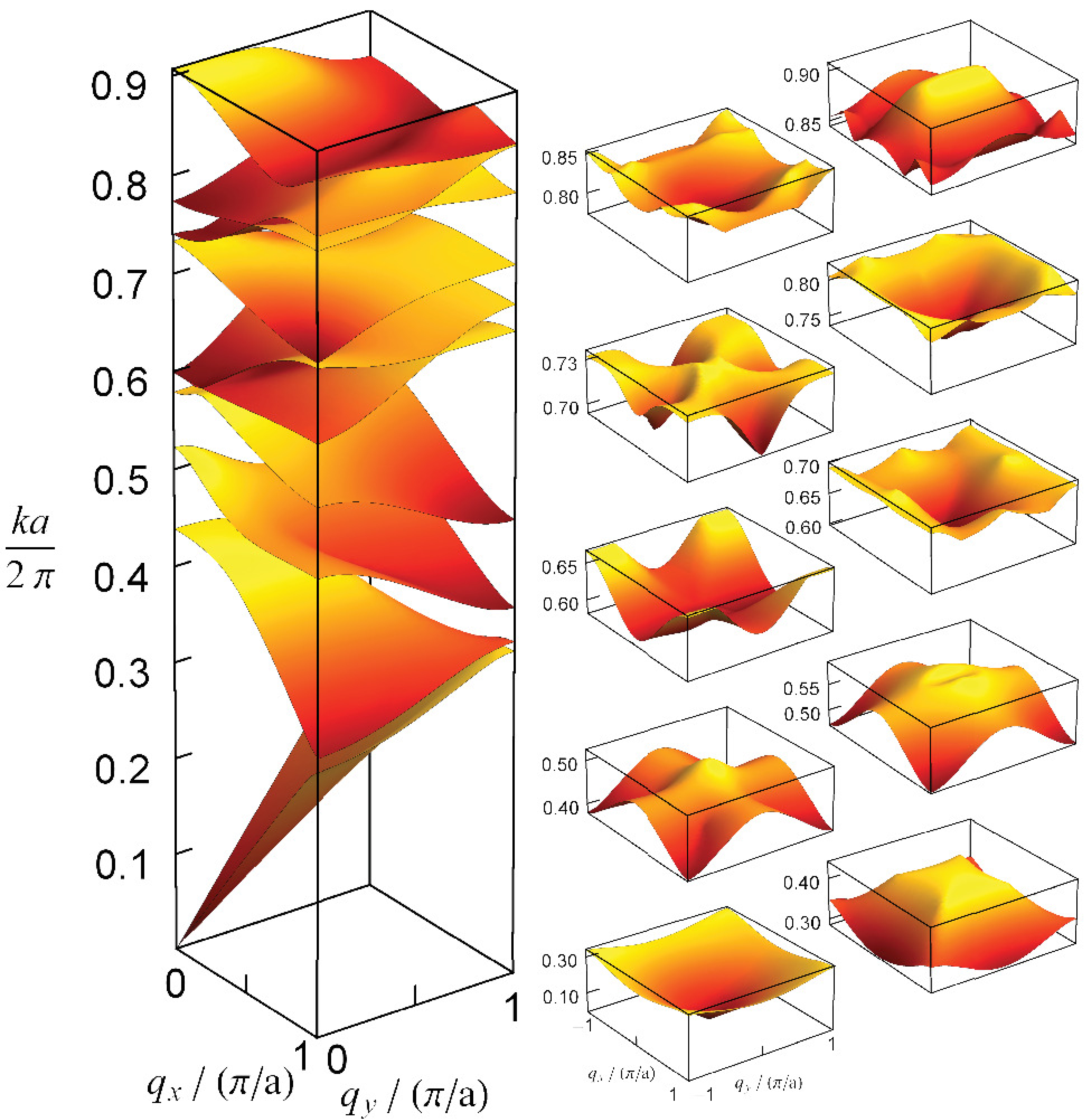}}\hspace{0.2in}
\subfloat[][\label{avsfig:TMCheckDisp}TM mode]
{\includegraphics[width=1.5in,height=1.8in]{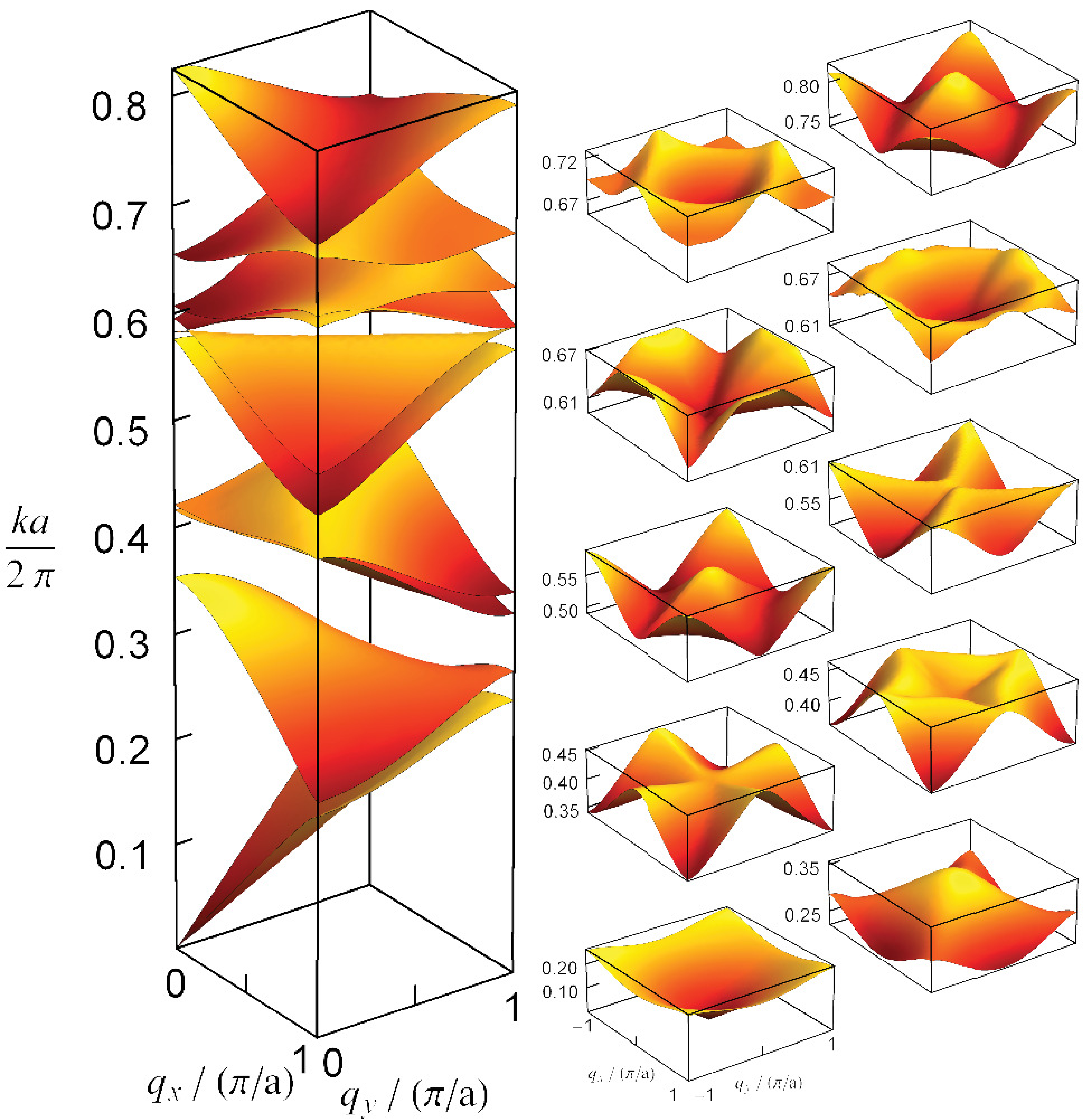}}
\caption{The eigenvalues of the transverse
(a) electric modes and (b) magnetic modes of the checkerboard lattice
of dielectric posts are plotted as surfaces in the first
Brillouin zone. On the left side of (a) an (b), the first ten eigenvalues
are shown in the irreducible part of the Brillouin zone for TE and TM modes,
respectively. On the right side, each eigenvalue has been separated from
the rest and extended to the full Brillouin zone through
symmetry operations.}
\end{center}
\end{figure}

We present results for a checkerboard superlattice of dielectric
posts which is readily solved using HFEM.   The   band  structure  for a checkerboard
lattice was computed using a mesh of $12355$ nodes, yielding a
matrix size of $74130\times 74130$.  The mesh was refined in the region
surrounding  the  edges within the checkerboard. Since the checkerboard
lattice has more internal boundaries per unit cell than the cylindrical
post geometry, a greater degree of mesh refinement was required, resulting
in a larger global matrix than that of the lattice of cylindrical posts.\\ 

The eigenvalues are plotted over a triangular path between the $\Gamma$,
$X$ and $S$ points. Compared to the dielectric posts, the checkerboard
shows much more activity and a denser band structure at low frequencies,
but it has a smaller band gap in the TM modes. Like the cylindrical posts,
this checkerboard has no TE band gap. The corresponding eigenfunctions
for the lowest modes at the high-symmetry points are plotted in
Figs.~\ref{avsfig:TEG12}-\ref{avsfig:TMG22}.\\

The vectors represent the electric  field in TE modes and the magnetic
field in TM modes, while  the shading of the background represents the
intensity of the magnetic field in  TE modes and the electric field in
TM  modes, with  lighter shades  corresponding to  regions  of greater
field  magnitude.    Note  that  the  eigenfunction   for  the  lowest
eigenvalue  is  omitted for  the  $\Gamma$-point  for both  modes  of
propagation. This is because those lowest eigenvalues approach zero at
the  $\Gamma$-point,  causing the  corresponding eigenfunctions  to be
trivial (zero everywhere).\\

The dispersion relations calculated for the irreducible Brillouin zone
and then their full
reconstruction over the entire zone is performed. The lowest few
TE and TM dispersions 
are shown in Figs.~\ref{avsfig:TECheckDisp}-\ref{avsfig:TMCheckDisp}.

\begin{figure}[ht] 
\vspace*{-0.2in}
\begin{center}
\subfloat[][\label{avsfig:Escher1}]{\includegraphics[width=1.8in]{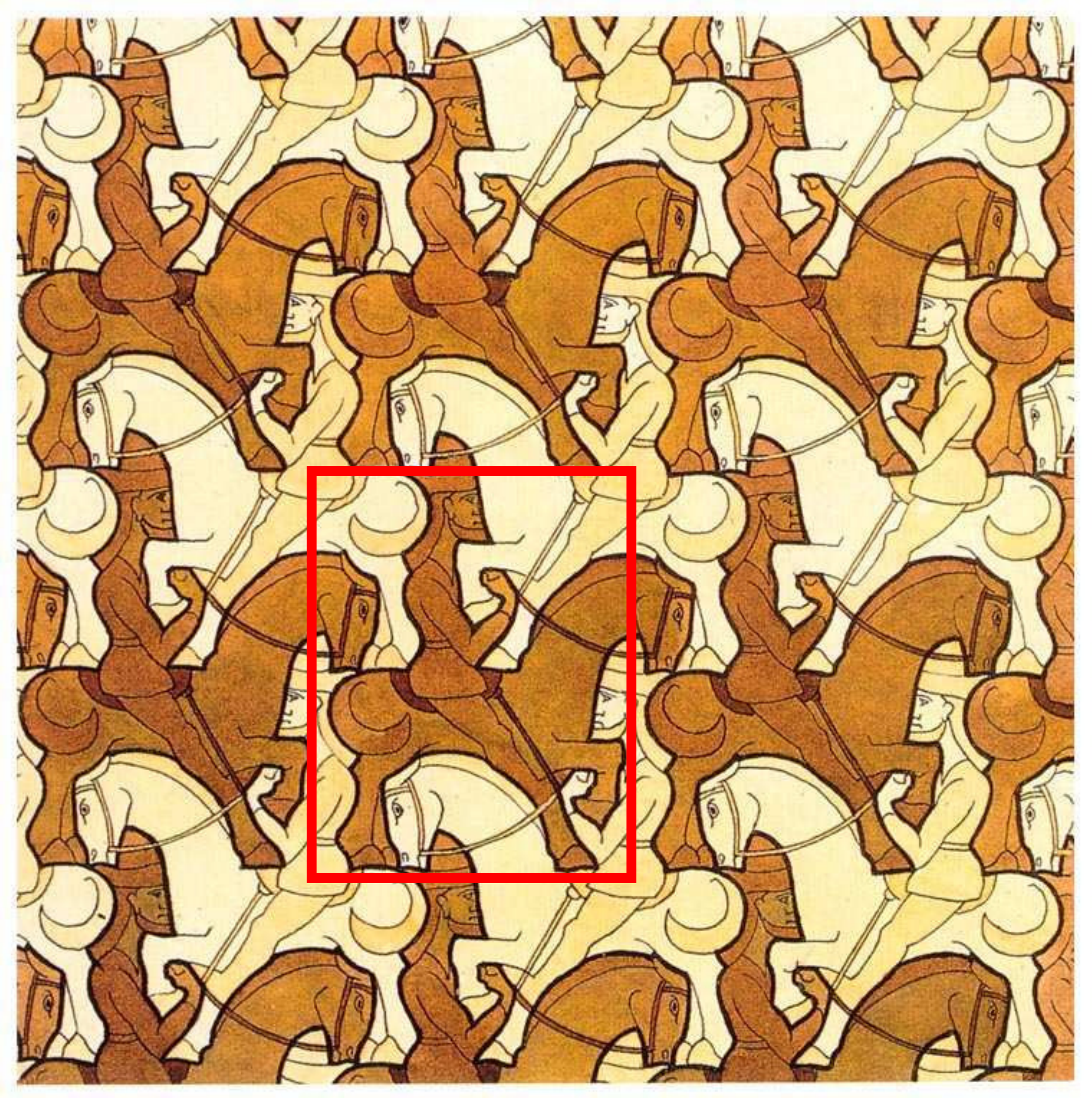}}\hspace{0.2in}
\subfloat[][\label{avsfig:Eschermesh}]{\includegraphics[width=1.6in]{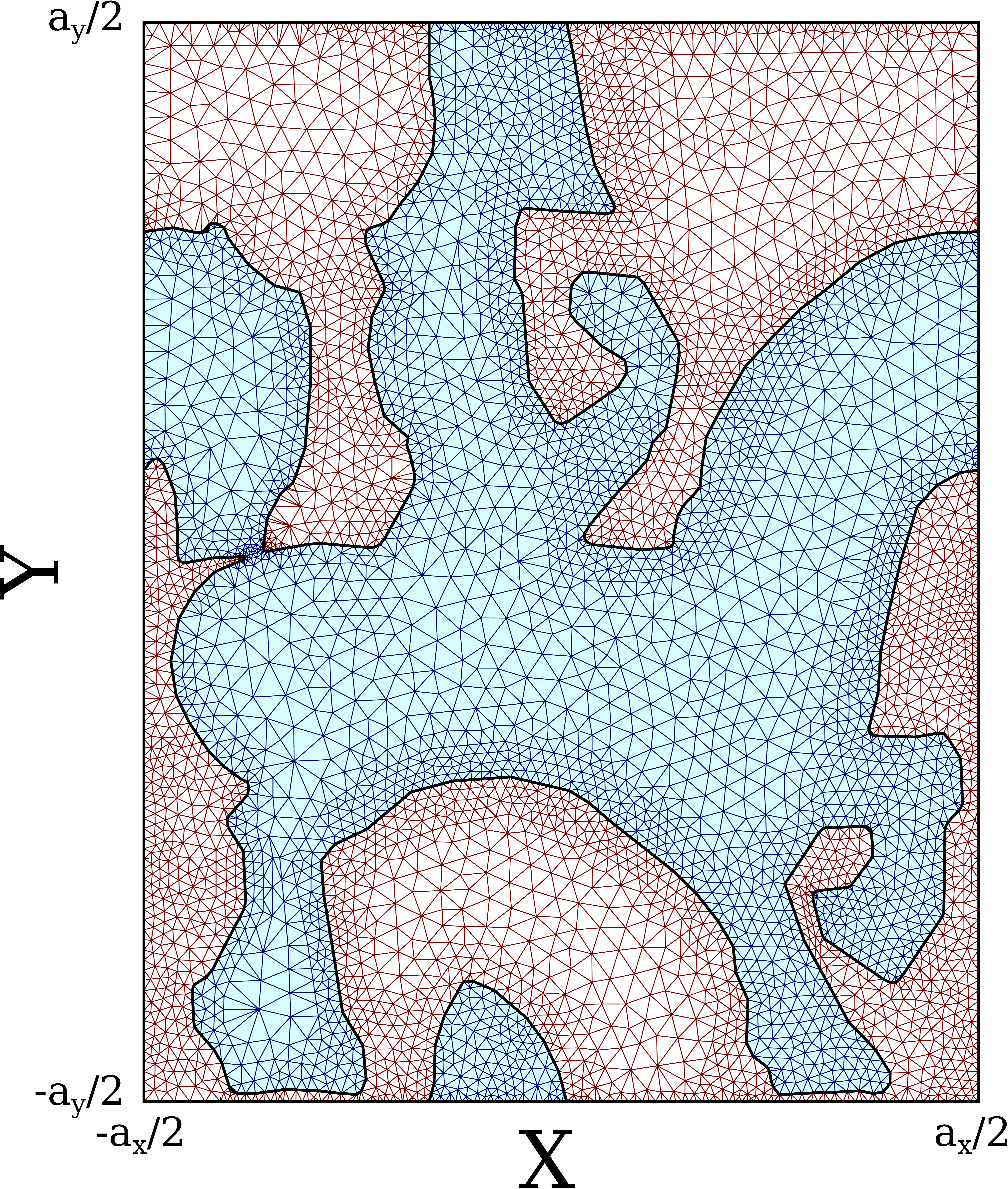}}\\
\subfloat[][\label{avsfig:escher2d}]{\includegraphics[width=2in]{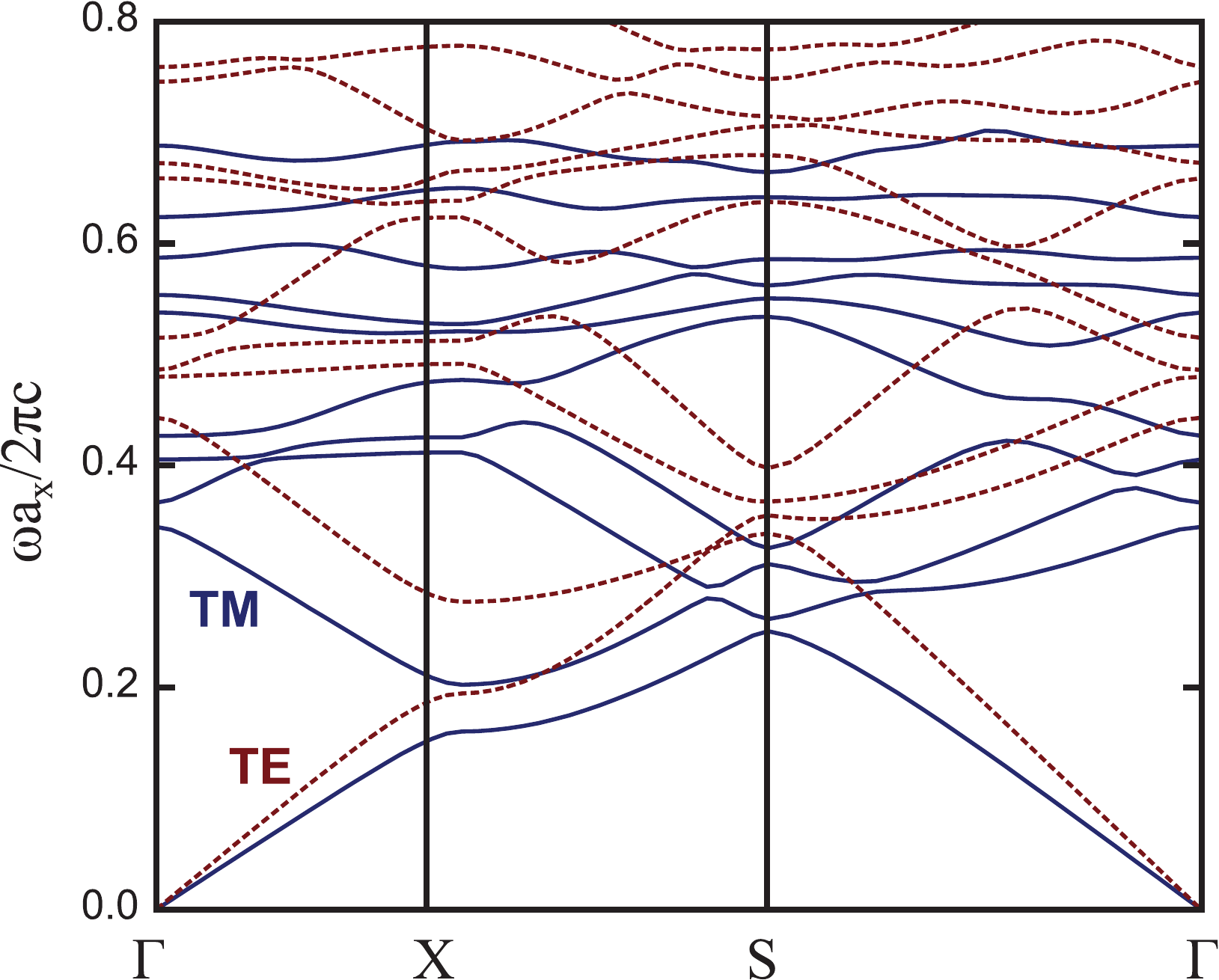}}
\end{center}
\vspace{-0.2in}
\caption[PC based on the work of M. C. Escher] {(a) The unit cell of a
  PC based  on ``The Horsemen'' by  M. C. Escher is  shown. The Escher
  tessellation was  chosen to illustrate  the capacity of the  HFEM to
  calculate the band structures  of complicated geometries with unique
  symmetry properties.\cite{avs:EscherFigure}  (b) A sample  mesh is given
  for  the  unit cell  of  an  Escher  tessellation. The  regions  are
  assigned $\epsilon_r$ $=$$8.9,1.0$ to form a 2D PC. Most calculations
  used more refined meshes than shown in (b), including $54,945$ nodes
  for a total of $329,670$ global degrees of freedom.  (c) TE (dashed)
  and TM (solid) eigenmodes for the associated 2D PC are shown. (Adapted from Boucher {\it et al.}, Ref.~\onlinecite{avs:CRB_PhCTs}.)}
\end{figure}

\begin{figure}[ht] 
\begin{center}
\subfloat[][\label{avsfig:Escher_TE_2} TE mode 2]
{\includegraphics[width=1.3in]{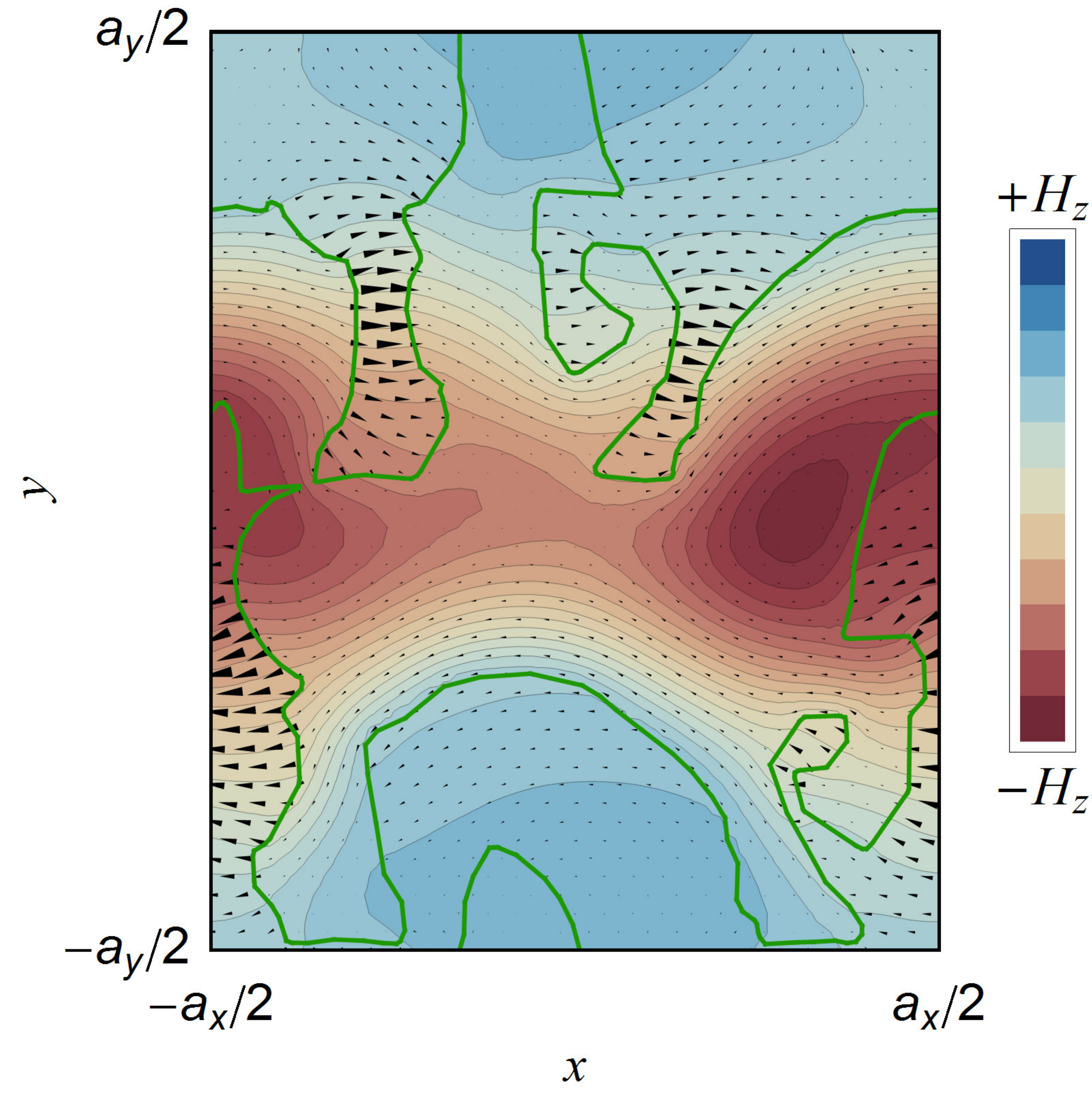}}
\subfloat[][\label{avsfig:Escher_TE_3} TE mode 3]
{\includegraphics[width=1.3in]{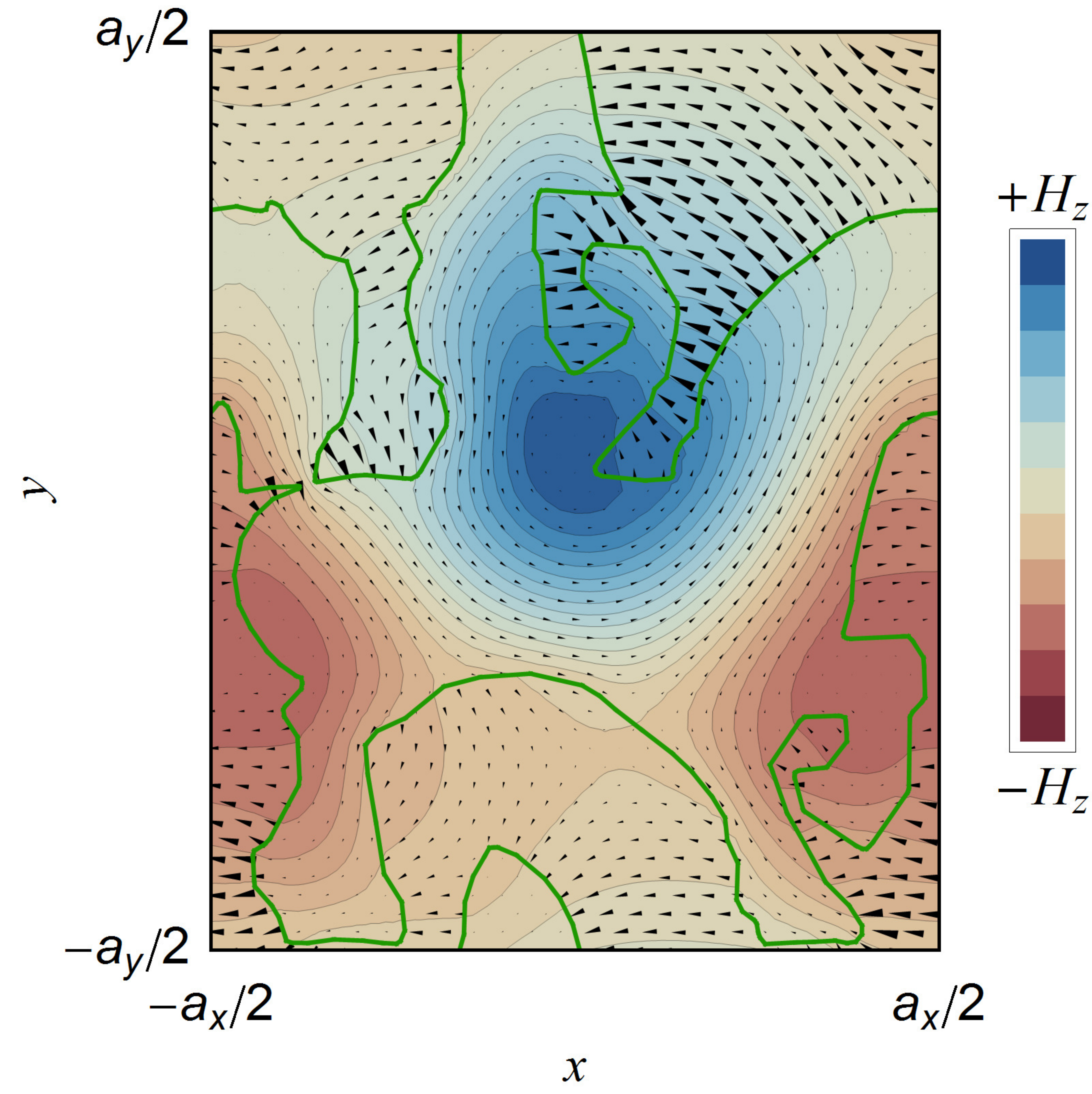}}\\
\subfloat[][\label{avsfig:Escher_TM_2} TM mode 2]
{\includegraphics[width=1.3in]{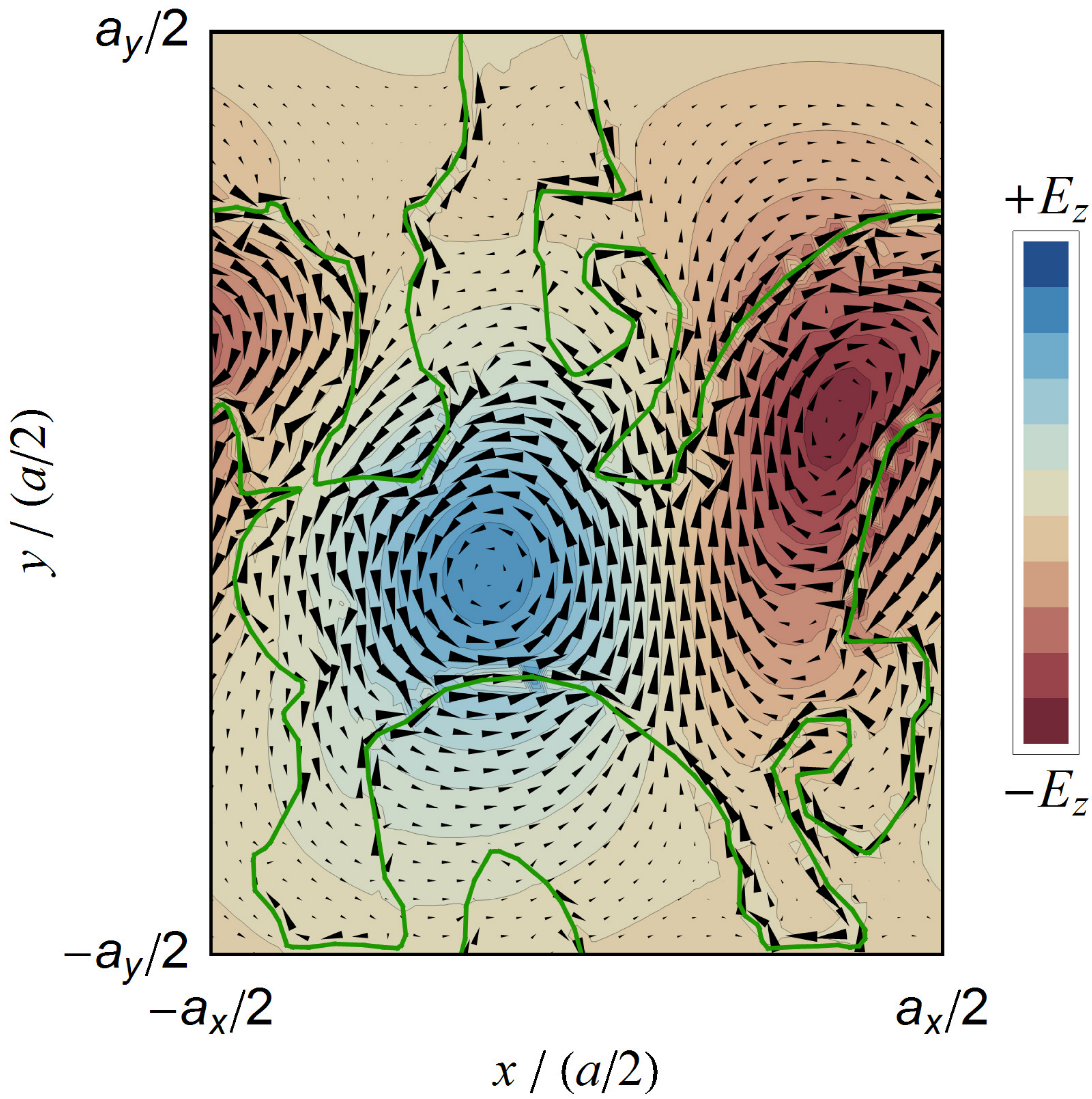}}
\subfloat[][\label{avsfig:Escher_TM_3} TM mode 3]
{\includegraphics[width=1.3in]{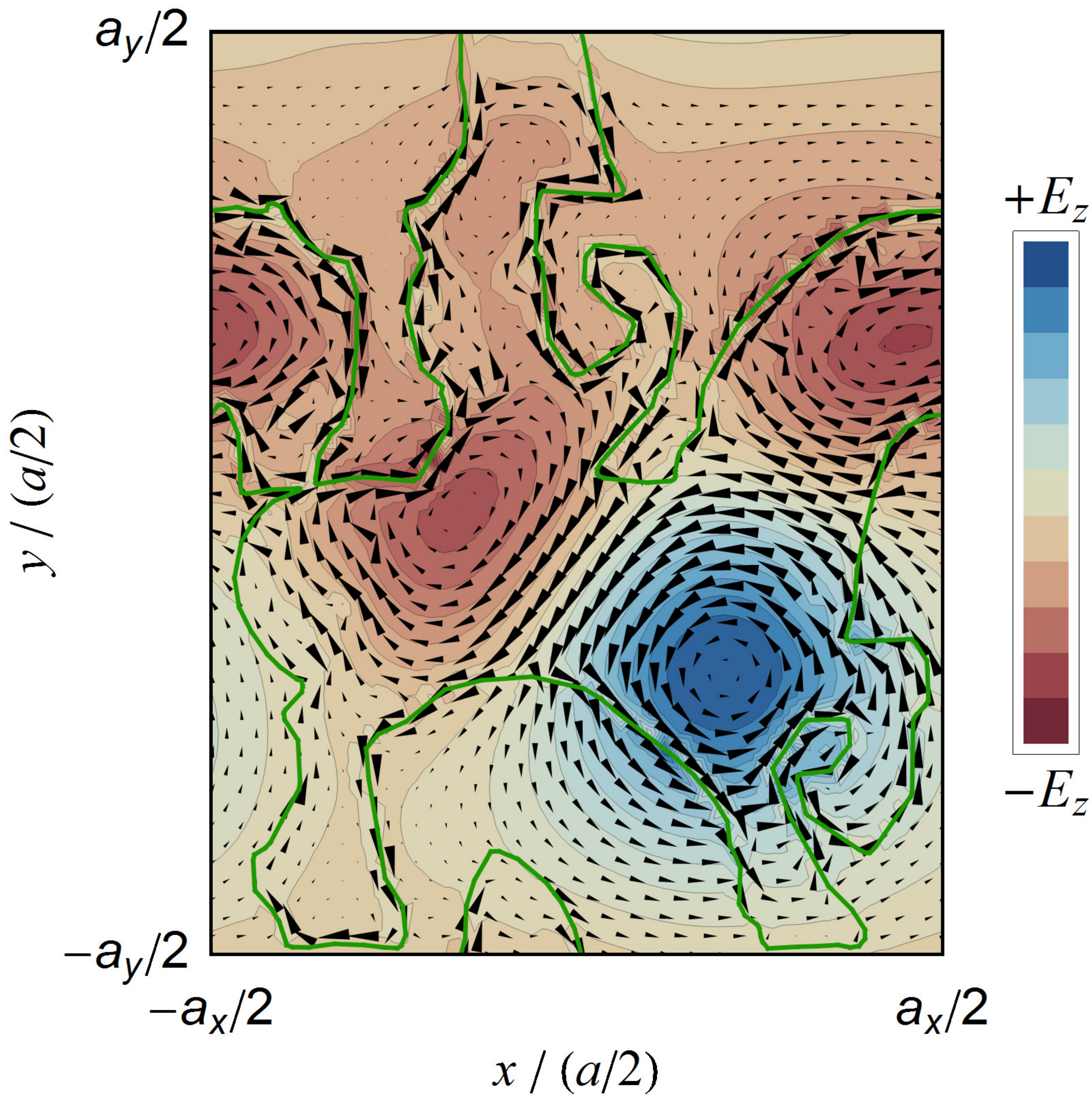}}
\caption[Escher:  Vector  fields,  $\Gamma$-point] {The  ${\bf E}$  and
  ${\bf H}$ fields  of the second and  third modes at  $\Gamma$ for the
  Escher lattice  are shown. For  (a,b), $E_{x,y}$ are  represented by
  vectors  and  $H_z$  by  the  contours.  For  (c,d),  $H_{x,y}$  are
  represented by vectors and $E_z$ by the contours. (Adapted from Boucher {\it et al.}, Ref.~\onlinecite{avs:CRB_PhCTs}.)}
\end{center}
\end{figure}
\subsection{Eigenstates for an Escher tessellation}
\begin{figure}[b!] 
\begin{center}
\subfloat[][\label{avsfig:TEEscherDisp}TE mode]
{\includegraphics[width=1.8in]{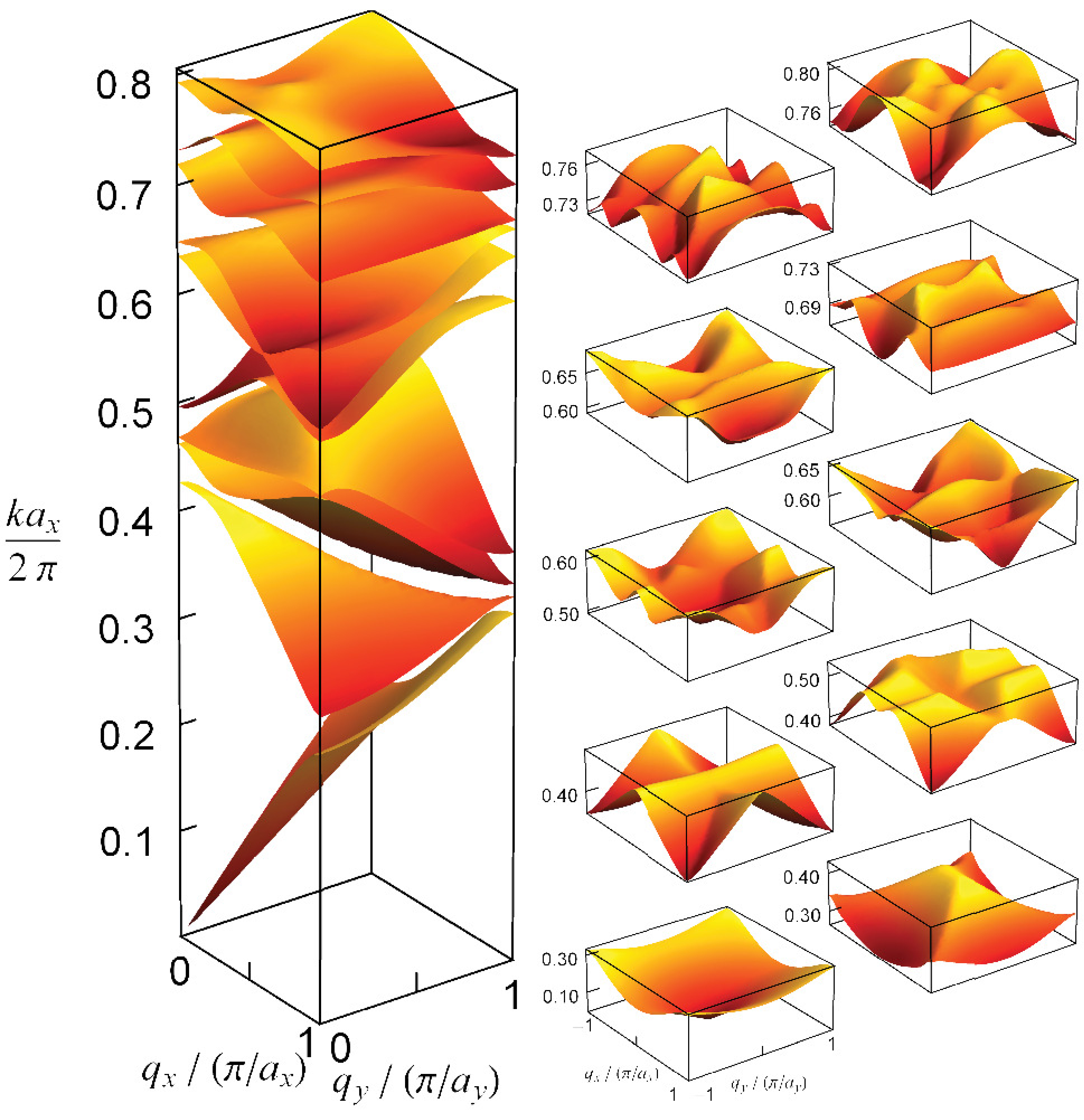}}\hspace{0.2in}
\subfloat[][\label{avsfig:TMEscherDisp}TM mode]
{\includegraphics[width=1.8in]{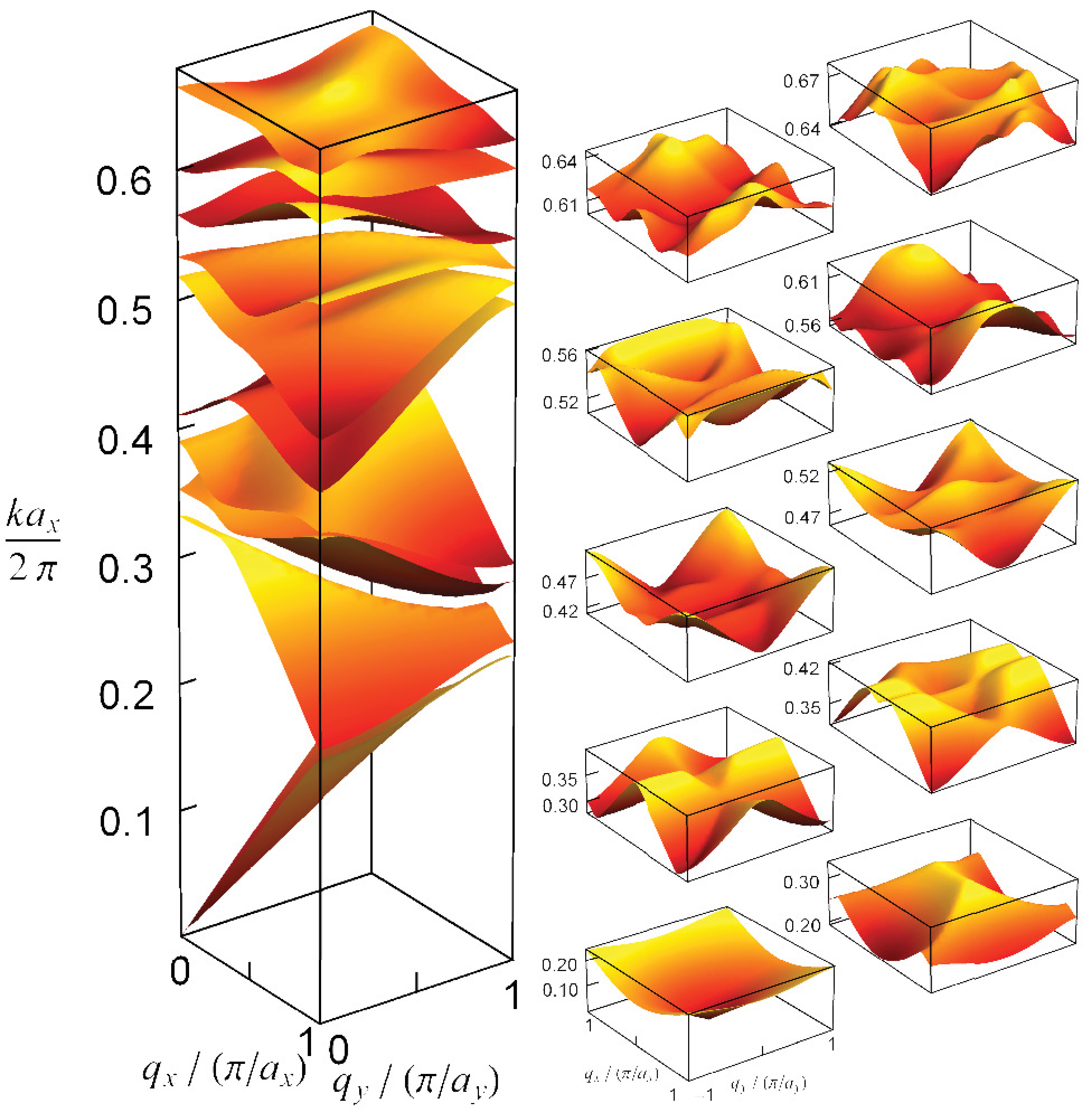}}
\caption{The eigenvalues of the transverse
(a) electric modes and (b) magnetic modes of the Escher superlattice
of dielectric posts are plotted as surfaces in the first
Brillouin zone. On the left side of (a) an (b), the first ten eigenvalues
are shown in the irreducible part of the Brillouin zone for TE and TM modes,
respectively. On the right side, each eigenvalue has been separated from
the rest and extended to the full Brillouin zone through
symmetry operations.}
\end{center}
\end{figure}

  In order to demonstrate the flexibility of the
Finite Element Method, a photonic crystal based on a tessellation by
M. C. Escher was simulated and its band structure was calculated.
The image used to produce the crystal was Escher's ``Horsemen''
as shown in Fig.~(\ref{avsfig:Escher1}). A sample mesh is given
in Fig.~(\ref{avsfig:Eschermesh}).

The   band  structure  for the Escher tessellation
was computed using a mesh of $54945$ nodes, yielding a
matrix size of $329670\times 329670$. Since the Escher unit cell
does not  have the same reflection symmetries as cylindrical
and checkerboard unit cells, the entire Brillouin zone was tested
instead of a small fraction of it.

The eigenvalues are plotted over a triangular path between the $\Gamma$,
$X$ and $S$ points in Fig.~(\ref{avsfig:escher2d}). The transverse
electric modes appear to converge to a band structure similar to that
of the cylindrical post, even featuring an anticrossing site
in approximately the same position. However, the transverse magnetic
modes fail to converge properly, even when using several tens of thousands
of nodes. This may be due to the high complexity of the dielectric
structure coupled with the slow error convergence of the action
formulation based on ${{\bf D}}$. The corresponding eigenfunctions
for the lowest modes at the high-symmetry points are plotted in
Figs.~\ref{avsfig:Escher_TE_2}-\ref{avsfig:Escher_TM_3}.

The vectors represent the electric  field in TE modes and the magnetic
field in TM modes, while  the shading of the background represents the
intensity of the magnetic field in  TE modes and the electric field in
TM  modes. As with the other crystal geometries, we omit the lowest
state at the $\Gamma$-point as a trivial solution.

The eigenvalues have also been plotted as surfaces over the 
first Brillouin zone, as shown in
Figs.~\ref{avsfig:TEEscherDisp} and \ref{avsfig:TMEscherDisp}. 

In conclusion, we anticipate  that the use of Hermite FEM will allow
the treatment of multiscale problems associated with photonic crystals
with embedded quantum dots, defects, and the like. The spatial
representation of the fields using Hermite triangular interpolation is
much more economical than employing plane-wave methods for such
structures allowing the deployment of more finite elements
strategically in specific regions as needed. The resulting global
matrices are still sparse and banded due to the local connectivity,
which leads to far more compact matrices than in other schemes with
the concomitant reduction in compute-time. While the transverse
magnetic modes continue to converge slowly in complex geometries,
the efficient calculation of the TE modes allows one to easily
determine which geometries have TE band gaps. Furthermore,
the extension to three-dimensional crystals, in which a separate
formulation based on ${{\bf D}}$ is no longer needed, may alleviate
this problem.


\section{Concluding remarks}

Electromagnetic  devices of  higher  frequencies  (e.g., mm-wave)  and
	increasing  complexity  are  being  employed  in  a  wide  variety  of
	industries. The design of  modern electronic includes electromagnetic
	components with  sophisticated interactions  both in isolation  and in
	integrated  combinations.   Designing  these  components   requires  a
	significant  amount  of  modeling  and simulation  and  these  demands
	continue  to  increase  with  higher levels  of  integration.  At  the
	nanoscale, similar circumstances are  faced for optical interconnects,
	quantum  well  laser  design,   and  in  plasmonics.  Again,  reliable
	simulations are essential  to ensure that each device  does not affect
	others near  it through electromagnetic cross-talk.  The novel effects
	exhibited  by  metamaterials   containing  negative  refractive  index
	components  are  all simulated  before  being  assembled in  order  to
	optimize their optical properties as desired.
	
	Commercial computational electromagnetic  modeling software relies  heavily on
	vector finite element, finite difference, and spectral methodologies. Here we
	focus  on  a  scalar  finite  element  approach  in  which  the  field
	components  are approximated  using  local  polynomials over  discrete
	subdomains. We  show that the  use of Hermite
	interpolation polynomials provide very accurate
	solutions   with  a   minimal   number  of   elements   used  in   the
	discretization. The ability to  reproduce smooth variational solutions
	for the  fields will  allow a  coarsely discretized  full-wave maxwell
	solver  to seamlessly  couple to  other solvers  for physically  small
	features, such  as small gate  geometries, quantum wells and  dots, or
	plasmonic structures which are all deeply subwavelength. 
	The Hermite interpolation polynomials are equally well suited to
	three-dimensional finite elements, e.g., 40 DOF or 56  DOF can be used to define
	the basis functions on tetrahedra. 
	
 We  have  shown  that  the  hermite  interpolation  polynomials  on  a
	triangular element are  able to eliminate the  spurious solutions that
	typically occur with Lagrange-type scalar shape functions. The results
	for the standard  rectangular waveguide with and  without a dielectric
	inhomogeneity directly  demonstrate the  efficacy of this  method. The
	eigenfrequency  for  the lowest  mode  for  the homogeneous  waveguide
	agrees  with  the  analytical  result   within  a  relative  error  of
	$10^{-15}$, which is superior to  hierarchical VFEM for about the same
	number of  DOF.\cite{avs:JFLeeHierarchical2003} 
	
	For the inhomogeneous waveguide, we have calculated the
	eigenmodes  that   are  trigonometric   in  both  regions   at  higher
	frequencies, and they  evolve into  solutions that  are sinusoidal  in the
	higher dielectric region  and hyperbolic in the region  with the lower
	dielectric as the  dielectric ratio $\epsilon_2/\epsilon_1$ increases.
	This behavior is analogous to  the development of above-barrier states
	in quantum wells that get localized and captured into the quantum well
	as  the  well  depth  is  increased. This  analogy  suggests  that  as
	$\epsilon_2$  is  increased more  modes  are  captured by  the  higher
	dielectric region  leading to  the sinusoidal  behavior in  the larger
	dielectric region and  an exponential decay into  the lower dielectric
	region. 


For cubic cavities, we  have  shown  that electromagnetic  simulations done  with  Hermite
elements  deliver   high  accuracy  and  smoother   representation  of
fields. We  have compared  our formalism  with analytical  results.  
Fewer finite elements  are needed to achieve comparable results
for eigenvalue calculations.

\begin{table*}[th!]
	\caption{We  present here  a summary, and contrast the properties between VFEM and HFEM calculations.}
	\label{tab:compare}
	\centering
	\vspace{-0.1in}
	{\hspace{-0.5cm}{\small\begin{tabular}{P{5.9cm}c||cP{5.9cm} }
				\hline
				\hspace{2cm}{\bf VFEM} &&& \hspace{2cm}{\bf HFEM} \\
				\hline\hline
				$\bullet$ In 2D, there are no spurious solutions with
				edge-elements. &&& $\bullet$ In 2D, there are no spurious solutions
				using Hermite finite elements with
				triangles.\cite{avs:pkass,avs:HFEM_JAP2016}          \\ 
				$\bullet$ Field directions are ill-defined at all the nodes. With
				increasing mesh density we have
				more area (2D) and volume (3D)
				around nodes where the field direction is not defined. Thus mesh refinement
				does not give improved results in applications.&&&
				$\bullet$ This  is
				a node-based FEM, and there are no  issues with field directions at nodes and
				throughout, including for 3D hexahedral Hermite elements. Mesh
				refinement allows the improved identification of spurious solutions
				in the positive spectrum.  
				\\
				$\bullet$ In  3D, all zero-frequency solutions are pushed  to the null-space
				through Nedelec compliant shape  functions. 
				
				Estimates are that for a matrix dimensions of 
				$10^3$ in typical EM calculations, there
				are  about   20\%--30\%  solutions   to  the  matrix   that  are   in  this
				category.\cite{avs:Peterson1994} They have to be calculated and then thrown away, being unusable solutions. Carrying this overhead 
				in the calculation is computationally expensive when considering sophisticated structures.&&&
				$\bullet$
				In 3D, a modest penalty factor $\lambda=1$ pushes  spurious solutions to
				the zero-frequency  sector and they do
                                 not appear in the calculated range of the spectrum.
				Some are left over, and  these affect the non-zero frequency spectrum.
				
				Now the  node-based  divergence   condition   is  
				super-imposed on the penalty calculation;   this leads  to  substantial  improvement in  the
				separation and identification  of the spurious solutions.       
				
				The key is the tag  provided by \mbox{$|\nabla\cdot \mathbf{E}|/|\nabla\times\mathbf{E}|$}. 
				This ratio keeps increasing for
				the  spurious solutions,  whereas it  decreases
				substantially for physically admissible solutions   
				as the mesh is refined. \\ 
				$\bullet$  The multi-scale  modeling for multi-physics systems
				cannot be performed: for example, modeling a vertical cavity surface
				emitting laser. &&& $\bullet$  Being node based, the modeling can
				accommodate multi-scale problems for multi-physics applications. \\
				\hline\hline
	\end{tabular}}}
\end{table*}

The divergence-free constraint  for the electromagnetic fields results
in spurious  solutions for the wave  equation.  Their eigenfrequencies
are pushed to  zero in the VFEM, either through  Nedelec compliance or
through their  removal at each iteration.  In either case, this  is an
expensive  numerical  procedure.   In  our approach,  we  imposed  the
divergence-free condition  by adding a  constant penalty term  (a 
Lagrange  multiplier,   set  to  unity).  In   addition,  through  the
derivative degrees  of freedom at each  node we have imposed  the same
constraint explicitly.  Now the remaining few non-zero frequency spurious solutions
are   eliminated    by   identifying   them   through    their   large
$|\nabla\cdot  {\bf E}|/|\nabla\times{\bf  E}|$ ratio.  This procedure
does not alter or influence the accuracy of the physical solutions.
Comparison of properties between VFEM and HFEM calculations are
presented in Table~\ref{tab:compare}. 

Group   theoretical   classification   of   eigenmodes   in   photonic
crystals,\cite{avs:sakodaPRB1995,avs:sakoda1997}  in               radio-frequency
cavities,\cite{avs:Sakanaka,avs:McIsaac}  and  in  metamaterials\cite{avs:Padilla}
were previously  discussed in the  literature. We have
considered the symmetries of a metallic cubic cavity, with and without
a dielectric  inclusion.  The  origin of  higher
degeneracy in the frequency spectrum in a cubic cavity are attributed to the existence of
accidental degeneracy. The operators  additional to those of the symmetry
group  $O_h$ have  been determined.  We have  derived a  coefficient
formula\cite{avs:RemAccDegen17}  which will  classify, and  project out  the symmetry  adapted
modes of  the corresponding irreducible representation.   The computed
field  distributions  are  symmetry-adapted as
predicted from group theory.

The accidental  degeneracy is lifted  with the insertion  of a
concentric cubic  dielectric of a smaller size.  The variation
of  the  spectrum  as  the  ratio  $\epsilon_2/\epsilon_1$  is
changed has been explored.  We have shown that this leads to a
reordering of some of the mode frequencies.

Since the FEM  is based  on geometry discretization, we are
now free to  change the shape of the cavity  and still obtain a
high accuracy using  HFEM.  This method is well  suited for 
mixed-physics applications, such as for quantum well lasers
in  electromagnetic cavities.   This is  because we  have \mbox{node-based} 
finite elements with scalar shape functions.

Applications to multiscale analysis is now feasible using the
present method. This option   is not open to VFEM due to the 
lack of directionality for fields at shared  nodes in the
finite element mesh. Very dense meshes lead to larger regions in which
field directions are ill-defined. Typically, quantum mechanical problems 
are solved with scalar basis functions, and the electromagnetic problems 
are dealt separately with VFEM or other techniques. HFEM formulation facilitates
an identical scheme for solving simultaneously both electromagnetic and quantum mechanical/acoustic
calculations.

We have shown  that the scalar  Hermite polynomials  have several
fundamental  advantages  for  obtaining the  band  structure  of  periodic
systems such as photonic crystals,  while compared with VFEM and other
plane wave expansion  methods.\cite{avs:CRB_PhCTs} Advantages are observed
in computational costs,  the ability to capture  spatial complexity in
the  dielectric   distributions,  a  substantially   higher  numerical
convergence with scaling, and  in obtaining variational eigenfunctions
free of  numerical artifacts.  We note that  the method  delineated in
this paper  is well  suited to model  and design  composite structures
such as  3D photonic band-gap crystals,  metamaterials and topological
photonic systems,  for applications in ultra-small  optical integrated
circuits. Hence, approaches reviewed here show great promise for the
simulation  of electrodyamics,  plasmonics, high  frequency circuitry,
and especially in mixed physics problems.

Finally, we note that the importance of accidental degeneracy and its
consideration is because the periodic table of elements and its
structure depend on it. The 
progression of elements in the table with the addition of more and
more electrons to the atoms is governed by the Pauli exclusion
principle and his ``aufbau prinzip.'' This then governs all of
chemistry and hence all of biology. May we be permitted to say that life itself depends
on accidental degeneracy?

Given the  periodic nature  of the  PCs, it is  natural that  the vast
majority of  published works  on photonic band  structure calculations
are analogous  to the  reciprocal space analysis  common to  the solid
state physics  analysis of propagating electronic  states in crystals.
The analogy  to solid-state has some drawbacks. In particular  the abrupt,
macroscopic discontinuities of dielectric regions make it difficult to
transform  the dielectric  function and  EM states  between reciprocal
space  and real  space  without artifacts.   In  practical use,  these
numerical artifacs can become sources of serious error when subsequent
calculations  in real  space are  required or  the system  symmetry is
lowered.  Interactions with  sub-wavelength features,  such  as quatum
structures that are most often at interfaces, can be very difficult to
resolve  if  the  EM  field  description is  coarse.  Other  demanding
examples are the computation of localized states (such as defects) and
slab geometries.\cite{avs:Jiang2012} Under  conditions of a periodic slab,
the predominant approach is to move away from the use of ~$10^6$ plane
waves (the 2D periodic system of a slab), plane waves corresponding to
the   third   dimension,   plus   supercells  corresponding   to   any
irregularity\cite{avs:FanJoannopoulos,avs:Johnson1999} toward  the time domain
where finite-difference  time-domain calculations are  now widely used
for the calculation of real-space EM fields.

VFEM  calculations  for the  eigenmodes  of  PCs and  VFEM-time-domain
calculations for  waveguide and defect geometries are  not common, but
examples  include  the real-space  construction  of localized  Wannier
basis  function  from  the   perfect  crystal  eigenmodes  to  compute
localized  defect modes.\cite{avs:SotirelisAlbrecht}  One  disadvantage of
such calculations is the pixelization of the resulting fields owing to
the lower order of normal field continuity in vector element formalism
or   the   spatial   gridding   of   finite-difference   time   domain
analysis. This  mixed order  real space description  results in  an EM
field that  is spatially coarser than the  quantum mechanical features
of embedded  solid state structures  such as quantum dots,  wells, and
other features common to modern semiconductor devices.

It is  ultimately more  desirable to obtain  field patterns  that have
continuous spatial derivatives within dielectric layers for convenient
calculation   of   quantum   mechanical   or   deeply   sub-wavelength
interactions. The  present HFEM approach provides  smoothly varying EM
wavefunctions   using  a   nodal  mesh   description   and  derivative
continuity,  yet  preserves  the  necessary  boundary  conditions  and
numerical constraints that have been demanded of VFEM. Accurate spatial
field description provided by HFEM will be important in inverse design 
schemes for PCs to engineer the topology of band structures.\cite{avs:WLi,avs:ZLi} 

For decades,  PC analysis by reciprocal space  techniques has produced
accurate  eignevalue results  as would  be expected  of  a variational
approach,  however  the  HFEM  offers  a  flexible,  robust  means  of
computing   the  eigenstates  of   a  PC   with  much   more  physical
eigenfunctions at  far lower computational cost.  In our calculations,
we  note that the  typical matrix  dimensions for  the PC  with square
geometry containing  cylindrical posts are  on the order  of $26\times
10^3$; however, the  local connectivity within HFEM leads  to a banded
matrix  with  0.158\%  occupancy.   This sparsity  is  a  demonstrable
advantage over  the plane-wave method.   The ability to  construct the
field   distributions   from    the   nodal   eigenvectors   with   no
discontinuities in  the reconstructed function and  its derivatives is
an additional benefit and allows for a high quality description of the
dual fields.  Furthermore, the  plane-waves are global  functions, and
the eigenfunctions  constructed using  these functions have  the usual
errors  on  the  order  of  the  square root  of  the  errors  in  the
eigenvalues.  However,   in  FEM,   this  error  can   be  distributed
nonuniformly by emphasizing areas (or volumes) of interest through the
redistribution of elements, putting  more elements in those areas that
are of particular interest and fewer elsewhere. The detailed agreement
with the published results for  the square lattice of dielectric posts
shows  that the  HFEM  provides accurate,  reliable  results that  are
derivable with banded, sparse matrices.

The ability of the FEM  to represent complex geometries is highlighted
by considering the use of an Escher tessellation to define a unit cell
of a PC. The same  example treated with reciprocal space methods would
require  an  enormous number  of  Fourier  components  to capture  the
details of the geometry.

For HFEM to  be fully extensible across a wider  range of PC modeling,
it  will  be  necessary  to  demonstrate its  applicability  in  three
dimensions. This is straightforward with Hermite tetrahedral and brick
elements. As a  node based  method, HFEM  is  suited for
domain decomposition and can be connected with existing frameworks for
treating the open  domains of finite systems as  well. A final feature
that  would be  required in  this context  is the  exploration  of the
time-domain  evolution of  solutions. The  finite  element time-domain
techniques that are already  prevalent in modeling such structures can
readily  incorporated with the methods we  have reviewed in  this
paper.

In conclusion, we  can anticipate that the use of  HFEM will allow the
treatment  of  multiscale  and  multi-physics  problems  that  require
detailed spatial  descriptions of EM fields. With  the solutions given
on the vertices of the triangle and the continuity guaranteed both for
the  normal  and tangential  derivatives  at  triangle interfaces,  it
becomes substantially simpler  to mix scalar-vector field calculations
involving curl  operators. This  a distinct advantage  over reciprocal
space methods  as we have shown  through cases designed  to stress the
sophistication of the spatial reconstruction of fields.


\section{Acknowledgments}
We  thank J. D. Albrecht, C. R. Boucher, and S. Pandey for valuable discussions.   DNP  thanks
Worcester  Polytechnic  Institute  for summer  undergraduate  research
fellowships.  Computational  resources for the calculations presented in this article are provided by the Center for
Computational NanoScience  at WPI.




\end{document}